\documentclass{aastex62}

\newcommand{\gaia}{\textit{Gaia }}

\usepackage{amsmath}
\usepackage{placeins}

\begin{document}

\title{The Milky Way's Shell Structure Reveals the Time of a Radial Collision}

\author{Thomas Donlon II}
\affiliation{Department of Physics, Applied Physics and Astronomy, Rensselaer Polytechnic Institute, Troy, NY 12180, USA}
\author{Heidi Jo Newberg}
\affiliation{Department of Physics, Applied Physics and Astronomy, Rensselaer Polytechnic Institute, Troy, NY 12180, USA}
\author{Robyn Sanderson}
\affiliation{Department of Physics and Astronomy, University of Pennsylvania, Philadelphia, PA 19104, USA}
\affiliation{Center for Computational Astrophysics, Flatiron Institute, New York, NY 10010, USA}
\author{Lawrence M. Widrow}
\affiliation{Department of Physics, Engineering Physics and Astronomy, Queen's University, Kingston, ON K7L 3N6, Canada}

\begin{abstract}
We identify shell structures in the Milky Way for the first time. We find 2 shells in the Virgo Overdensity (VOD) region and 2 shells in the Hercules Aquila Cloud (HAC) region using Sloan Digital Sky Survey, \textit{Gaia}, and LAMOST data. These shell stars are a subset of the substructure previously identified as the Virgo Radial Merger (VRM). Timing arguments for these shells indicate that their progenitor dwarf galaxy passed through the Galactic center 2.7 $\pm$ 0.2 Gyr ago. Based on the time of collision, it is also possible that the VRM is related to the phenomenon that created phase-space spirals in the vertical motion of the disk and/or the Splash, and could have caused a burst of star formation in the inner disk.

We analyze phase mixing in a collection of radial merger $N$-body simulations, and find that shell structure similar to that observed in Milky Way data disappears by 5 Gyr after collision with the Galactic center. The method used to calculate the merger time of the VRM was able to reliably recover the correct merger times for these simulations.

Previous work supports the idea that the VRM and the \gaia Sausage/\textit{Gaia}-Enceladus Merger are the same. However, the \gaia Sausage is widely believed to be 8--11 Gyr old. The disparate ages could be reconciled if the larger age is associated with an infall time when the progenitor crossed the virial radius; we do not constrain the time at which the progenitor became bound to the Milky Way. Alternatively, the \gaia Sausage could be younger than previously thought. 
\end{abstract}


\section{Introduction} \label{sec:intro}

Overdensities in the Milky Way's stellar halo provide information about the shape of the halo's gravitational potential, and indicate that the outer portions of the Galaxy are not in equilibrium \citep{Ivezic2012}. One such overdense feature in the halo is the Virgo Overdensity (VOD), which was originally identified by \cite{Vivas2001} as an overdensity of RR Lyrae stars (RRLs) in the Virgo constellation. The VOD contains a wealth of stellar substructure, including the Sagittarius Stream \citep{Ibata2001}, the Virgo Stellar Stream \citep{Duffau2006}, the Parallel Stream \citep{Sohn2016, Weiss2018b}, the Perpendicular Stream \citep{Weiss2018b}, associated moving groups \citep{Duffau2014, Vivas2016}, and other minor structure such as the Cocytos Stream \citep{Grillmair2009, Donlon2019}. A thorough history and background of the VOD can be found in the introduction of \cite{Donlon2019}.

The Hercules Aquila Cloud \citep[HAC,][]{Belokurov2007} is another overdensity in the Milky Way's stellar halo. A common origin for the VOD and the HAC was proposed by \cite{Simion2019}, who noted that the VOD and the southern portion of the HAC were on opposite sides of the Galaxy and had similar kinematics. \cite{Simion2019} claimed that the VOD and the HAC are connected to the \gaia Sausage, a structure in Galactocentric velocity space that is characterized by a wide dispersion in radial velocity and a narrow dispersion in rotational velocity \citep{Belokurov2018b}. This signature velocity structure is thought to have been caused by a major merger in the Milky Way's early history. The two overdensities would then be the result of a massive, ancient, and highly radial merger event that has since been dubbed the \gaia Sausage Merger \citep[GSM,][]{Simion2019}, also known as the \textit{Gaia}-Enceladus Merger \citep{Helmi2018}. We choose to refer to the merger event as the GSM for the remainder of this work, though it should be acknowledged that the merger event was independently discovered by multiple groups. While \cite{Helmi2018} characterized the \textit{Gaia}-Enceladus Merger as a retrograde halo structure, \cite{Belokurov2019} classified the retrograde portion of the material in the halo as the ``Sequoia'', and not belonging to the Sausage.

The GSM is thought to have occurred between 8 and 11 Gyr ago \citep{Belokurov2018b, Helmi2018, Simion2019}. The main argument for this age of the GSM is that the merger is responsible for the creation of the thick disk, which is dated by an end of star formation in thick disk stars between 8 and 11 Gyr ago. If the GSM is indeed responsible for the puffing up of the thick disk, then the time that the thick disk was quenched would correspond to the time at which the GSM heated the disk. This timeline suggests a quiescent Milky Way, where our Galaxy experienced a few massive mergers early on in its history, but was mostly collision-free until recent times.

Around the same time that the VOD was connected to the GSM, \cite{Donlon2019} showed that a single orbit passed through the two largest moving groups identified by \cite{Duffau2014} in the VOD. This orbit was highly radial, with an apogalacticon of 26 kpc and a perigalacticon of 0.3 kpc. \cite{Donlon2019} evolved an $N$-body simulation of a single Sagittarius-sized dwarf galaxy along this radial orbit for 2 Gyr, and found that it simultaneously fit material previously attributed to the Perpendicular Stream, the Parallel Stream, the Virgo Stellar Stream, all of the associated moving groups, and some material previously thought to belong to the Sagittarius Stream. Thus, a single radial structure explained the majority of the substructure in the VOD. \cite{Donlon2019} named the radial merger event that created the VOD the Virgo Radial Merger (VRM), which is characterized as being responsible for a collection of stars in the VOD on radial orbits with a wide range of energies. In this work, we use the term ``VRM'' to refer to the merger event between the progenitor of the VOD and the Milky Way.

The VRM has been connected with other halo substructure besides the VOD. \cite{Li2016} identified the Eridanus-Phoenix Overdensity (EPO) in the south Galactic halo, and noted that the VOD, the HAC, and the EPO all lie on a single polar orbit. \cite{Donlon2019} found that a simulated VRM left debris in the regions of the VOD, the HAC, and the EPO, and proposed that a single radial merger could be responsible for all three overdensities. In that case, the overdensities would not share a single polar orbital plane, but would lie along the three ``spokes" of a single trefoil structure with perigalacticons within a kpc of the Galactic center. 

Figure 5 of \cite{Donlon2019} shows that the $N$-body simulation of the VRM left material in the local Solar region that was nearly identical to the characteristic shape of the \gaia Sausage. Additionally, the VRM left debris in the VOD and HAC, which is the same material that \cite{Simion2019} claims to be GSM debris. The halo is thought to be composed primarily of debris from a single radial merger event \citep{Deason2013, Belokurov2018b, Deason2019}. If this is the case, it is possible that the VRM and the GSM are the same, and that the single merger event is responsible for the majority of mass in the stellar halo. The counterargument comes from the timeline; \cite{Donlon2019} used simulations to show that the VRM could recreate the observed debris in the local Solar Neighborhood, the VOD, and the HAC if it occurred just 2 Gyr ago. This is 6--9 Gyr after the hypothesized time of the GSM. In this work, we show that the coherent structures associated with the VRM could not have survived for 8--11 Gyr. Thus, if the VRM and GSM are the same, the latter must have occurred more recently than originally thought.

Motivated by the disparity between the proposed ages of the VRM and the GSM, we seek a new method to identify the age of the VRM using shell substructure. Shells are common in elliptical galaxies, and are widely thought to be the artifacts of major radial merger events \citep{Hernquist1988, Sanderson2013}. They are named for their appearance as thin, extended ``umbrella''-like groups of stars at uniform Galactocentric radius. Shells arise at the turning points in the orbits of stars in the debris field of a radial merger event, and the stars in shells should therefore have near-zero Galactocentric radial velocity. We aren't able to measure the velocities of material in these shells outside of the Milky Way, which would help confirm this interpretation. However, such a full kinematic survey of shell stars is possible within the Milky Way. 

In this work we identify shell substructure in the Milky Way for the first time, and we argue that these shells are indeed associated with the VRM and therefore a radial merger event. Through analysis of $N$-body simulations of radial mergers, we develop a metric to describe how radial mergers evolve, and the timescales over which this occurs. We find that phase mixing places an upper limit of 5 Gyr for locating shell substructure in this radial merger; after that, the shells cannot be isolated in any of the simulations. Rewinding the particles in the VRM's shell substructure back to the time when the progenitor fell through the center of the Milky Way allows us to calculate an infall time of 2.7 $\pm$ 0.2 Gyr ago. This result is similar to the previously hypothesized age of the VRM from \cite{Donlon2019}.

\section{Data} \label{sec:data}

We construct two sets of observational data in the Milky Way halo; one for RRLs, and the other for Blue Horizontal Branch stars (BHBs). Each dataset contains full 6-dimensional phase space information for each star. This 6D information is calculated from distances derived from the presumed absolute magnitudes for the standard candles, proper motions obtained from \gaia Data Release 2 \citep[DR2,][]{Gaiamission, Gaiadr2contents}, and radial velocities determined from spectra obtained from the Sloan Digital Sky Survey Data Release 14 \citep[SDSS DR14,][]{SDSSIV, SDSSDR14} or from the LAMOST Experiment for Galactic Understanding and Exploration \citep[LEGUE,][]{LEGUE}. We utilized the method outlined in \cite{JohnsonSoderblom1987} to calculate 3D Galactocentric velocities from this information.

Many recent analyses of the stellar halo were performed in the local solar region using only \gaia data. By utilizing SDSS, we restrict ourselves to a much smaller region of the sky (notably, missing the EPO), but gain more accurate radial velocities at larger distances than would be possible using \gaia data alone. Additionally, using SDSS as well as \gaia allows us to utilize BHBs as tracers; selecting BHBs and calculating their distances requires spectroscopy and accurate \textit{ugriz} photometry.

The RRL dataset consists of objects classified as RRL stars in \gaia DR2 plus stars classified as RR Lyrae in the \cite{Liu2020} LEGUE catalog. We follow the process outlined in \cite{Iorio2019} in order to obtain the most RRLs possible and ensure clean data. This process consists of taking stars from the \gaia DR2 \verb!vari_rrlyrae! and \verb!vari_classifier_result! tables combined with data from the general \verb!gaia_source! table based on the \verb!source_id! of each star. The apparent magnitudes of the RRL stars in the variable star tables were calculated by modelling the light curves with a truncated Fourier series in order to determine the intensity-averaged mean magnitudes \citep{Neeley2019, Clementini2019}. The apparent magnitudes were then used to calculate the distances to each star, where the absolute magnitudes of all RRLs were assumed to be 0.63 in the \gaia G-band \citep{Muraveva2018}.  

Following the procedure used in \cite{Iorio2019}, we restrict our dataset based on two selection criteria: we enforce that \verb!astrometric_excess_noise! $<$ 0.25, and \verb!phot_bp_rp_excess_factor! $<$ 1.5. The former, \verb!astrometric_excess_noise!, describes the disagreement between observations of a source and the \gaia astrometric model, and is likely insignificant for values below 2 \citep{Lindegren2012}. The latter, \verb!phot_bp_rp_excess_factor!, is an estimation of background and contamination issues affecting the \gaia BP and RP photometry \citep{Riello2018}. Both of these criteria are found in the main \verb!gaia_source! table. \cite{Iorio2019} also enforce a constraint that the reddening must be low ($E(B-V) <$ 0.8). We do not enforce this cut, as we did not find any portion of our two fields to have an average reddening $E(B-V) > 0.9$ due to the fact that our fields are not positioned near the disk. We then matched our \gaia sample to all stars with spectra in SDSS DR14, using the TOPCAT software \citep{TOPCAT} to perform an on-sky match with a maximum tolerance of 1$''$ separation. 

\begin{figure}
\center
\includegraphics[width=\linewidth]{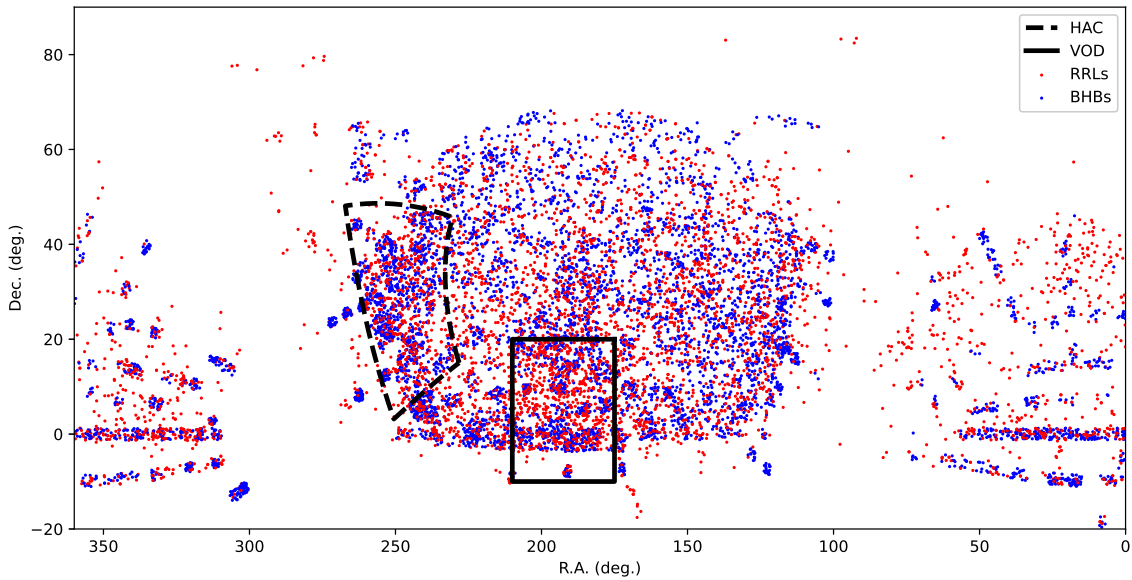}
\caption{An on-sky plot of our full 6D data set. Red points represent RRL stars, and blue points represent BHB stars. Extents of the VOD (solid line) and HAC (dashed line) fields are indicated. There are 6023 RRL stars and 5743 BHB stars in this figure, for a total of 11,766 stars between the two datasets. There are 1553 stars in the VOD region and 1480 stars in the HAC region. \label{fig:sky}}
\end{figure}

In order to maximize the number of available RRL stars, we also include stars that were identified as RRLs by \cite{Liu2020} using data from the LEGUE and SEGUE surveys. In the case of a star being identified as an RRL in both the \gaia data and the \cite{Liu2020} catalog, we opted to use the \gaia data. The \cite{Liu2020} catalog adds 3244 unique RRL stars to our dataset. Figure \ref{fig:sky} shows our final RRL dataset, containing 6023 stars.

We select BHB stars for our second dataset from SDSS DR14 photometry and spectroscopy. We use extinction-corrected color cuts of -0.25 $<$ (g - r)$_0$ $<$ 0, 0.8 $<$ (u - g)$_0$ $<$ 1.5 to identify BHBs \citep{Yanny2000}.  The color cuts select a specific temperature range of BHBs, and eliminate white dwarfs and QSOs. We also employ a surface gravity cut of 0 $<$ log$\, g_{WBG}$ $<$ 3.5 in order to eliminate blue straggler contamination \citep{Newberg2009}. The distance to each star was calculated using the absolute magnitude relation for BHBs \citep{Deason2011}: 
\begin{equation}
M_{g(BHB)} = 0.434 - 0.619(g-r)_0 + 2.319(g-r)_0^2 + 20.449(g-r)_0^3 + 94.517(g-r)_0^4.
\end{equation} We matched this data to \gaia DR2 proper motions in the same way that we matched the RRL dataset. Note that the color selection we used for BHB stars overlaps with the possible range of colors of RRL stars, so several stars classified as RRLs in Gaia also satisfied the BHB color cut. If a star was found in both datasets, the RRL data was retained, as the distance estimate for an RRL variable star ($\sim4\%$ distance error at a heliocentric distance of 20 kpc) is more accurate than the photometric distance estimation for a BHB star ($\sim10\%$ distance error). After removing the duplicate stars, our BHB dataset contained 5743 stars, for a total of 11,766 stars between the two datasets.

Figure \ref{fig:sky} shows the VOD and HAC regions, where we expect to find primarily VRM debris, in relation to the SDSS footprint. The two regions are selected based on visible overdensities in RRL and BHB populations in the north Galactic cap and previous literature. We define the VOD region to be 175$^\circ$ $<$ R.A. $<$ 210$^\circ$ and -10$^\circ$ $<$ Dec. $<$ 20$^\circ$, similar to the extent of the region given by \cite{Vivas2016} but extending to slightly higher declination. The HAC canonically exists in two parts: the HAC in the north (b $>$ 0$^\circ$) and the HAC in the south (b $<$ 0$^\circ$). Our data is limited to the SDSS footprint in the north Galactic cap, so we will only be considering the HAC region in the north. We define the extent of the HAC in the north to be 20$^\circ$ $<$ l $<$ 75$^\circ$ and 30$^\circ$ $<$ b $<$ 55$^\circ$. This is similar to the HAC region defined by \cite{Martin2018}. In order to minimize thick disk contamination we chose to only include data above $b=30^\circ$ , whereas \cite{Martin2018} included data above $b=20^\circ$. From this point forward, we will refer to the ``north portion of the HAC'' as the HAC region.

There are 908 RRLs (8\% of total RRLs in the sample) and 645 BHBs (5\% of sample BHBs) in the VOD region. There is a similar number of stars in the HAC region, with 675 RRLs (7\% of sample RRLs) and 805 BHBs (7\% of sample BHBs). This gives us 1553 stars in total in our VOD dataset and 1480 stars in total in our HAC dataset.

We derive our 6D phase space information for each star from standard candle estimates of distances, position on the sky, SDSS or LEGUE radial velocity, and \gaia proper motions. Error in radial velocity and \gaia proper motions are given in the survey data. We assume negligible errors in position on the sky ($\sigma_{\alpha}$ = $\sigma_\delta = 0$). Distance errors in RRLs are calculated using an absolute magnitude error of $\pm 0.08$ in the \gaia G-band  \citep{Muraveva2018}, and distance errors in BHBs as measured using SDSS photometry were approximated as 10\% of the distance value \citep{Martin2018}. We calculate errors in Galactocentric radius $r$, Galactocentric radial velocity $v_r$, and angular momentum $L_z$ with standard error propagation techniques. The average errors over both datasets are: $\sigma_r = \pm 0.6$ kpc, $\sigma_{v_r} = \pm 21$ km s$^{-1}$, and $\sigma_{L_z} = \pm 222$ kpc km s$^{-1}$. 

\section{Identifying Shells in the Milky Way} \label{sec:shells}

Radial collisions result in ``shells'' on opposite sides of the host galaxy's center at the location of each apogalacticon \citep{Hernquist1988}. In the idealized case where the host potential is spherical and the progenitor's orbit is exactly radial, the corresponding shells will be spherically symmetric, or ``umbrella shaped.'' In galaxies such as the Milky Way, the disk
is enough to break spherical symmetry. It is also widely believed that
dark halos are aspherical. The shells that arise in systems such as
these are, in general, less sharply defined and no longer lie at
surfaces of constant Galactocentric radius \citep{Hernquist1988, Sanderson2013}. Caustic structures that form in asymmetric potentials or from progenitors with nonzero angular momentum will typically have thicker ``blurred'' shells, but nevertheless share similar kinematic properties with their spherical counterparts \citep{Sanderson2013}. In the Milky Way, we expect to identify shells that are not spherically symmetric because the Milky Way's potential is not spherical, as is obvious from the presence of a significant disk component.

In this work we use the terminology laid out in \cite{Sanderson2013} regarding shell structure. The terms are defined from a top-down perspective: ``Radial merger'' describes a merger event with low angular momentum that will create shell structure. ``Caustic structure'' describes the entirety of the material with similar energy values in a radial merger that will cause a particular shell. A caustic structure is not only limited to material in a shell, but also includes material that is falling from a shell back towards the host galaxy, or is moving away from the host galaxy to form a shell. ``Caustic surface'' refers to the $r$--$v_r$ phase space surface along which material in a caustic structure lies. ``Shell,'' or ``shell (sub)structure,'' refers to the portion of the caustic surface with radial velocities near zero where stars bunch up to form the characteristic ``umbrella'' in position space.

One way to identify shell substructure is to plot the $r-v_r$ phase space distribution of stars, where $r$ and $v_r$ are the Galactocentric radius and Galactocentric radial velocity, respectively. In these coordinates, shells appear as parabolic or ``candy corn'' shaped structures \citep{Hernquist1988, Sanderson2013}. However, phase mixing on the order of even a few Gyr wraps the merger debris enough that the shells begin to overlap in phase space. Based on the velocity errors in our datasets, we do not have the necessary resolution to identify caustic substructure in the Milky Way using a phase space density approach. 

We will instead identify caustic structure in stars with near zero radial velocity, $v_r$, as measured from the Galactic center. By selecting stars in a small $v_r$ range around zero, we preferentially select stars at the surface of shells and remove stars in other structures at the same distances, as well as eliminate the interior portions of the radial merger debris, where the stars are not currently in shells.  This technique allows us to avoid fitting models to our velocity data, which has larger errors than the position data. In a histogram of the radial position from the Galactic center, $r$, of merger stars, peaks arise at the surface of each shell \citep{Hernquist1988}. This works in a spherically symmetric potential, but would be problematic in an axisymmetric potential, as any particular shell will not necessarily be located at the same $r$ everywhere in an axisymmetric potential \citep{Hernquist1988, Sanderson2013}. However, by restricting our analysis to a small region of the sky, any particular shell in that region will be located within a small range of Galactocentric radii, which allows us to recover the density peak. Since shells arise where the member stars are bunched up at apogalacticon, the $v_r$ of stars in a shell will be near zero. 

Other stellar structures with small $v_r$ may be located at the same distances as shells in the sky, notably structures on nearly circular orbits. Shells must have small angular momentum, because they are radial structures. In an axisymmetric potential we typically only consider $L_z$ angular momentum, as $L_x$, $L_y$, and $L$ are not true integrals of the motion. By cutting the data so that $|L_z| <$ 500 kpc km s$^{-1}$, we eliminate structures on non-radial orbits while retaining shell structure. In the VOD region, this eliminates Sagittarius Stream member stars from our data at higher distances. It should be noted that further angular momentum cuts may be required in datasets in other parts of the sky, as objects on highly polar, non-radial orbits will still have small $L_z$, but are not in shells.

\subsection{Developing a Shell Model} \label{sec:cau-model}

We utilize the analytical phase space density model for a caustic structure derived in \cite{Sanderson2013}, 
\begin{equation} \label{eq:psdensity}
f(r,v_r) \propto \frac{1}{r_s^{5/2} \Omega_s}\sqrt{\frac{\kappa}{\delta_r^2}} \: \exp \Big\lbrace \frac{-[r_s - r - \kappa v_r^2]^2}{2\delta_r^2} \Big\rbrace,
\end{equation} where $r_s$ is the distance of the shell from the Galactic center, $\kappa$ describes the curvature of the caustic surface, $\delta_r$ is the characteristic width of the shell, and $\Omega_s$ is the solid angle spanned by the shell. \cite{Sanderson2013} also include a term for the radial velocity of the debris at the surface of the shell in their model, $v_s$. This term is directly proportional to the total angular momentum of the progenitor galaxy. The total angular momentum of the VRM debris was shown to be very small by \cite{Donlon2019}, so we ignore the $v_s$ term in our analysis. It should be noted that we expect our shells to be relatively thick due to the axisymmetric potential of the Milky Way. The curvature of the caustic surface approximately depends on the strength of the underlying potential, \begin{equation}\label{eq:kappa}
\kappa \approx \frac{1}{|2g(\vec{r})|},
\end{equation} where $g(\vec{r})$ is the gravitational force towards the Galactic center at some position $\vec{r}$ with respect to the center of the Galaxy. 

In order to isolate shell stars, we wish to describe the density of stars with some $|v_r| < v_c$ in a caustic surface, where $v_c$ is chosen to select as large and as pure a sample of shell stars as possible. This selection will allow us to determine the corresponding shape of the shells in our radial density histograms. We integrate over our velocity range ($-v_c < v_r < v_c$) to find our density function:
\begin{equation} \label{eq:psd_integral}
\rho (r) = \int_{-v_c}^{v_c} f(r,v_r)dv_r
\end{equation} We do not require a complete computation of this expression, as we simply wish to determine its approximate form. Assuming that $v_c$ is small, we approximate the integral as a rectangle of width $2v_c$ and height $f(r,0)$,
\begin{equation}
\rho (r) \approx 2v_c \cdot f(r,0).
\end{equation} Evaluating this expression yields
\begin{equation} \label{eq:gaussian}
\rho (r) \propto \frac{2v_c}{r_s^{5/2}}\sqrt{\frac{\kappa}{\delta_r^2}} \exp \Big\lbrace \frac{-[r_s - r]^2}{2\delta_r^2} \Big\rbrace,
\end{equation} which is a Gaussian distribution, where the overall amplitude of the density peak depends linearly on our value of $v_c$. Thus, it is reasonable to approximate the radial density of stars in a caustic structure as a Gaussian for sufficiently small values of $v_c$. 

\begin{figure}
\center
\includegraphics[width=\linewidth]{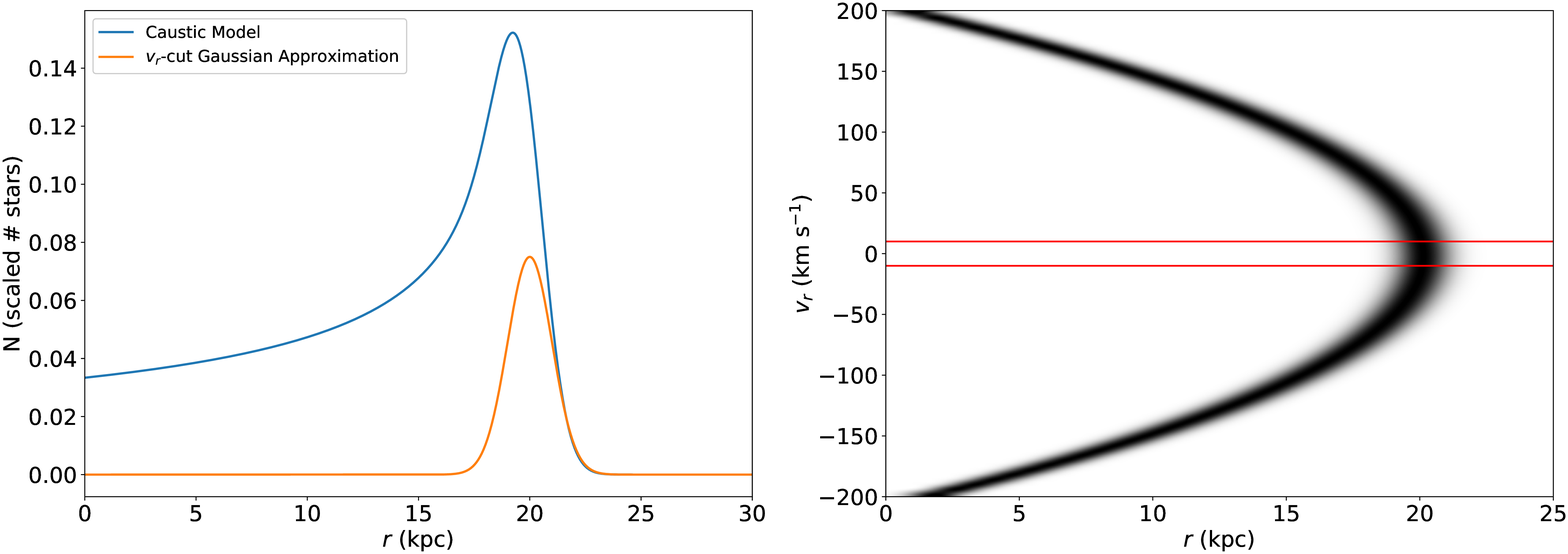}
\caption{Selection of shell structure from a caustic structure. \textit{LEFT:} We compare the radial density of the caustic model from \cite{Sanderson2017} (blue) and a Gaussian approximation to the portion of the caustic that is in a shell, as selected using $|v_r| < 10$ km s$^{-1}$ (orange). In this example, our model parameters are $r_s$ = 20 kpc, $\delta_r$ = 1 kpc, and $\kappa$ = 4.74$\times$10$^{-4}$ kpc s$^{2}$ km$^{-2}$. Smaller values of $\kappa$ (a steeper phase space curve) result in larger amplitudes in the corresponding Gaussian distributions. The center of the Gaussian is slightly offset to larger values of $r$ compared to the peak of the caustic distribution. Note that the Gaussian approximation has a smaller amplitude than the caustic model. This is due to removal of stars near apogalacticon in the caustic structure, but with velocities outside of $|v_r| < 10$ km s$^{-1}$. \textit{RIGHT:} We show the $v_r$-$r$ phase space distribution for the caustic model given in the left panel. The subset of the phase space used in the Gaussian approximation lies between the red lines. The maximum value of the approximation is less than the maximum value of the caustic model because stars at similar $r$ in phase space are excluded from the approximation by the $v_r$ cut. \label{fig:model}}
\end{figure}

Figure \ref{fig:model} shows a comparison of this Gaussian approximation and the original model. Note the slight difference in the radial location of the maximum of each model, and that the Gaussian distribution has a smaller maximum than the original model. This means that we have fewer stars available when looking at shells compared to entire caustic surfaces; the trade-off is that shells are substantially easier to isolate.

\subsection{Fitting the Model to Observed Data} \label{sec:fitting}

The requirement for the approximation performed in Section \ref{sec:cau-model} requires that $\kappa v_r^2 \ll |r_s - r|$. We determine that $\kappa$ = 4.74$\times$10$^{-4}$ kpc s$^{2}$ km$^{-2}$ after evaluating Equation \ref{eq:kappa} for the model potential we use for the Milky Way in this work (see Section \ref{sec:simulations}) at a distance of 25 kpc from the Galactic center and a distance of 25/$\sqrt{2}$ kpc above the plane of the disk. This is roughly the location of the VOD. A reasonable shell width, $\delta_r$, is on the order of 1 kpc \citep{Sanderson2013}, which corresponds to a typical separation of $|r_s - r| <$ 0.5 kpc. Solving for the requirement of the approximation, we find that $|v_r| <$ 30 km s$^{-1}$. Due to the size of the velocity errors in our dataset ($\sigma_{v_r} = \pm 21$ km s$^{-1}$), we require that $|v_r| <$ 10 km s$^{-1}$ in order to reduce contamination from material that is not actually in shells.

\begin{figure}
\center
\includegraphics[width=\linewidth]{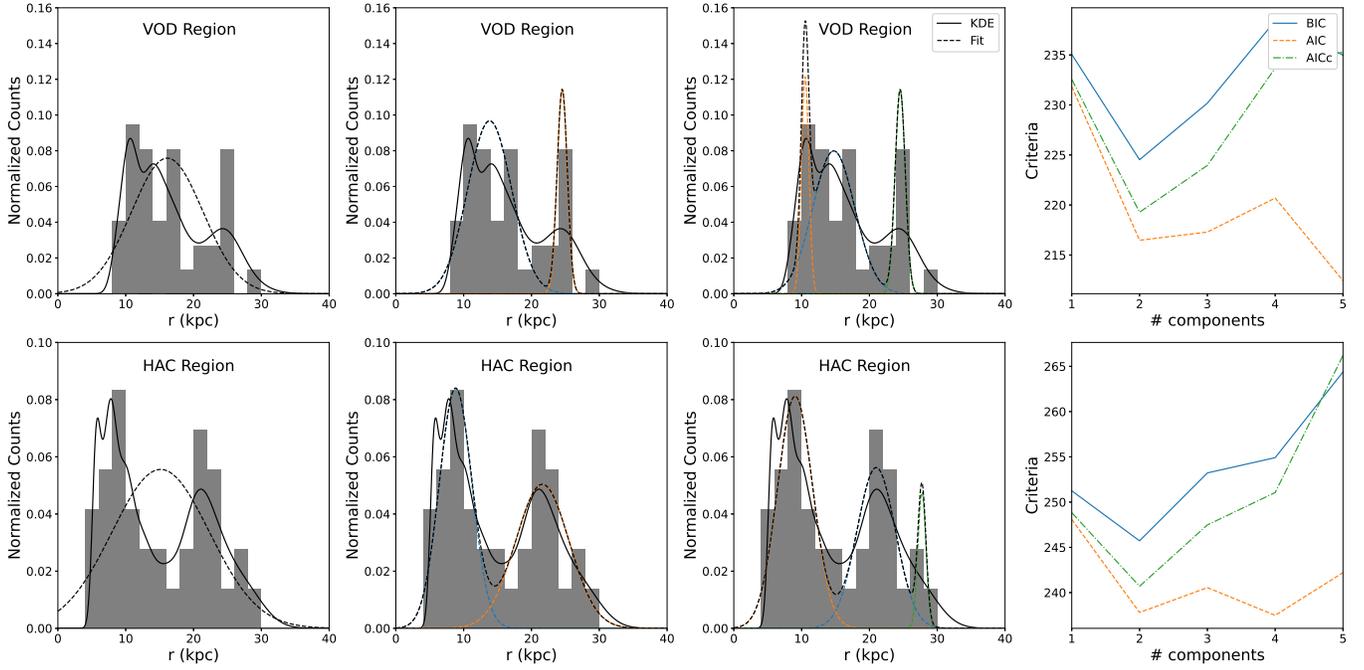}
\caption{Gaussian mixture model fits for the data in the VOD region (top row) and the HAC region (bottom row) using the EM algorithm. The first three columns show the one, two, and three component Gaussian models for both regions plotted over a histogram of the data (dashed lines). The data was not binned when computing the EM algorithm. A KDE of the data is shown in solid black lines, calculated by placing a Gaussian at the distance of each star with a standard deviation equal to the uncertainty in the distance value for that star. The KDE is scaled to match the normalization of the histogram. The rightmost column shows provides the BIC, AIC, and AICc values for models containing up to five components. Stars with $r>30$ kpc are not shown here, as they were not considered in the EM algorithm fits. }  \label{fig:gmm}
\end{figure}

Figure \ref{fig:gmm} shows histograms of Galactocentric radius, $r$, for the observed data, including all of the sample stars in both the VOD and HAC fields with $|v_r| < 10$ km s$^{-1}$ and $|L_z| < 500$ kpc km s$^{-1}$. We acknowledge that our radial velocity cut likely removes many stars that are actually in shells from the datasets due to our large radial velocity errors; the entire range of velocity in the selected data is about the same as the one sigma velocity error. Cutting in $L_z$ is less problematic, as our error in angular momentum is less than a quarter of our cut range. The candidate shell star data can be found in Tables \ref{tab:voddata} and \ref{tab:hacdata}.

In Figure \ref{fig:gmm}, Gaussian mixture models are fit to the cut candidate shell data using the \textbf{scikit-learn} \\ \verb!sklearn.mixture.GaussianMixture! implementation \citep{scikit-learn} of the expectation-maximization (EM) algorithm \citep{Dempster1977}. A kernel density estimation (KDE) of the data is shown in Figure \ref{fig:gmm} to provides a smooth approximation of the underlying density of the sample that is not dependent on binning, and is a better representation of what the fitting algorithm ``sees''. The radial density of each Gaussian is modelled as \begin{equation}
f(r) = a\exp\Big\lbrace \frac{-(r-b)^2}{2c^2} \Big\rbrace,
\end{equation} where the values of $a, b$, and $c$ are determined by the fitting algorithm.

Since the true number of Gaussian components in our data is not clear, we use the EM algorithm to fit Gaussian mixture models with up to 5 components to the data. Each quality of each fit was then assessed with its associated Bayesian Information Criterion \citep[BIC,][]{bic} and Akaike Information Criterion \citep[AIC,][]{Akaike1974}. These information criteria help prevent overfitting or underfitting the data. By adding another Gaussian component to a Gaussian mixture model, one will generally find that the model is a better fit to the data. The BIC and the AIC weigh the improvement in the likelihood of the fit against a penalty for adding additional parameters. The model where the information criteria are minimized is then the most statistically significant result.

Both the BIC and the AIC suffer from the assumption that the number of data points is much larger than the number of parameters in the model. The small number of data points in our sample impacts the validity of the BIC and AIC values, particularly when the number of Gaussians is large. The corrected Akaike Information Criterion \citep[AICc,][]{Hurvich1989} is an adjustment of the AIC for small numbers of data points. We find that this value is a better indicator of the quality of a model than the AIC. The BIC is less affected by the small number of data points, and we find that the BIC agrees with the AICc for models with fewer than five Gaussian components.

The EM algorithm preferentially fit Gaussians to individual stars at large distances instead of the interior structures with many stars. Finding a shell with only one star in it is not reasonable, so we omitted datapoints with $r \geq 30$ kpc while fitting the Gaussian mixture models.

Comparing the information criteria of the data for Gaussian mixture models with varying numbers of components, we find that both the VOD and HAC datasets are best modeled by only two Gaussians. This corresponds to four statistically significant shell structures in total. Our Gaussian mixture model fit to the candidate shell star data suggests the existence of shells in the VOD region at $r$ = 13.8 kpc and 24.5 kpc, and in the HAC region at $r$ = 8.8 kpc and 21.6 kpc. These results show that the shells in the HAC region are closer to the center of the Galaxy than the shells in the VOD region. This supports the idea that the shells are all formed from debris of the same merger event; one expects shells to arise at different distances on opposite sides of the Galaxy, since the material in one shell must have a different energy than the material in the other shells.

In the case of a smooth halo with no substructure, we would expect to see a single Gaussian distribution. If enough cuts are made to the data, the single Gaussian distribution could be quite noisy due to small number statistics. The EM algorithm produces a best fit single Gaussian at a distance of 15 kpc in the VOD region, and 16 kpc in the HAC region. While the information criteria of the data are minimized for a two component model, we wish to strongly rule out the possibility that the data comes from a single Gaussian. To this end we use two tests: the Anderson-Darling Test \citep{adtest, adtest2} and Hartigan's Dip Test \citep{diptest}. Both of these tests are designed to evaluate the likelihood that a given distribution is derived from a single normal distribution. The Anderson-Darling test yielded $p<0.01$ for both regions, meaning that we can confidently reject the null hypothesis that either region is composed of data derived from a single Gaussian. The Hartigan's Dip Test produced $p=0.16$ for the VOD region data, and $p=0.15$ for the HAC region, providing a $\sim 85\%$ chance that either region is derived from more than one Gaussian distribution. Between these two statistics, we claim that the shell candidate stars make up substructure in the halo, and are not simply a smooth halo background.

\begin{figure}
\center
\includegraphics[width=\linewidth]{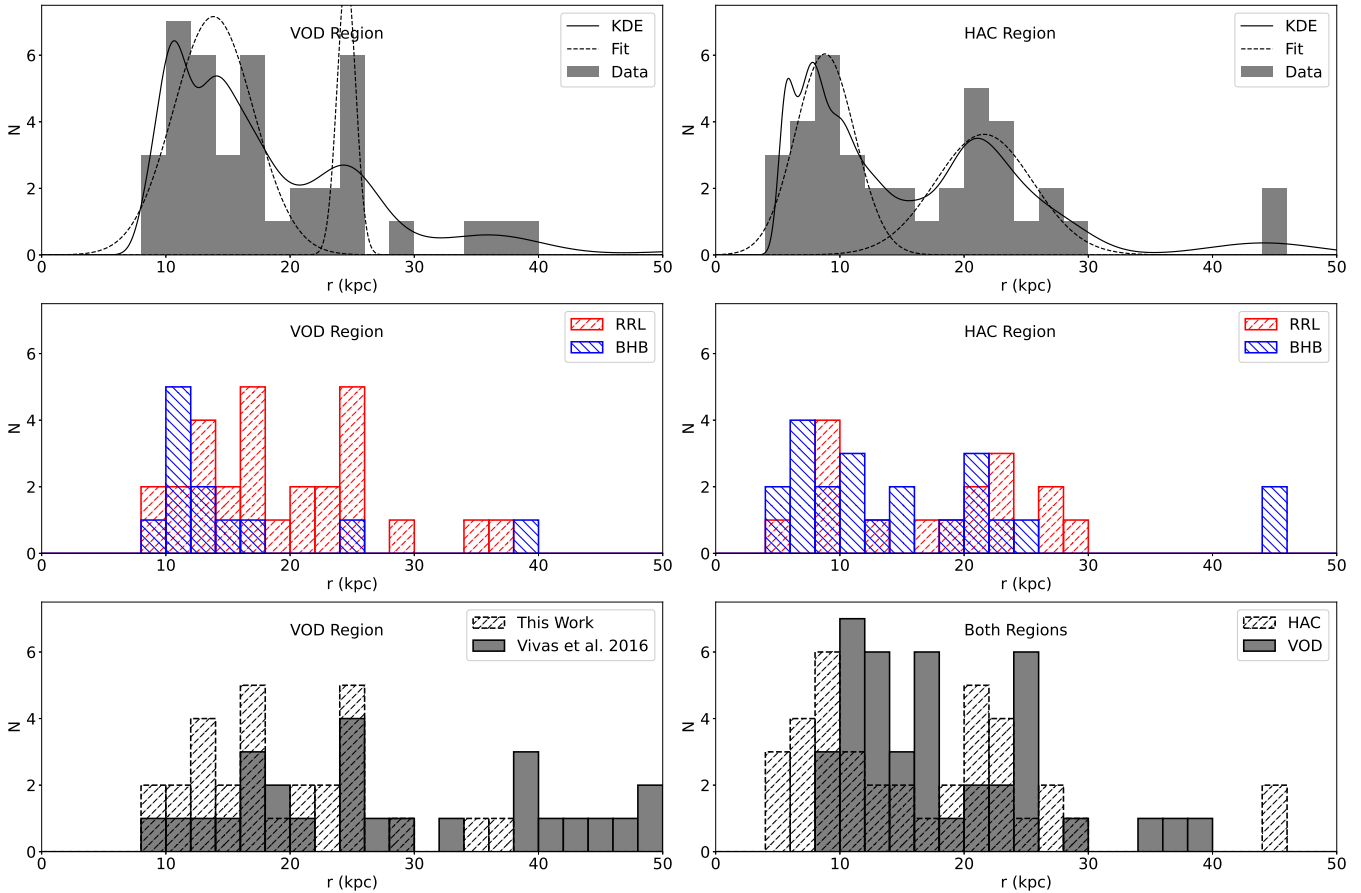}
\caption{\textit{TOP:} Histograms of the number of candidate shell stars in the VOD (left) and the HAC (right) regions, as a function of Galactocentric radius (solid gray). The data has been cut to satisfy $|v_r| < 10$ km s$^{-1}$ and $|L_z| < 500$ kpc km s$^{-1}$. Plotted on top of this data in dashed lines are our best Gaussian mixture model fits for shells in the data. The left panel shows the two fit (binned) Gaussians with parameters $(a, b, c)$ = (7.2 stars kpc$^{-1}$, 13.8 kpc, 3.2 kpc) and (8.4 stars kpc$^{-1}$, 24.5 kpc, 0.7 kpc). The right panel shows the two fit Gaussians with parameters $(a, b, c)$ = (6.2 stars kpc$^{-1}$, 8.8 kpc, 2.4 kpc) and (3.8 stars kpc$^{-1}$, 21.6 kpc, 3.93 kpc). A KDE of the data is shown in solid black lines, calculated as in Figure \ref{fig:gmm}. \textit{MIDDLE:} Histograms of the data separated into RRL (red) and BHB (blue) populations.  \textit{BOTTOM:} On the right, the RRL distibution for the VOD region (dashed white bars) are compared to the RRL data from the \cite{Vivas2016} catalogue after identical velocity cuts. On the left, the data for the two regions are superimposed.\label{fig:data}}
\end{figure}

Figure \ref{fig:data} shows histograms of the candidate shell stars against the best fits from the EM algorithm. The middle row of Figure \ref{fig:data} shows the data after velocity cuts have been applied to select only shell stars, split into separate distributions of RRL and BHB stars. Although BHB stars  make up a much smaller percentage of the sample in the VOD region than in the HAC region, the distributions for different types of stars in each region appear to be similarly distributed in Galactocentric radius. Slight differences in the distributions of the different types of stars may be due to distance errors, which are about the size of a bin width beyond 20 kpc for BHB stars, and around half a bin width for RRL stars. Curiously, the shell at $\sim25$ kpc in the VOD region appears to be composed almost entirely of RRL stars. A comparison of the candidate shell stars in both regions shows that the shells in the VOD region do not lie at the same distance as the shells in the HAC region. We note that BHB stars have distance errors of around 10\%. At 20 kpc from the Galactic center, a 10\% error can move the stars over in the histogram by as much as an entire bin. The slight differences in the distances in the peaks of the RRL and BHB distributions in the HAC region may be due to the errors in distance, or simply due to small number statistics.

Some stars in the datasets that were identified as RRL stars also satisfied the constraints for BHB stars (Section \ref{sec:data}). We chose to take the RRL distance data for these stars, as they are probably more accurate than the corresponding BHB distance calculation. However, we want to identify how our analysis changes if these stars are actually BHBs and not RRLs. There were a total of 1994 stars in our data (16.9\% of all stars) that were identified as both RRL and BHB stars. Out of these stars, 371 (3.2\% of all stars) had a difference between the RRL distance calculation and the BHB distance calculation of greater than 0.1 kpc. For these 371 stars, the mean difference between the calculated distances from the Galactic center is 3.9 kpc, and the standard deviation in the distance differences is 3.4 kpc.

We then recalculated which of these stars are located in shells if we use the BHB distances instead of the RRL distances. In the VOD region, two stars in the 18-20 kpc bin and one star in the 14-16 kpc bin are removed. The VOD region gains one star in the 14-16 kpc bin and two stars in the 12-14 kpc bin. In the HAC region, one star is removed from the 24-26 kpc bin and two stars are removed from the 26-28 kpc bins. The HAC region gains one star in the 20-22 kpc bin and two stars in the 18-20 kpc bin. If we use the BHB distances for the stars in this work that are identified as both types of star, the interior shell of the VOD and the exterior shell of the HAC become more pronounced. Since it seems more likely that stars identified as variable are RRL rather than BHBs, we choose to adopt the RRL distance values even though it makes the shells slighly less obvious.

\cite{Donlon2019} showed an excess of material at 40 kpc from the Sun in the VOD, and \cite{Fardal2019} shows an ``Outer Virgo Overdensity'' at $\sim$ 75 kpc from the Sun. The data from \cite{Vivas2016} in the bottom right panel of Figure \ref{fig:data} shows stars with small $v_r$ at $\sim$40 kpc from the Galactic center that extend out to larger distances. These structures may be outer shells formed from the VRM in the VOD region, but we are not able to analyze them in this work due to the lack of 6D information for those stars. $L_z$ and $v_r$ errors are large at these distances, which inhibits our ability to properly locate distant shell substructure in observed data. These distant shells are not incompatible with our current models, as the VRM $N$-body simulation from \cite{Donlon2019} placed merger material out at 40-70 kpc from the Sun in the VOD region, and shells are seen at out to 60 kpc from the Galactic center in the simulations described later in this work. We predict that simulations of a progenitor falling in from the virial radius would also form shells at large $r$ values from material in its trailing tidal tail.

\subsection{Radial Velocity Errors of RR Lyrae Variable Stars}\label{sec:rrlvelocities}

When considering stellar velocities, it is important that we recognize that RRL stars are pulsating variable stars; light from these stars will have Doppler shifts due to both translational motion and the time dependent expansion and contraction of the surface of the star. These pulsational velocities can be as large as 120 km s$^{-1}$ (see Figure 2 of \citealt{Sesar2012}, Figure 5 of \citealt{Duffau2014}, and Figure 2 of \citealt{Vivas2016}). The radial velocities of the RRL stars in our catalog that are not from the \cite{Liu2020} catalog are derived from SDSS stack spectra. Since RRL stars are typically targeted for SDSS spectra as potential BHB or quasar candidates based on color, SDSS radial velocities are not calculated with pulsating variable stars in mind. This means that our measured velocities, based only on the Doppler shift, could be wildly different than the actual velocities of the stars. 

\cite{Vivas2016} created a data set of RRL stars in the VOD in a way that cuts out stars with wildly erroneous radial velocity measurements. In this dataset, the majority of RRL stars had radial velocity uncertainties below 10 km s$^{-1}$. We matched this data set with \gaia data in order to obtain 3D velocities, and then cut it so that $|v_r| < 10$ km s$^{-1}$ and $|L_z| < 500$ kpc km s$^{-1}$ in order to isolate shell structure. This is compared with our RRL data in the VOD region in the bottom left panel of Figure \ref{fig:data}. In this figure, the \cite{Vivas2016} data implies the existence of shells in the VOD region at the same distances as the two statistically significant shells that were identified in RRL and BHB stars.

One might be concerned that large radial velocity errors would confuse the selection of shell stars using Galactocentric radial velocities. However, the velocity errors just from measurement are already quite large, of the order of the random velocity errors due to pulsation of the RRLs. Many stars that are actually in the shell are not selected due to random error or RRL pulsation, reducing our shell signal. Stars that are actually up to 120 km s$^{-1}$ away from $v_r = 0$ could be erroneously included in the sample. However, on average, the radial velocity measured from stack spectra will not be 120 km s$^{-1}$ off from the actual radial velocity measurement, since the pulsational velocity is only that large for a small portion of the RRL oscillation period, and stack spectra usually include measurements from several different epochs. In practice, the maximum difference between a stack spectra with multiple measurement epochs and the actual radial velocity is closer to 50 km s$^{-1}$, and the actual error is usually even smaller (see Figure 3 of \citealt{Vivas2016}). The right panel of Figure \ref{fig:model} shows that within the same caustic structure, stars that actual have Galactocentric radial velocities of $v_r = \pm$50 km s$^{-1}$ would only be $\sim$2 kpc closer than the actual shell structure, which is within the systematic distance error for stars in our dataset beyond 20 kpc from the Sun. 

The radial velocities of RRL stars from the \cite{Liu2020} catalog are calculated by fitting the observed velocity as a function of phase to an empirical pulsating velocity template curve. It is expected that this will reduce the error in radial velocity due to pulsational velocity compared to calculations of radial velocity from stack spectra. In order to test the differences between radial velocities that are calculated using stack spectra and velocities that are calculated by fitting observed velocity to an empirical velocity curve, we looked at the sample of 369 RRL stars that had measured spectral radial velocities in both the \textit{Gaia} RRL catalog and the \cite{Liu2020} catalog. The radial velocities of both catalogs are generally consistent; the mean difference between the two catalogs' radial velocities is 21 km s$^{-1}$, and the standard deviation of the differences is 15 km s$^{-1}$. The mean difference between the catalogs' radial velocities is roughly the same size as the upper limit for the estimated error in the radial velocity from either catalog. This suggests that a sizeable fraction of the differences between the two catalogs' radial velocities are due to measurement uncertainties and not necessarily due to pulsational velocities at the surface of a star. While radial velocities of RRL stars derived from stack spectra are likely less accurate than radial velocities derived from empirical velocity curves, the typical departure of the radial velocity value derived from stack spectra from the ``correct'' value is in practice much smaller than the 120 km s$^{-1}$ upper limit for the magnitude of the pulsational velocity. We maintain that radial velocities derived from stack spectra for RRL stars are fairly good estimates for the actual radial velocities of those stars, although they should be used with caution.

RRLs that are not associated with any VRM caustic structure and also have low $L_z$ angular momentum are relatively uncommon, especially in the VOD and HAC regions of the sky. This explains why we are able to recover the shells even though many RRLs could have large radial velocity errors. The similarity between the BHB and RRL data, and the positions of shells from RRLs in \cite{Vivas2016} data in the VOD support the conclusion that the RRL radial velocity errors that are derived from SDSS spectra do not significantly impact the outcomes of this work.

It should be noted that using the radial velocity values from the \cite{Liu2020} catalog in place of the radial velocity values from the \textit{Gaia}/SDSS catalog did not substantially change the histogram of stars (Figure \ref{fig:data}) that satisfy the Galactocentric radial velocity and angular momentum cuts. This is consistent with our analysis that the radial velocities of RRL stars derived from stack spectra that we use in this work are typically not very far off from the ``correct'' radial velocity measurements, and that the errors in the measured radial velocity values are often just as large as the difference between the stack spectra radial velocity measurement and the velocity curve fit radial velocity measurement. This may not be the case for all datasets, as the quality of the radial velocities derived from stack spectra depend on several factors, such as the number of observations, the quality of the spectra, and the times at which the observations are made.

\section{Timing the VRM: Phase Mixing Constraints} \label{sec:constraint}

\subsection{Radial Merger Simulations} \label{sec:simulations}

In order to explore phase-mixing in radial mergers, we create a suite of 120 $N$-body simulations of radial mergers. These simulations are all initialized with a single Plummer profile progenitor dwarf galaxy \citep{Plummer1911}. A third of these simulations use a dynamical mass of $10^9 M_\odot$ and a scale radius of 3 kpc, which is consistent with measurements of typical dwarf galaxies such as the Fornax dSph \citep{Penarrubia2008}. These parameters have been used previously as likely parameters for the progenitor of the VRM by \cite{Donlon2019}, who claimed that an $N$-body simulation with this mass and scale radius more accurately recovered the kinematics of the VRM than simulations with a mass of $10^7 M_\odot$ and a scale radius of 0.4 kpc. In order to explore the effect of progenitor mass on radial merger phase mixing, we also perform simulations with a dynamical mass of $10^8 M_\odot$ and a scale radius of 1 kpc, as well as simulations with a dynamical mass of $10^{10} M_\odot$ and a scale radius of 10 kpc. The progenitors in these simulations are given zero initial velocity, so that the initial energy of the dwarf galaxy is determined solely by its initial distance from the Galactic center, and the merger is guaranteed to be nearly radial. 

Since we are starting the dwarf galaxy at rest in an axisymmetric potential, there are only two parameters required to create the full range of substructure for a particular dwarf model. These are the initial inclination angle ($i$) from the Galactic disk as viewed from the Galactic center, and the initial distance from the Galactic center ($r_0$). The inclination angle is defined so that $i$ = 0$^\circ$ in the disk, and $i=90^\circ$ when the dwarf galaxy is positioned along the Galactic Z-axis. The values we explored for $i$ range from 0$^\circ$ to 90$^\circ$ in increments of 10$^\circ$, and the values we explored for $r_0$ are 20, 30, 45, and 60 kpc. The observed VRM debris has an $L_z$ distribution centered on zero \citep{Donlon2019}, so we did not run simulations with nonzero angular momenta. We used a value of $Y$ = 0 for the initial position of the dwarf galaxy, as the azimuthal angle of impact does not change the simulation due to the axisymmetric symmetry of the Galaxy.

The simulations were run on the MilkyWay@home $N$-body software \citep{SheltonThesis} in a static gravitational potential consisting of a Hernquist bulge, Miyamoto-Nagai disk, and a logarithmic halo using potential parameters from Orphan Stream Model 5 in \cite{Newberg2010}. Each simulation contained 20,000 bodies, and was integrated forwards in time for 10 Gyr. For the rest of this work, we will refer to each simulation solely based on its progenitor's mass, initial inclination angle, and initial distance from the Galactic center. When necessary, the results of the simulation are also identified by a timestep.

Note that it is likely that the progenitor of the VRM was originally located at or near the Milky Way's virial radius, and then fell into the Milky Way along a decaying orbit. As the orbit decayed, each apogalacticon of the dwarf galaxy's orbit would be located closer to the Galactic center than the previous apogalacticon. Eventually, the dwarf galaxy collided with the Milky Way; this occurred after the final apogalacticon on the dwarf's orbit, which determines the final orbital energy of the progenitor dwarf galaxy. After the dwarf galaxy passed through the Milky Way, its constituent stars became bound solely to the Milky Way, and their individual energies were locked in. We need only to constrain the distance of the final apogalacticon pass of the progenitor of the VRM in order to characterize the present day kinematics of the structure, which corresponds to $r_0$.

\subsection{Causticality: A Metric for Phase Mixing} \label{sec:causticality}

\begin{figure}
\center
\includegraphics[width=\linewidth]{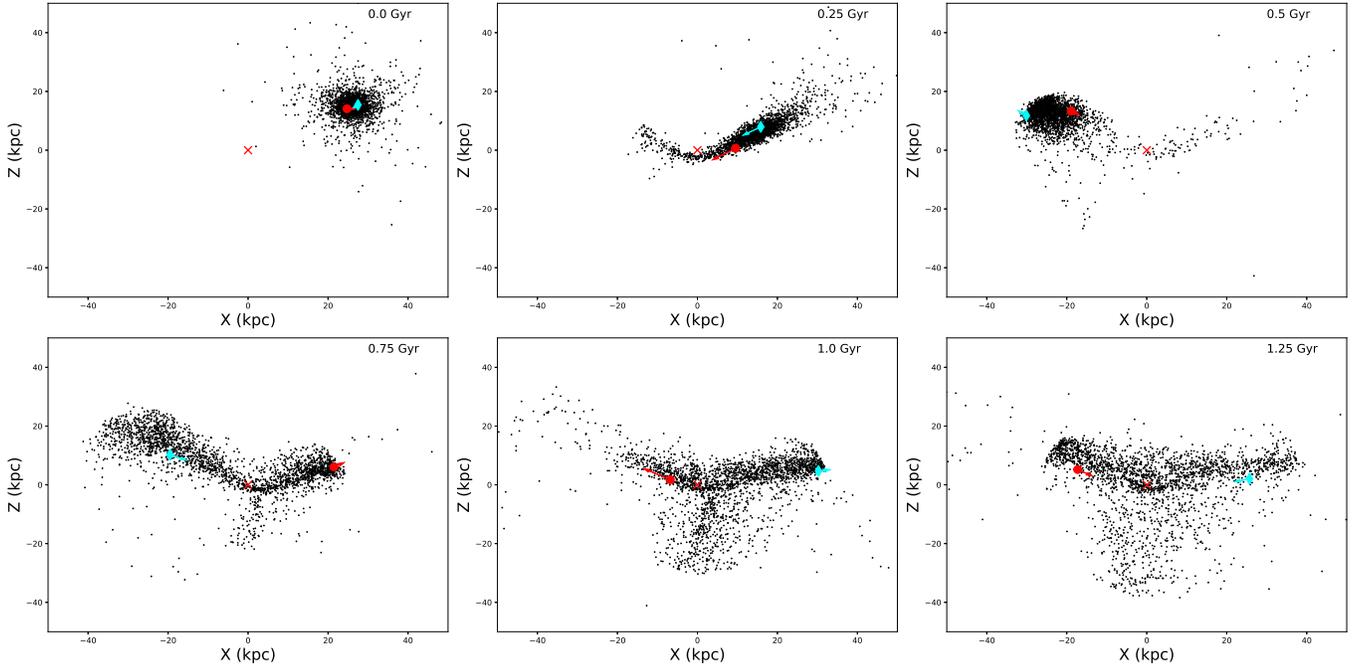}
\caption{Galactocentric $Z$ vs. $X$ for six timesteps of an $N$-body simulation of a radial merger in a model Milky Way potential. The six panels show a progression of timesteps spanning 1.25 Gyr. The initial parameters for this simulation are $i$ = 30$^\circ$, $r_0$ = 30 kpc, and a mass of $10^9 M_\odot$ (see Section \ref{sec:simulations}). The black dots show a random sample of one quarter of the bodies in the simulation. The red ``x'' marks the location of the Galactic center. We follow two particles, a red circle and a cyan diamond, throughout time;  the relative magnitudes and directions of the velocities of these particles are shown with arrows. At the beginning of the simulation, the red particle is at a lower initial energy than the cyan particle. In the next panel, they are both falling inwards towards the Galactic center. By the top right panel, the red particle has reached apogalacticon and begun falling inwards, while the cyan particle is moving outwards towards apogalacticon (in a shell). In the bottom left panel, one sees that the red particle is located in a shell on the right while the cyan particle is now beginning its second infall. By this point, there are already two shells formed in the simulation. In the bottom panels, the two particles are located in different shells. The period of oscillation for the cyan particle is longer than that of the red particle, since the cyan particle had a larger initial energy. Note how two shells are located above the Galactic plane, corresponding to our HAC and VOD regions, and that this simulation predicts a third location for shells below the Galactic plane, which could correspond to the EPO. This simulation demonstrates how shells form, and explains how the VOD, HAC, and EPO could be related through a single ``trefoil'' structure. Not all particles oscillate between all three locations (for example, the red and cyan model particles do not move into the shell located at negative $Z$ values). The shells corresponding to the VOD and HAC are both above the plane due to the influence of the disk; in a spherically symmetric potential, two sets of shells would be diametrically opposed. \label{fig:model_particles}}
\end{figure}

Simulated radial mergers experience phase mixing of the dwarf galaxy debris over large timescales.  Figure \ref{fig:model_particles} shows how small variations in the initial energies of stars in a radial merger will cause the material in shells to segregate based on energy. This phase-wrapping causes more shells to form as the age of the radial merger increases. Phase mixing is also caused by slight tangential velocities of stars in each shell due to variations in angular momentum. This causes shells to grow in surface area over time. Eventually, debris from a merger reaches a phase-mixed equilibrium within the Galaxy.

\begin{figure}
\center
\includegraphics[width=\linewidth]{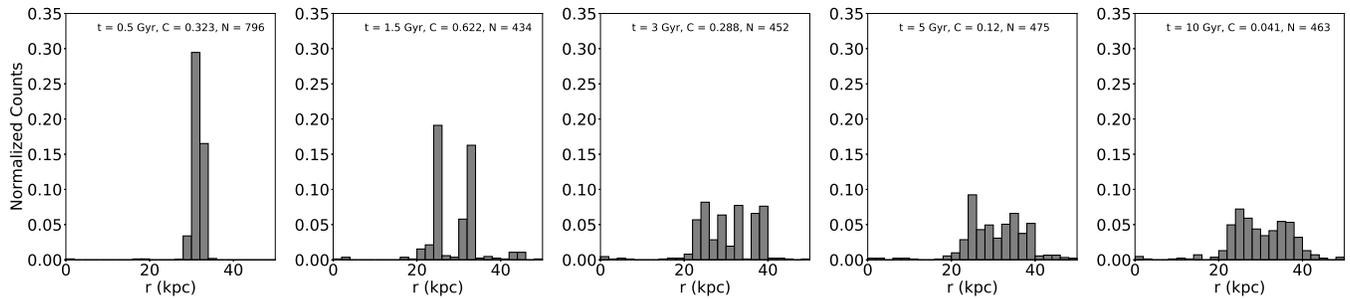}
\caption{Radial density histograms for the radial merger simulation with $i$ = 30$^\circ$, $r_0$ = 30 kpc, and a mass of $10^9 M_\odot$. The plotted data includes only particles with $|v_r| < 10$ km s$^{-1}$, $|L_z| < 500$ kpc km s$^{-1}$, and located within a 50 degree by 50 degree region of the sky in order to mimic the observed data. Each panel shows the data at a fixed time in the simulation; the times range from 0.5 Gyr to 10 Gyr. Note the strong peak at 0.5 Gyr. By 1.5 gyr, there are two peaks, corresponding to two shells. After 3 Gyr, the number of shells has grown large enough that individual peaks are no longer apparent. The value of the causticality decreases over time as the peaks become less distinct. By 5 Gyr, the simulation is fairly phase mixed, with a small causticality. \label{fig:sim}}
\end{figure}

Figure \ref{fig:sim} shows the simulation of a radial merger with initial inclination angle $i$ = 30$^\circ$, $r_0 = 30$ kpc, and a mass of $10^9 M_\odot$. The data for this simulation was cut to mimic our data shown in Figure \ref{fig:data}. This simulation shows that over time, the strong peaks demonstrating shells phase mix into a smooth distribution. When a merger has phase mixed completely, shells will no longer be present (that is, the number of caustic surfaces eventually approaches infinity and it is impossible to detect any one shell in particular). 

We seek an objective way to describe the amount that a radial merger has phase mixed from its initial state. To this end we develop a numerical metric for a histogram $h$, which we call the ``causticality'' $C$ of that histogram. With $n$ bins in $h$, and $N_i$ counts in the $i^{\textrm{th}}$ bin of $h$, we define causticality as

\begin{equation}\label{eq:causticality}
C = \frac{\sum\limits_{i=2}^n (N_i - N_{i-1})^2}{\sum\limits_{i=2}^n (N_i + N_{i-1})^2}.
\end{equation} Causticality consists of a ratio of the sum of squared differences between each bin and its previous neighbor and the sum of squared sums of each bin and its previous neighbor. The denominator in Eq. \ref{eq:causticality} imposes that $C = C(m \cdot h)$ for any constant $m>0$, which ensures that the causticality only depends on the shape of the input histogram and not the total number of data points in the histogram.

Causticality can take on all values between zero (completely mixed) and unity (completely unmixed). Consider a distribution with particles in only one bin, or a distribution that has undergone no phase mixing; the measured causticality of this system would be equal to unity. This is also the case for a system where every other bin is empty. On the other end, a uniform distribution has a causticality of zero. $C$ is larger for distributions with sharp peaks than distributions with gradual, smooth peaks: Figure \ref{fig:sim} shows how phase mixing lowers the measured value of the causticality over time in a simulation of a radial merger.

In practice, the measured causticality of our systems do not ever actually become zero. This is because the debris of radial mergers relax into a smooth Gaussian distribution, not a uniform distribution. Additionally, the invariance of causticality with respect to the number of particles assumes that a constant bin size is being used. Typically, bin size decreases as more objects are added to a histogram, which can alter the measured value of causticality. Thus, it is important that the bin sizes for the observed data and the simulated data are identical.

If observable shells are present in our data, we would expect to see a few strong, sharp peaks. As radial mergers age, these few thin shells become wider and more numerous until the overall radial distribution approaches a smooth distribution. We therefore expect the causticality to decrease over time for each simulation of a radial merger, until the merger is completely phase mixed, at which point the causticality assumes some small constant value.

Appendix \ref{app:cau_uncer} contains an analysis of the uncertainty in causticality, as well as the effects of bin size on measured causticality values. We determine that 2 kpc bin sizes are best for this data.

This is the first time that causticality has been introduced, and we recognize that it is beneficial to perform a similar analysis of phase mixing using a widely-studied statistic. In Appendix \ref{app:kld}, we use Kullback-Leibler Divergence \citep[KLD,][]{Kullback1951} instead of causticality to evaluate how phase mixed a distribution is. We find that it is more difficult to get an estimate on the merger time of the VRM from measurements of the KLD, but we are still able to determine that the VRM must have occurred within the last 5 Gyr. One advantage of causticality over the KLD is that causticality does not require comparison to some assumed baseline distribution: while the KLD as it is applied in this work only approaches zero when the distribution is uniform, the causticality rapidly approaches zero as a distribution becomes smooth regardless of its shape.

\subsection{Constraining the VRM Merger Time}\label{sec:pmconstraints}

\begin{figure}
\center
\includegraphics[width=\linewidth]{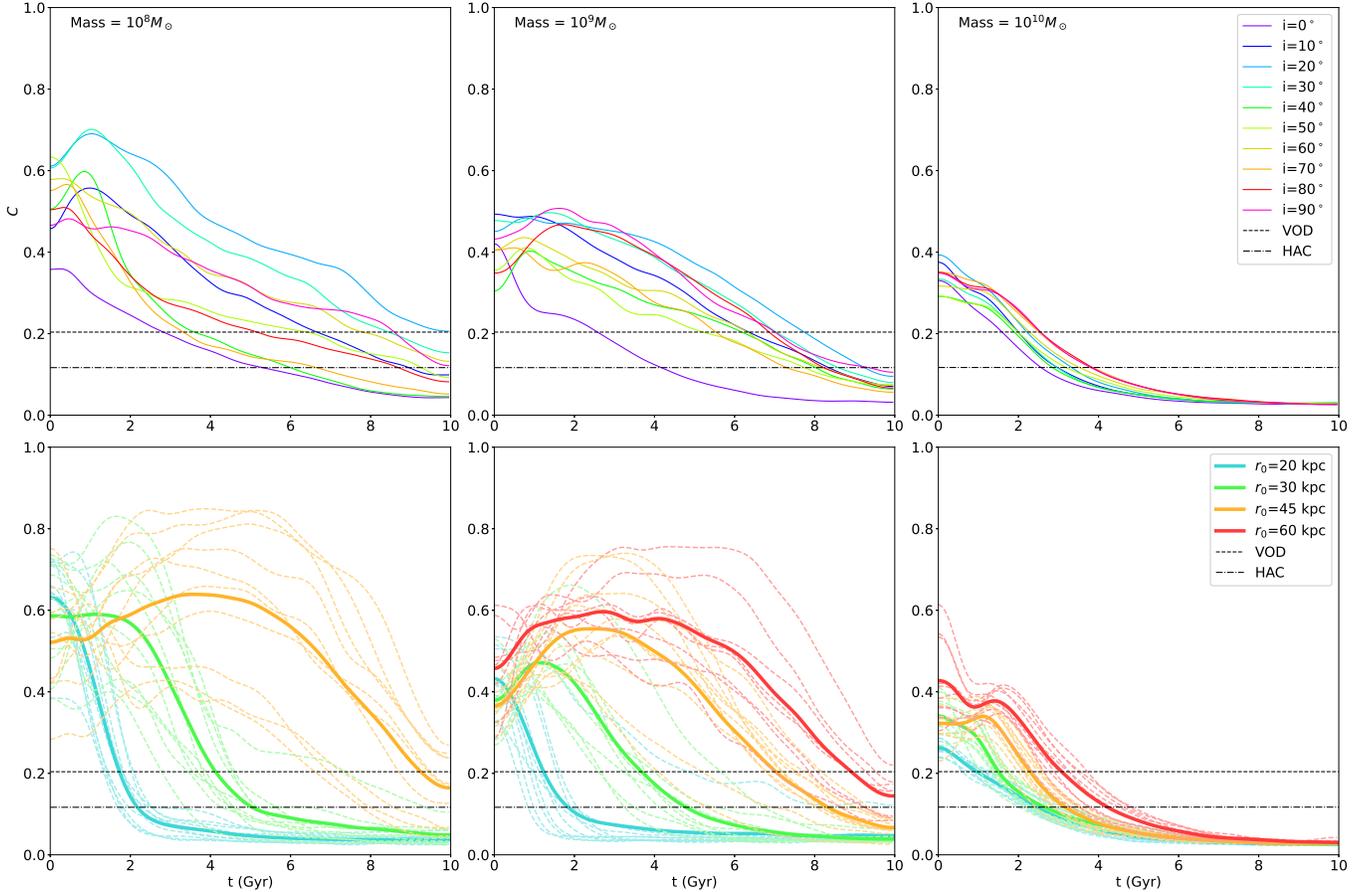}
\caption{Values of causticality vs. time for all 120 simulated radial mergers. The causticality was calculated for each timestep of a radial histogram of all stars in the simulation with $|v_r| < 10$ km s$^{-1}$, $|L_z| < 500$ kpc km s$^{-1}$, and a roughly 50 degree by 50 degree portion of the sky, which was manually selected to contain the majority of the shell(s) for each simulation. The causticality lines have been interpolated in order for trends to be more visible. Dashed horizontal lines give the values of causticality for the observed data in the VOD region (0.20) and the HAC region (0.12). Observational errors will move the measured causticality of the observed data downwards, making the observed data appear older than it actually is. Each column contains data for a given progenitor dwarf galaxy mass, given in the top left corner of the upper panel. On average, a smaller progenitor mass means that radial merger will take longer to phase mix. The top panels show mean causticality for simulations with a given initial inclination angle, averaged over the four initial distances.  Inclination angle does not appear to have a monotonic effect on how long shells take to phase mix; increasing the inclination angle could increase or decrease the mixing time. The bottom panels show values of the causticality for simulations with a given initial distance. Mean values of the causticality are shown with thick solid lines, and are averaged over all inclination angles. Individual simulations are shown in thin dashed lines, and are colored by their initial distance. Simulations with larger initial distances take longer to phase mix on average than simulations with smaller initial distances.  \label{fig:causticality}}
\end{figure}

Figure \ref{fig:causticality} shows plots of the causticality over time for all 120 radial merger simulations, with varying inclination angle, radius of initial infall, and mass. Evaluating the causticality for our observed data, we find that the value of the causticality is 0.20$\pm$0.09 for the VOD region and 0.12$\pm$0.06 for the HAC region. Any background halo stars or observational errors in these datasets will decrease the measured values of the causticality in the observed data, so these calculated values of the causticality of the observed data are lower limits. Background stars or observational errors will make the observed data appear to be older than it actually is, since the simulated data does not suffer from these issues.

Figure \ref{fig:causticality} shows that simulations with larger initial distances take longer to phase mix on average than simulations with smaller initial distances. This means that the approximate time it took the VRM debris to phase mix to its current state depends on what the distance of the final apogalacticon ($r_0$) of the VRM progenitor's orbit was. In order to get a good estimate of the age of the VRM from this data, we need to determine a likely value of $r_0$ for the progenitor of the VRM.

Assume that the progenitor dwarf galaxy remains largely intact prior to reaching the final apogalacticon. The individual stars in the dwarf galaxy have a range of energies that are limited by the fact that they are gravitationally bound together. When the dwarf galaxy is tidally disrupted as it passes through the Galactic center, the individual stars have a range of energies near the energy per mass of the original dwarf galaxy, and will therefore populate orbits with a range of apogalacticons that are near the final apogalacticon of the dwarf galaxy. For example, Figure \ref{fig:sim} shows that the debris of a simulated radial merger with $r_0 = 30$ kpc primarily populates the halo at distances between 20 and 40 kpc.

If the radial merger event has enough material to dominate the stellar halo, which is true for the VRM, then we expect to see a stellar break in the halo in the range of distances that the radial merger populates. Various works place the stellar breaks at a range of 18 to 28 kpc \citep{Watkins2009, Deason2011, Sesar2011, Pila-Diez2015, Xue2015, Deason2018}. This range in the stellar break radius may be due to measurements in different regions of the sky; certain portions of the Galaxy, such as the VOD and the HAC, will have proportionally more VRM debris than regions without visible overdensities. Measurements of the stellar break radius in the overdense regions will produce values similar to $r_0$ for the VRM progenitor; measurements in other areas of the sky will produce values of the stellar break that are smaller than $r_0$ for the VRM progenitor, due to the majority of VRM material in that region not being near apogalacticon, and therefore populating regions with smaller Galactocentric radii. A measurement of the stellar break over the entire halo would then be an average of the actual $r_0$ value for the VRM progenitor and other regions of the sky, and would be expected to yield a value somewhat smaller than the actual value of $r_0$ for the VRM progenitor. 

These values for the stellar break suggest that the value of $r_0$ for the VRM must be between 20 and 30 kpc in order to generate a stellar break near those distances. This value of $r_0$ for the VRM is supported by \cite{Villalobos2008}, who showed that simulations of dwarf galaxies falling into the Milky Way from distances near the virial radius tend to have a final apogalacticon around $r_0 \sim 20$ kpc from the Galactic center before colliding with the Galaxy.

In Figure \ref{fig:causticality}, we find that simulations with $r_0$ = 20 kpc have similar causticalities to the observed data between 1 and 2 Gyr after the simulation is started. The simulations with $r_0$ = 30 kpc have causticalities that are similar to the observed data between 3 and 5 Gyr. Since we expect the progenitor of the VRM to have a value of $r_0$ between 20 and 30 kpc, the time at which the VRM has values of causticality that are similar to the observed data is expected to be within the last 5 Gyr.

This is again consistent with \cite{Villalobos2008}, who state that shell structure is present for only $\sim$2 Gyr after dwarf galaxies collide with the Milky Way in their simulations. Further, we were unable to locate identifiable shell structure beyond 5 Gyr after collision in our simulations with $r_0 < 60$ kpc. These points suggest that after 5 Gyr, VRM-like mergers have phase mixed beyond what we see in the observed data, and that the VRM cannot have happened that long ago, as we still see shell substructure.

Note that as the mass of the progenitor dwarf galaxy increases, less time is required to phase mix the debris to observed levels. For progenitors with a mass of $10^{10}M_\odot$, this mixing time is very consistent, and does not appear to strongly depend on initial inclination angle or initial distance. No simulations with this mass had a value of causticality similar to the observed data after 5 Gyr.

For simulations with masses of $10^9M_\odot$ and $10^{10}M_\odot$, the time to phase mix increases monotonically with initial distance. However, simulations with a mass of $10^8 M_\odot$ and $r_0$ = 60 kpc appeared to phase mix more quickly than simulations with identical mass and smaller initial distances. This is due to the distance cut that we made on our simulations when calculating causticality. Since the observed data did not extend beyond 50 kpc, we cut the simulated data to only include objects within 50 kpc. The $10^8M_\odot$, $r_0$ = 60 kpc simulations place the majority of their material beyond 50 kpc from the Galactic center, so the measured causticalities only include the small number of objects that remain within 50 kpc. This deflates the measured value of causticality. However, a structure that places almost all of its debris beyond 50 kpc cannot possibly be responsible for shells within 30 kpc of the Galactic center, so simulations with $10^8M_\odot$ and $r_0$ = 60 kpc are not viable models for the VRM. Data from the $10^8M_\odot$ and $r_0$ = 60 kpc simulations is omitted from Figure \ref{fig:causticality} to avoid potentially misleading the reader.

\section{Timing Radial Mergers in Simulations} \label{sec:dating}

It is difficult to determine the orbit of a radial merger's progenitor through the usual methods used to fit orbits to tidal streams because there is no visible progenitor and no coherent motion of member stars on the sky. Without an orbit, we cannot properly simulate the progenitor of the VRM in order to explore how the merger event evolves over time. \cite{Donlon2019} provided a possible orbit for the progenitor of the VRM based on the motion of moving groups in the VOD, but they did not determine whether that orbit properly constrained the infall of the VRM progenitor or its debris in other regions. Here we derive a new method for determining the merger time of a radial merger without knowing its orbit.

Stars with similar energies in a radial merger event will oscillate back and forth with similar periods, producing shells wherever the stars turn around on their orbits. This period is analytically calculable for spherically symmetric systems. However, in an axisymmetric potential, the period for each group of stars will depend on the initial energy of the stars and also the inclination of the material with respect to the Galactic disk. The initial energies of the stars and their motions also depend on the details of the Milky Way potential. We do not know the initial orbital parameters of the progenitor of the VRM for certain, so it is not possible to derive an analytic expression for the position of a caustic surface in this case. We therefore take a numerical approach using the data that we have available to us. 

With non-radial mergers, the progenitor typically survives as a bound object for many orbits while material is tidally stripped away from the progenitor. It might therefore be necessary to account for dynamical friction, which could substantially change the orbit of the progenitor. On the other hand, with a radial merger, the system becomes unbound after its plunge through the Galactic center and dynamical friction is not a factor in the orbits of the merger debris. For this reason, we do not consider dynamical friction in our models.

\subsection{Modeling Oscillations of Shells} \label{sec:osc-model}

Stars in a shell will possess similar energies, and will have moved similarly throughout the lifetime of the radial merger. Thus, we can expect a single star in some shell to oscillate with the same period as the rest of the stars in the shell. This means that we can model caustic oscillation in the Galactic potential as a free oscillation of a single model mass at the location of a shell. At any given point in time, we expect to see multiple shells at different radii and position on the sky for a given radial merger. By assigning model particles with zero initial velocity to each shell and then tracing these particles' motions back in time, we can compare the relative positions of each caustic surface throughout time. At some time in the past, the particles on all of the shells must have been at the same place, bound to the dwarf galaxy. The difference in each shell's distance from one another can then be used to pinpoint the time that a radial merger was still self-bound.

We begin by placing a model particle at each shell found by the BIC fitting algorithm for some radial merger (either simulated or observed data). We must be able to allow motion from one side of the Galaxy to the other, so we arbitrarily select the shells on one side of the Galaxy to be located at positive $r$ and the shells on the other side to be located at negative $r$. Particles will then oscillate between positive and negative $r$ values. Using Galpy version 1.4.1 \citep[][http://github.com/jobovy/galpy]{Bovy2015}, we calculate the forces on each model particle at each time step in order to integrate the particles backwards in time for 10 Gyr. By integrating each model particle back in time, we expect to find a point in time where each caustic structure is located at the same point in space. Additionally, if each caustic structure is moving in the same direction (have the same signs of $dr/dt$) at that point, then we claim that we have found the point in time that the progenitor of the radial merger was still self-bound. This point in time will be the ``merger time'' of the radial merger.

\subsection{Constraining Likely Merger Times} \label{sec:mergertimes}


In order to measure the overall difference in the location of shells, we introduce $\delta (t)$ as a metric to measure the total distance between model particles, 

\begin{equation} \label{eq:delta} 
\delta (t) = \sum\limits_{i<j}^N |\vec{r}_i(t) - \vec{r}_j(t)|.
\end{equation} Here, we use vector subtraction in order to ensure that the model particles are spatially close to one another, and not simply at similar Galactocentric distances. A small $\delta (t)$ means the caustic structures are clumped together, while a high $\delta (t)$ corresponds to a large separation of the shells. Times with a small $\delta (t)$ are more likely to be a time when the progenitor was still bound. 

As the progenitor of the radial merger collides with the host galaxy, all of the caustic structures are initially falling inwards, and then transition to moving outwards from the host galaxy. We count the number of 1st derivatives of the model particles that are positive ($dr/dt > 0$) in order to determine the number of caustic structures moving in the same direction. The passage of the dwarf galaxy through the Galactic center would cause the number of positive first derivatives to go directly from 0 to the total number of shells. If a minimum of $\delta(t)$ corresponds to a time where the number of model particles with $dr/dt > 0$ is either 0 or the number of shells in the data, then that is a likely time that the dwarf galaxy passed through the Galactic center and was tidally disrupted. This is especially true if the number of model particles with $dr/dt > 0$ goes directly from 0 to the number of shells or vice versa at that time, as this corresponds to the progenitor passing through the Galaxy.

\subsection{Recovering Merger Times from Radial Merger Simulations} \label{sec:recovery}

We tested the method outlined in Sections \ref{sec:osc-model} \& \ref{sec:mergertimes} by using it to measure merger times for the series of simulations of radial mergers in a Milky Way-like galaxy described in Section \ref{sec:simulations}. These simulations had a known evolution time, which allowed us to verify whether or not the method was able to recover the correct time of collision for each different simulation. 

We created datasets similar to the ones used for the analysis of the VRM debris by cutting our simulation data in $v_r$ and $L_z$ identically to the way it was cut in the observed data. Additionally, we only looked at data out to $r$ = 50 kpc, in order to ensure that our method could recover the correct infall time despite our limitations in distance in the observed datasets. The majority ($> 75\%$) of the simulation data was typically found in this distance range.

Next, we cut the simulated data based on sky position. Our observed data is cut to only two regions of the sky, the VOD and HAC regions. In order to emulate this, we looked at the radial merger simulations in R.A. and Dec., and selected the data from 2 regions that were chosen to be similar in size to the VOD and HAC regions, have roughly the same positions in the sky (within $\sim$ 20$^\circ$), and to contain an overdensity of the simulated bodies. The regions had to be chosen by hand, since the simulations did not all place shell structure in the same places on the sky. We had to cut out regions around the shell overdensities because the Galactic potential was not spherically symmetric, so the shells are not at a constant radius; by selecting a limited region of the sky, the shells were at approximately constant radius over the selected data. We made these cuts in the forty radial merger simulations with a mass of $10^9M_\odot$ for a range of evolve times between 1 and 5 Gyr. The correct merger time was calculated for each simulation as when the center of mass of the dwarf galaxy passed through the Galactic center. Only the simulations with a mass of $10^9M_\odot$ were used, since this mass is similar to the predicted mass of the VRM (see Section \ref{sec:oldSausage}), and still has a wide range of possible phase mixing times compared to more massive merger events (Figure \ref{fig:causticality}). Using only a third of the simulations also allowed us to cut down on the number of simulations that had to be analyzed manually, while still retaining an idea of the capabilities and limitations of the method.

The next step was to locate the shells in the data. First, we used an EM algorithm to fit a Gaussian mixture model to a histogram of the data and optimized the number of Gaussians using a BIC method, similar to the process outlined in Section \ref{sec:fitting}. However, the EM algorithm often preferred to fit Gaussian components to outlier peaks with small star counts if the interior shells were located close to one another. As it was not possible to supervise every fit in order to avoid this behavior, we chose to use a more robust algorithm utilizing a residual sum of squares as the goodness-of-fit.

This fitting algorithm minimizes a binned residual sum of squares, 
\begin{equation}
\textrm{RSS} = \sum_i^m \frac{1}{\eta} \Big[ \int_{r_i}^{r_{i+1}}f(r)dr - n_{i,obs} \Big]^2,
\end{equation} where $m$ is the number of bins, $f(r)$ is the fit Gaussian mixture model, $r_i$ and $r_{i+1}$ describe the bounds of the $i^\textrm{th}$ bin, and $n_{i,obs}$ is the value of the $i^\textrm{th}$ bin of the histogram of observed data. The observed data is compared to the integral of the fit model over each bin instead of the value of the model at the center of each bin; this will constrain the fit model to the number of stars in the data, as well as improve the estimate of the width of each shell. We also use a normalization constant
\begin{equation}\label{eq:eta}
\eta = N - k - 1,
\end{equation} where $N$ is the total number of data points in the histogram and $k$ is the total number of parameters being fit. We used the differential evolution algorithm \citep{Storn1997} to minimize the residual sum of squares of our data, which provides the optimal fit for our data. Specifically, we use the implementation of the differential evolution algorithm from the \textbf{scikit-learn} python package \citep{scikit-learn}.

We employ the BIC in order to determine how many shells are statistically significant. For this case the BIC is given by
\begin{equation}
\textrm{BIC} = k\ln (N) + 2 \ln (\textrm{RSS}),
\end{equation} where $N$ and $k$ are the same as in Equation \ref{eq:eta}. As before, the BIC prevents overfitting our observed data. We fit up to 3 Gaussians to both regions individually, and compare the corresponding BICs; the fit with the lowest BIC was taken as the most statistically significant result. The positions of the peaks of the best fit Gaussian Mixture Model were taken to be the distances of the shells. 

We then used the model particle rewinding method on the adjusted simulation data to recover the most likely merger time for each simulation. We placed model particles in the centers of the selected regions at the distances of the shells. These model particles were allowed to move freely in our Milky Way potential, resulting in oscillations in their distance from the Galactic center. Tracing these oscillations throughout this period identified times when the model particles were all close to one another, corresponding to a likely time where the progenitor of the merger was still coherent.  The best fit merger time from the simulation was then selected from the results and compared to the correct merger time.

\begin{figure}
\center
\includegraphics[width=\linewidth]{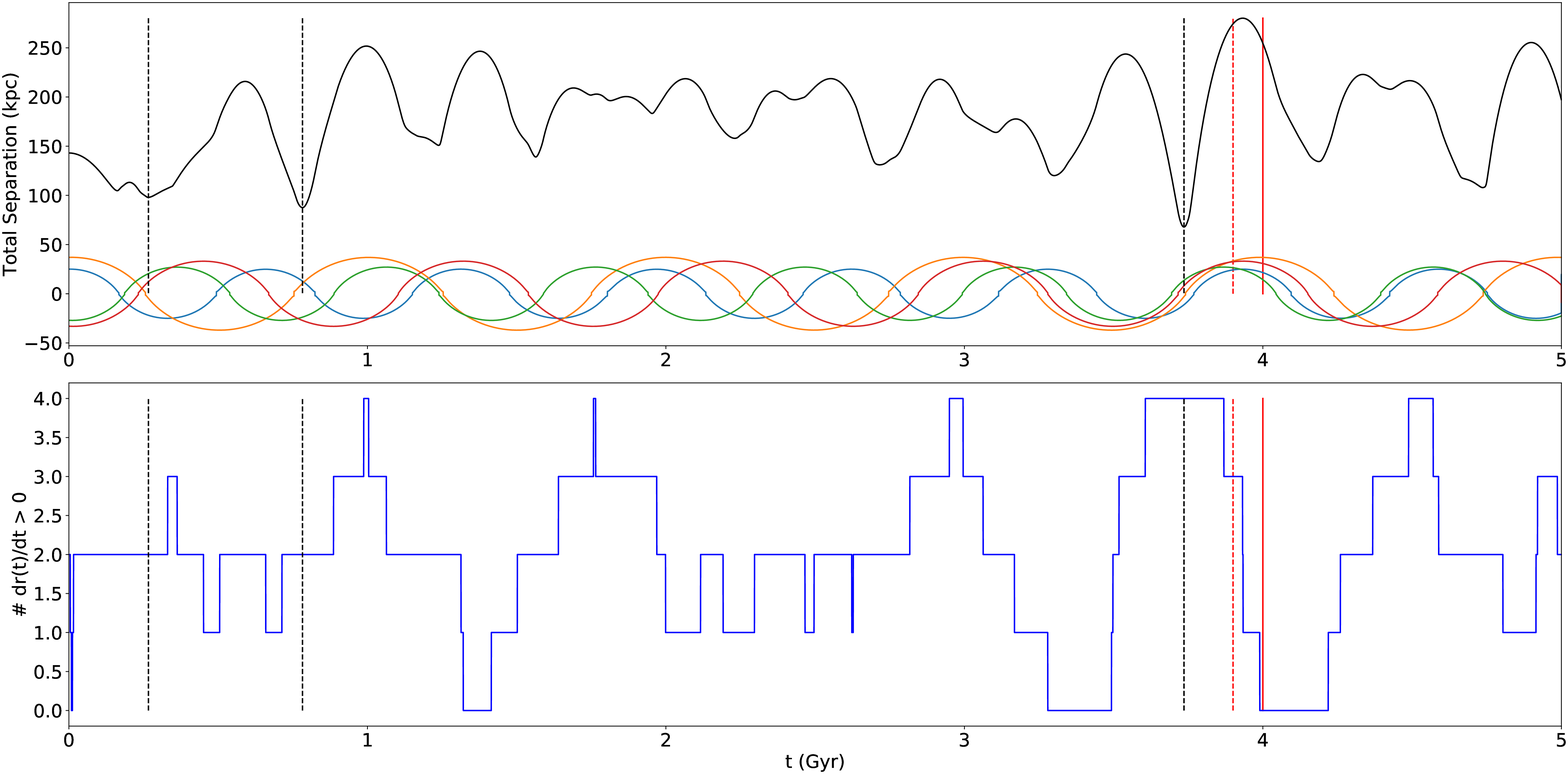}
\caption{\textit{TOP}: Plot of total separation, or $\delta (t)$, over 5 Gyr for the oscillating caustic surface model outlined in Section \ref{sec:mergertimes}. This figure uses a simulation with initial inclination angle $i$ = 30$^\circ$, initial distance $r_0$ = 30 kpc, a mass of $10^9M_\odot$, an evolve time of 4 Gyr (solid red line), and a merger time, defined as the time since the center of mass of the dwarf galaxy passed through the Galactic center, of 3.9 Gyr (dashed red line). The distance of each shell from the Galactic center is shown with colored solid lines, where positive $r$ values correspond to the model particle being in the VOD region, and negative $r$ values correspond to the model particle being in the HAC region. In order to be considered for the best merger time, minima were required to have a value of $\delta(t)$ at least 2 standard deviations below the mean of the overall $\delta(t)$ distribution. \textit{BOTTOM}: Plot of the number of caustic surfaces with $dr/dt > 0$ over the same time period as the top panel. Times where $\delta (t)$ is at a local minima and is at least two standard deviations below the mean are marked with dashed vertical lines. Local minima of $\delta(t)$ that coincide with times where the number of model particles with $dr/dt > 0$ is zero or 4 likely indicate the time of the merger.  Note that although there are 3 local minima of $\delta(t)$, the only one that lines up with a spot where the number of model particles with $dr/dt > 0$ is zero or 4 is 3.7 Gyr. This also corresponds to the shells all lining up in their oscillations in the top panel.  \label{fig:sim_osc}}
\end{figure}

Figure \ref{fig:sim_osc} shows the result of our method on one such simulated radial merger. The Galactocentric distances of the model particles are provided in order to show how $\delta(t)$ depends on the positions of the model particles. It is clear that the method is able to recover the time at which the oscillating model particles line up, as this is on average within a few tenths of a Gyr of the actual merger time of the simulation. 

\begin{figure}
\center
\includegraphics[width=0.5\linewidth]{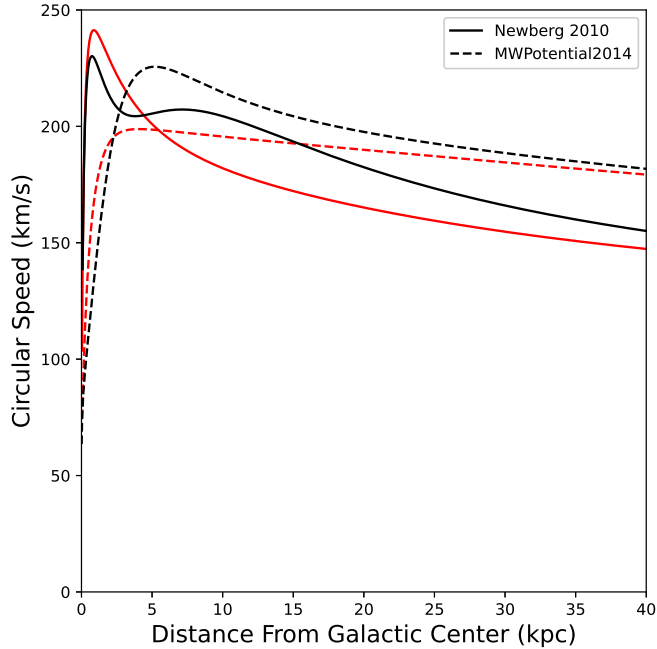}
\caption{The two model potentials that are used in this work. The model potential fit using the orbit of the Orphan Stream in \cite{Newberg2010} is plotted in a solid line, while the Galpy built-in \textit{MWPotential2014} is plotted in a dashed line. The circular speed curves in the plane of the disk are shown in black, while the circular speed curves along the axis of symmetry ($Z$-axis, perpendicular to the disk) are plotted in red. These potentials are similar, as they are both fit to the Milky Way, but have different shapes. \label{fig:potshapes}}
\end{figure}

In order to examine the dependence of the method on the shape of the Galactic potential, we also rewound the model particles in the Galpy built-in potential \textit{MWPotential2014} \citep{Bovy2015}. \textit{MWPotential2014} uses a spherical power law potential for the central bulge, a less massive Miyamoto-Nagai disk than the \cite{Newberg2010} model, and a double power law spherical halo potential. The two model potentials are compared in Figure \ref{fig:potshapes}. Calculating merger times with this second potential helps us determine whether or not an ``incorrect'' model potential impacts the results, as the radial merger simulations were only run in the \cite{Newberg2010} model potential. This helps us to understand the limitations of our method, as the Milky Way's actual gravitational potential will not be truly identical to any model potential that we choose.

\begin{figure}
\center
\includegraphics[width=\linewidth]{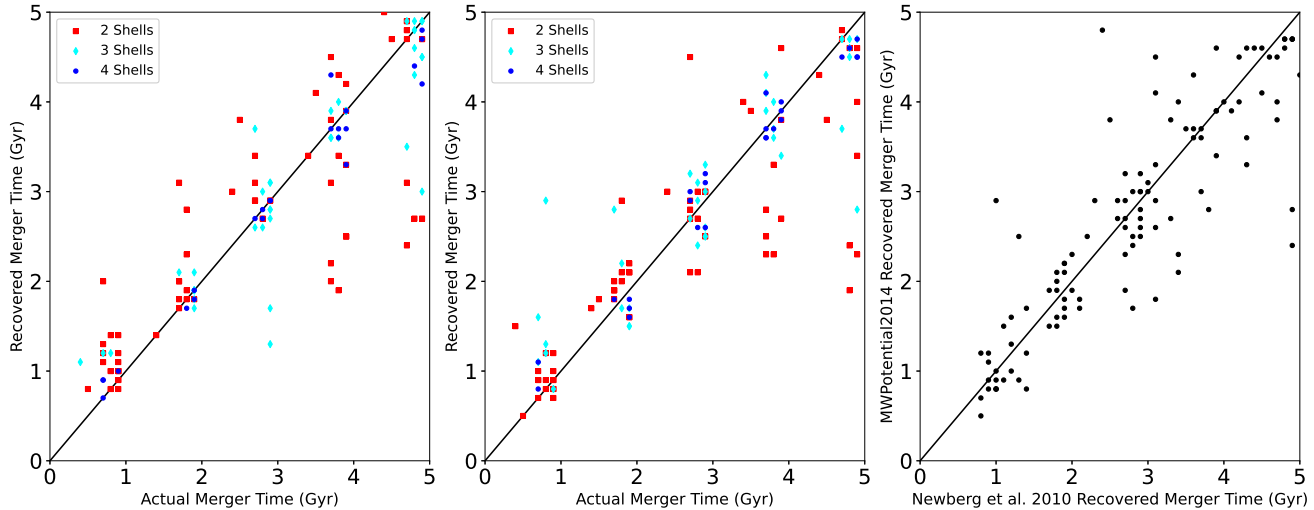}
\caption{Recovered merger time vs. actual merger time for all the results given in Table \ref{tab:recovered}. The left panel shows merger times recovered with the \cite{Newberg2010} potential, and the middle panel shows merger times recovered with the \textit{MWPotential2014} potential. The right panel shows a comparison of the recovered merger times for each simulation using the two model potentials. The closer a result is to the solid black line, the more accurate the result. Red squares denote simulations with 2 identified shells, cyan diamonds correspond to 3 shells, and blue circles correspond to 4 identified shells. As merger time increases, it becomes more difficult to accurately recover the time of merger of a simulation. Simulations with more shells are more accurate. Considering all simulations, the standard deviation of the differences between the calculated values and the correct values is 0.52 Gyr for the \cite{Newberg2010} model potential, and 0.53 for the \textit{MWPotential2014} model. The standard deviation of the differences drops to 0.20 Gyr for the \cite{Newberg2010} model potential, and 0.12 Gyr for the \textit{MWPotential2014} model, if only simulations where the algorithm recovered 4 shells are considered. The mean difference in recovered merger times between the two model potentials over all simulations is 0.42 Gyr. \label{fig:recovery}}
\end{figure}

Table \ref{tab:recovered} and Figure \ref{fig:recovery} show the results of our method on simulated data using both the ``correct'' and ``incorrect'' potentials. It is clear that the method is able to recover the infall times of the simulated radial mergers for a range of initial distances and evolve times. Including simulations that recovered 2 or more shells in the data, the standard deviation of the differences between the calculated value and the correct value is 0.52 Gyr for the ``correct'' \cite{Newberg2010} model potential, and 0.53 for the ``incorrect'' \textit{MWPotential2014} model. The standard deviation of the differences decreases to 0.38 Gyr in the \cite{Newberg2010} model and 0.39 Gyr in the \textit{MWPotential2014} model if only simulations with 3 or more shells are considered, and drops even further to 0.20 Gyr in the \cite{Newberg2010} model and 0.12 in the \textit{MWPotential2014} model if only simulations with 4 shells are considered. There are 4 shells identified in our observed data, so we estimate the error in infall time to be the larger of the two standard deviations of the differences: $\sigma_t$ = $\pm$0.2 Gyr. 

This value is only the approximate error contributed by the recovery method. Other factors, such as using an incorrect shape of the Galaxy potential, might increase the actual error in our calculated value.  However, the two model potentials used in this work did not seem to have a large impact on our method. This leads us to believe that any reasonably shaped model potential would likely produce similar results.

Some combinations of initial angles and initial distances of the simulated dwarf galaxy progenitors resulted in little to no shell structure. This is why, for example, there are no simulations with $i=0^\circ$, $r_0$ = 30 kpc and a mass of $10^9M_\odot$ in Table \ref{tab:recovered}, as this simulation did not produce identifiable shell substructure.  The simulations with inclination angle $i = 30^\circ$ appeared similar in R.A. and Dec. to the simulations of individual components of the VRM debris in \cite{Donlon2019}. It is possible that this inclination angle may be close to the correct value of the inclination angle of the progenitor of the VRM.

 We only tested simulated evolve times up to 5 Gyr, as that is near the age of radial mergers at which shells stop being easily located. A more thorough suite of tests would include tests  exploring the impacts of different progenitor profiles. Even without this exhaustive testing, we believe that our method is able to determine the correct time of collision of the VRM progenitor and the Milky Way, as the method shows clear success in recovering the correct values for a variety of radial merger simulations.

\section{Timing the VRM Revisited: Shell Oscillations} \label{sec:age_of_vrm}

\begin{figure}
\center
\includegraphics[width=\linewidth]{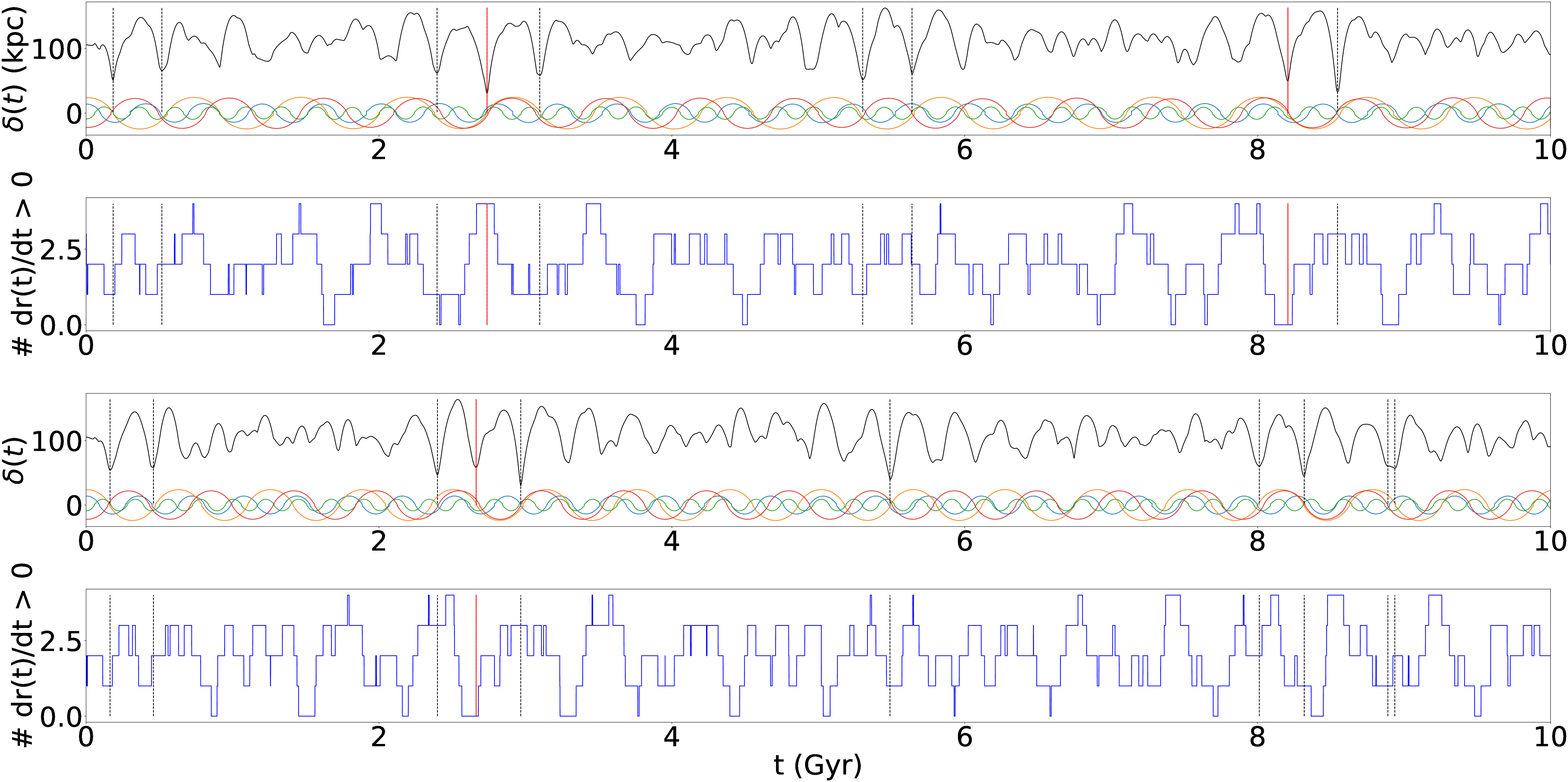}
\caption{Plots of total separation $\delta (t)$ over 10 Gyr for the oscillating caustic surface model outlined in Section \ref{sec:mergertimes}, for the observed data. Below each plot of $\delta(t)$ vs. time is a plot of the number of caustic surfaces with $dr/dt > 0$ over the same time period. The top two panels are for the Orphan Stream Model 5 potential, and the bottom two panels are for the \textit{MWPotential2014}. Times where $\delta (t)$ is at a significant local minima are marked with dashed vertical lines. A minimum is considered to be significant if it is at least two standard deviations below the mean value of $\delta(t)$. In the case of the Orphan Stream Model 5 potential, this corresponds to values less than $\delta(t)$ = 67 kpc, and for \textit{MWPotential2014}, it corresponds to values less than 61 kpc. Times marked with dashed lines that coincide with times where the number of model particles with $dr/dt > 0$ = 0 or 4 are marked in solid red vertical lines, and likely merger times. Corresponding values for each local minima are given in Table \ref{tab:times}. We find that 2.7 Gyr is a likely merger time in both models, so we conclude that 2.7 Gyr ago the progenitor of the VRM collided with the Galactic center. \label{fig:osc}}
\end{figure}

Figure \ref{fig:osc} shows the results of our oscillating model particle method on the observed data. The top two panels show the value of $\delta(t)$ and the number of particles with $dr/dt>0$ for the Milky Way potential from \cite{Newberg2010} that was used to generate the simulations in Section \ref{sec:constraint}. A minimum was considered significant if it was at least two standard deviations below the mean of $\delta(t)$. Nine significant minima were identified, of which only two corresponded to a time where the number of model particles with $dr/dt>0$ was either 0 or 4. These minimima are for merger times of 2.7 Gyr and 8.2 Gyr. It is unlikely that the 8.2 Gyr merger time is realistic, due to the phase mxing constraints provided in Section \ref{sec:constraint}.

The bottom two panels of Figure \ref{fig:osc} show the same values, but calculated with the \textit{MWPotential2014} model. If we instead use the \textit{MWPotential2014} to approximate the actual Milky Way then we find ten significant minima, of which only one corresponds to a time when all the particles are moving in the same direction. This time is 2.7 Gyr ago, which is consistent with the first minimum of the Orphan Stream Model 5 potential. The 8.2 Gyr merger time is not recovered with the second potential. This fact and the phase mixing constraints on the merger time of the VRM suggest that the 8.2 Gyr merger time is a false positive. Table \ref{tab:times} shows the positions and velocities of the model particles at local minima for these calculations. From these results, we propose that the progenitor of the VRM collided with the Galactic center 2.7 $\pm$0.2 Gyr ago.

A merger time of 2.7 Gyr ago is consistent with the conclusions from Section \ref{sec:constraint}: phase mixing constraints on the VRM debris leads one to believe that the progenitor of the VRM collided with the Galactic center between 1 and 5 Gyr ago. A merger time of 2.7 Gyr ago corresponds with the progenitor of the VRM having a final apogalacticon before collision between 20 and 30 kpc, and agrees with our previous analysis.

The motion of particles in the halo strongly depends on the shape of the Galactic potential. It is not clear at the moment what shape the Galactic halo potential actually takes. Further complicating the Galactic potential, the Large Magellanic Cloud (LMC) was recently found to be approximately 10\% of the total mass of the Milky Way \citep{Erkal2019}, and therefore has a large impact on halo substructure. The LMC is then expected to produce a substantial time-variable torque on the VRM debris throughout its evolution, which would generate precession of the structure and if anything accelerate phase mixing. However, the similar collision times we recover from two different potentials is a good sign that the shape of the potential is not a dominant factor.

%
%

\section{Relationship with Previously Discovered Substructure} \label{substructure}

\subsection{The Argument for an Ancient \textit{Gaia} Sausage} \label{sec:oldSausage}

The concept of the ``ancient last major merger'' of the Milky Way has occupied the literature on Galactic structure for decades \citep{Kuijken1989, Gilmore2002}. \cite{Gilmore2002} identified a group of disk stars rotating slower than expected in the local Solar Neighborhood and claimed that this group was evidence of the last major merger, which had occurred 10-12 Gyr ago. Further, \cite{Deason2013} claimed that a massive merger approximately 10 Gyr ago would explain the break in the density profile of the Milky Way halo at 30 kpc. The theory suggests that in the time since this massive merger event, there has been a period of Galactic quiescence, up until the Sagittarius dSph and Magellanic Cloud mergers that are currently underway. 

The last major merger is thought to have formed the Milky Way's thick disk, which has stellar ages older than 10 Gyr. One theory for the creation of a thick disk is that the Milky Way's proto-disk was heated by a collision with a large dwarf galaxy merger \citep{Quinn1986, Velazquez1999}. This would have caused the existing disk stars to be kicked up on orbits with larger vertical action. Gas in the disk would radiatively cool, so stars formed after this merger event could remain in a cold, younger thin disk (effectively ``quenching'' the thick disk). 

With the release of \gaia DR2, evidence suggested that the last major merger was finally identified. The GSM \citep{Simion2019} and the \textit{Gaia}-Enceladus Merger \citep{Helmi2018} were independently discovered, and are interpreted to be the same ``last major merger of the Milky Way.'' The age of this merger event was reported to correspond to the age of the thick disk -- between 8 and 11 Gyr ago. \cite{Helmi2018} found that the youngest age of stars in this merger were 8 to 10 Gyr old, which is consistent with this massive merger indeed having created the thick disk. This agrees with the results of \cite{Gallart2019}, who claim that star formation in the Milky Way transitioned from thick disk star formation to thin disk star formation around 9.5 Gyr ago, corresponding to the \gaia Sausage Merger. \cite{Gallart2019} also claim that both the \gaia Sausage progenitor and the Milky Way's in-situ halo had finished the bulk of their star formation by 10 Gyr ago. \cite{Belokurov2019} claim that star formation in the inner halo ended around the same era, and assuming that the cause for the quenching of star formation was the merger event itself, this is additional evidence for an ancient merger.

The presumed merger age determines a preferred mass of the ancient merger progenitor. Comparison of the metallicity of the RRL stars in the \gaia Sausage to RRL stars in the LMC suggests that the progenitor of the \gaia Sausage had a similar mass as the LMC did 10 Gyr ago \citep{Belokurov2017, Zinn2020}. \cite{Mackereth2019} analyzed the chemical and orbital properties of the \gaia Sausage in comparison to EAGLE simulations, and determined that it likely had a mass of at least 10$^9 M_\odot$, and was accreted around 9 Gyr ago.

This large mass suggests that this merger is the main component of the stellar halo; \cite{Deason2019} showed that the total stellar mass of the halo is on the order of 10$^9 M_\odot$. Simulations suggested that most of the stars on radial orbits in the halo came from massive satellites approximately 6-10 Gyr ago, and that these stars would dominate the halo between 10 and 30 kpc of the Galactic center \citep{Belokurov2018b}. The inner halo is indeed dominated by stars on highly eccentric orbits \citep{Iorio2019}. This is all consistent with the halo being primarily formed from a single merger event. 

For all of these reasons, the present view of the \gaia Sausage is that it is an ancient merger event with a stellar mass between  $10^8 M_\odot$ and $10^9 M_\odot$ that is responsible for the formation of the thick disk, and makes up the majority of the stellar mass of the halo. We will now attempt to reconcile this viewpoint with the merger time of the VRM progenitor that was calculated in this work.

\subsection{The Argument for a Young \textit{Gaia} Sausage} \label{sec:youngSausage}
The \gaia Sausage is characterized by a structure in velocity space with high dispersion in $v_r$ and small rotational velocities in the local Solar Neighborhood \citep{Belokurov2018b}. It has been shown that the VRM debris looks very similar to the \gaia Sausage in the local Solar Neighborhood, so much so that the two structures are likely identical \citep{Donlon2019}. If the VRM debris and \gaia Sausage are identical, then how do we explain the gap between the 2.7 Gyr ago infall time of the progenitor of the VRM and the 8-10 Gyr age of the \gaia Sausage?

In Section \ref{sec:constraint}, we developed an argument that the VRM debris must populate the stellar halo at the distance ranges in which we identify shells. Since the VRM debris dominates the stellar halo, the debris must also be located near and inside the stellar break at $\sim20$ kpc. To satisfy these conditions, the progenitor of the VRM must have had a final apogalacticon before collision between 20 and 30 kpc. The measured values of the causticality in simulations with $r_0$ = 20 kpc and $r_0$ = 30 kpc in Figure \ref{fig:causticality} match the measured causticality values for the observed data around 1 Gyr and 5 Gyr, respectively. This suggests that the merger time of the VRM is within the last 5 Gyr. Additionally, by 5 Gyr after collision, our method's ability to recover the merger time of a radial merger dropped substantially (Figure \ref{fig:recovery}). Finally, this work recovered a merger time of only 2.7 Gyr for the VRM progenitor. All of this suggests that the progenitor of the VRM collided with the Milky Way more recently than 8-11 Gyr ago.

If the dynamical mass of the VRM is actually closer to $10^{10}M_\odot$, then even large initial distances constrain its merger time to be within the last 5 Gyr (Figure \ref{fig:causticality}). The GSM is expected to dominate the halo, and would then have a dynamical mass of around $10^{9}$-$10^{10}M_\odot$. This suggests two possible scenarios: If the GSM is indeed ancient, it either cannot dominate the material in the halo, or it cannot be responsible for the relatively unmixed stellar overdensities in the halo. In this case, the VRM and the GSM cannot be identical events, as the VRM debris populates shells in the VOD and the HAC. If the VRM and the GSM are not the same, then the Milky Way halo is composed of more than one major merger event. On the other hand, if the GSM is responsible for the VOD and the HAC and dominates the halo, then it is unlikely that the merger event occurred more than 5 Gyr ago. If this is the case, the VRM and the GSM probably describe the same merger event. 

\cite{Villalobos2008} ran simulations of dwarf galaxy mergers that collided with a model Galactic disk after falling in from near the virial radius. After an orbital decay over a period of $\sim1.5$ Gyr, the dwarf galaxy finally collides with the Galaxy from a distance of $\sim$20 kpc from the Galactic center. Thick disks are shown to form approximately 0.25 Gyr after these collisions. The work states that thick disk and satellite disruption reach an equilibrium after $\sim$ 2 Gyr after the collision occurs, and that major shell structure is only visible for these 2 Gyr. If the thick disk was formed in this way 8-11 Gyr ago, then shells in the resulting merger debris would not be visible at the present day.

In this work, we use the term ``merger time'' to describe when the progenitor of the VRM's center of mass passes through the Galactic center. This is a collision between the dwarf galaxy and the Milky Way. If the progenitor for the \gaia Sausage crossed the Milky Way's virial radius sometime around 10 Gyr ago (``infall time''), it could have spent the last 8 Gyr in a prolonged orbit about the Galaxy while it shed its initial energy, until finally colliding with the Galaxy 2.7 Gyr ago. This would have resulted in the progenitor of the \gaia Sausage making many passes through the Milky Way over its lifetime, which could cause tidal shocks in both structures. We speculate that these tidal shocks from perigalacticon passes of the \gaia Sausage progenitor could potentially kick up thick disk stars from the Milky Way protodisk through large-scale tidal torques, as well as generate star formation through the collapse of gas clouds in the disk. This could explain the findings that the star formation history of the thick disk oscillates with a period of $\sim$3 Gyr \citep{Gallart2019}, which is a resonable time between passes for a distant halo object with a highly eccentric orbit.

We did not include dynamical friction in our models of the VRM in this work, because it does not play a role during and after the disruption of the satellite as it passes through the center of the Milky Way. However, dynamical friction would allow the VRM progenitor to shed energy as it orbits and falls inwards from the virial radius. A reduction in energy is required for a dwarf galaxy formed outside or near the Milky Way's virial radius to eventually become substantially more bound to the Milky Way, and would cause the VRM progenitor to slow down from earlier non-radial passes of the Milky Way before its final radial collision.

An initial tidal shock from a close encounter with the Milky Way could be responsible for the quenching of star formation in the \gaia Sausage 8 to 11 Gyr ago (stellar ``age'' of the progenitor). However, according to \cite{Brown2014}, many dwarf galaxies in the local group stopped star formation around the same time 12 Gyr ago during reionization of the Universe, so alternatively it could be that the \gaia Sausage had already stopped active star formation even before it became gravitationally bound to the Milky Way. This could also explain why the star formation in both the Milky Way in-situ halo and the \gaia Sausage progenitor appear to slow down around 12 Gyr ago \citep{Gallart2019}, even before the proposed accretion time of the progenitor of the \gaia Sausage around 8-11 Gyr ago. If the accretion event were the primary underlying cause of the drop in star formation rates in both structures as \cite{Gallart2019} claim, then it is not expected for star formation to begin slowing down before the accretion event occurs. It is possible that the previous metallicity arguments for the age of the \gaia Sausage were not dating the time of the collision between the dwarf progenitor and the Milky Way, but instead were measuring other properties of the infall event. Note, however, that our method for determining merger time only measures the time since the collision between the progenitor and the Milky Way, and does not constrain when the progenitor became bound to the Milky Way, when star formation was quenched in the Milky Way or the progenitor dwarf galaxy, or when the thick disk was created.

If the \gaia Sausage didn't collide with the Milky Way 8-11 Gyr ago, then what caused the thick disk? \cite{Amarante2020}, \cite{BeraldoeSilvia2020}, and \cite{Clarke2019} all show that it is possible to form a thick disk in Milky Way analogues in isolation, provided that there is a period of lumpy disk activity early in the Galaxy's life. Additionally, \cite{Ma2017} show a cosmological simulation of a Milky Way-mass galaxy forming a bifurcated disk without a collision. These systems are able to generate both the kinematic and chemical properties of the thick and thin disks without merger events. The thick disk may form itself without perturbation from a satellite, in which case there is no need for the \gaia Sausage to be the perturber of the thick disk. Additionally, \cite{RodriguezWimberly2019} claim that it is unlikely that star formation in a thick disk would be quenched by the infall of a large dwarf galaxy. If either of these situations is the case, then the age of the GSM is not required to line up with the time of the quenching of star formation in the thick disk. 

\subsection{Additional Possibly Related Substructure} \label{sec:other-sub}

The EPO is a recently discovered overdensity with a lower surface brightness than either the VOD or the HAC. It has been suggested that the EPO is associated with the VOD and the HAC, and that they share a common origin \citep{Li2016, Donlon2019}. The results of this work are consistent with this idea; Figure \ref{fig:model_particles} shows ejecta in a direction 120$^\circ$ from the two regions containing the majority of the shell substructure, as viewed from the Galactic center. This work suggests a single ``trefoil'' structure connecting the three overdensities, instead of a model where material oscillates only between the HAC in the south and the VOD.

We note that the VOD appears to visually extend to higher declinations, tracing the Perpendicular Stream \citep{Weiss2018b}. It is possible that the Perpendicular Stream is associated with one or more shells, or that the Perpendicular Stream traces a long radial tail of infalling or outgoing material. It may be beneficial to look for shell structure in the upper portion of the Perpendicular Stream in the future in order to further constrain the behavior of the progenitor of the VRM. A study of the southern portion of the HAC would also be beneficial in exploring the full extent of the VRM debris, as well as continuing the search for shells in other regions of the Milky Way.

The phase space spiral in the disk is a structure in $z$--$v_z$ phase space that has been characterized as ongoing phase mixing in the local Solar Neighborhood from a recent perturbation in the disk \citep{Li2019}. The disk is known to not be in equilibrium \citep{Widrow2012, Yanny2013, Williams2013, Carlin2013, Xu2015}, and the vertical disk disequilibrium presents locally as the phase space spiral \citep{Li2019}. The phenomenon that caused the phase space spiral occurred at least 500 Myr ago, and it is suggested that the structure could only exist in the Milky Way for $\sim$ 4 Gyr \citep{Li2019}. Our merger time for the progenitor of the VRM lies within these constraints, and we suggest that it is possible that the cause of the phase space spiral is indeed the VRM. Perhaps the VRM ``wobbled'' the Milky Way's central bulge in a way that would propagate waves in the disk and generate vertical motion in disk stars. Further exploration is required to determine the relationship between the VRM and the phase space spiral, if any exists.

The ``Splash'' \citep{Belokurov2019} is another substructure of the Galactic disk in the local solar region. The Splash is characterized by a large population of metal-rich stars on highly radial orbits in the inner halo. While the formation origin of stars in the Splash is not yet known, it has been hypothesized that the Splash consists of stars thrown out of the Milky Way's (proto)disk from the GSM 9.5 Gyr ago \citep{Belokurov2019}. Since the VRM is a radial merger, it is likely that metal rich material in the Galactic center would be disrupted from the initial impact of the progenitor of the VRM with the Galactic center, which could explain the high-metallicity material in the Splash. \cite{Amarante2020} shows that the Splash could instead have been created through another disk process, such as a lumpy disk early in the Milky Way's formation history. Either way, there is no need for an ancient infall event to have created this substructure. 

\cite{Snaith2014} shows an oscillation in star formation of the Milky Way inner disk from 6 Gyr ago until approximately 2 Gyr ago. There is a strong peak in star formation around 2.5 Gyr ago. It is possible that this oscillation in star formation rate could be due to tidal shocks from the passes of the VRM progenitor as it fell in from the virial radius, and that the burst of star formation at 2.5 Gyr ago was due to the infall of gas from the VRM and perturbations in the disk due to the collision event. It is also possible that the increase in star formation around 2.5 Gyr ago could be caused at least in part by an impact of the Sagittarius dSph progenitor with the disk; \cite{Ruiz-Lara2020} attribute a peak in the star formation history of the local solar region around 2.0 Gyr ago to such an event. 

\section{Conclusions} \label{sec:con}

The VRM is a recently discovered radial merger in the Galactic halo \citep{Donlon2019}. It is known that radial mergers cause shell substructure \citep{Hernquist1988, Sanderson2013}. In this work we successfully locate these shells and use them to determine an elapsed time of 2.7$\pm$0.2 Gyr ago since the progenitor passed through the center of the Milky Way.

In order to achieve this result, we used \gaia, SDSS, and LAMOST data to build two 6D phase space datasets: one in the VOD region, and another in the HAC region. These regions were chosen due to their high concentration of VRM debris. Each dataset contains both RRLs and BHBs in order to maximize the amount of available data. In order to isolate shell substructure, this data was limited to $|v_r| < 10$ km s$^{-1}$ and $|L_z| < 500$ kpc km s$^{-1}$. These cuts only allow stars that are on radial orbits and are at apogalacticon, where the shells form.

The surface of a shell has small Galactocentric radial velocity, and the density as a function of Galactocentric radius has a Gaussian distribution. We fit shell models to the data using a Gaussian mixture model with Bayesian and corrected Aikaike information criteria to determine the most likely number of shells in our dataset in each region. We were able to identify four statistically significant shells in our data -- two in the VOD region, and two in the HAC region. A simulation of the VRM shows that it can connect the VOD and HAC with the EPO as a single trefoil structure in the halo.

In order to populate the stellar halo inside the stellar break and produce shells at the observed distances, the progenitor of the VRM had to have a final apogalacticon before collision between 20 and 30 kpc.  We utilize $N$-body simulations to analyze phase mixing in radial mergers. Using causticality as a measurement of the phase mixing a radial merger has gone through, we find that radial mergers with $r_0$ between 20 and 30 kpc have similar measured causticality as the observed data earlier than 5 Gyr after collision. Identifiable shell structures are not seen in these simulations after 5 Gyr at the distances where shells are observed in the Milky Way. 

Simulations of radial mergers with large masses ($\sim 10^{10}M_\odot$) become phased mixed to the levels of the VOD and the HAC within 5 Gyr. If the GSM is indeed ancient, it either cannot dominate the material in the halo, or it cannot be responsible for the relatively unmixed halo substructure. In this case, the VRM and the GSM cannot be identical events, as the VRM was discovered in VOD debris, which is populated with shell structure. If the VRM and the GSM are not the same, then the Milky Way halo is composed of more than one major merger event. On the other hand, if the GSM is responsible for the VOD and the HAC and dominates the halo, then it is unlikely that the merger event occurred more than 5 Gyr ago. If this is the case, the VRM and the GSM probably describe the same merger event. 

Caustic surfaces contain stars that oscillate back and forth from one side of the Galaxy to the other. We modeled the four caustic surfaces as point particles and rewound their orbits in a model Milky Way potential. By measuring the total separation of these model particles over a reverse integration of 10 Gyr, and by requiring that all the particles be moving in the same direction, we were able to recover the infall times for a collection of simulated radial mergers. Using this technique on the observed data, we determined that the caustic surfaces making up the VRM were all located in the same place with similar velocities approximately 2.7 Gyr ago. This corresponds to a collision between the progenitor of the VRM and the Galactic center 2.7$\pm$0.2 Gyr ago.

Based on this age, it is possible that the VRM is responsible for the Splash, the phase space spiral in the disk, and possibly a burst of star formation in the inner disk through perturbation of the Galactic center during the infall collision.  

Although we believe the VRM and the GSM represent the same radial merger event, both our phase mixing argument and our calculated collision time show that observed levels of phase mixing are too recent to have been caused by a collision 8--11 Gyr ago. This result produces tension between our merger time of the VRM and the published age of the GSM. However, we do not constrain the time at which the progenitor became gravitationally bound to the Milky Way. This apparent conflict could be resolved if the GSM is younger than previously thought, or if its published age is closer to the time of gravitational capture than the time of collision with the Milky Way.

\acknowledgments

We wish to express our deep gratitude to Yang Huang, from Yunnan University, who provided us with a copy of the catalog (Liu et al. 2020) of 5290 RR Lyrae stars spectra of the LAMOST Experiment for Galactic Understanding and Exploration (LEGUE; Deng et al. 2012) and the Sloan Extension for Galactic Understanding and Exploration (SEGUE; Yanny et al. 2009) surveys before it was publicly available. We would like to thank the anonymous referee and statistical editor for their insightful comments that improved this work. This work was supported by NSF grant AST 19-08653; contributions made by the Marvin Clan, Babette Josephs, and Manit Limlamai; and the 2015 Crowd Funding Campaign to Support Milky Way Research. Lawrence M. Widrow was supported by a Discovery Grant with the Natural Sciences and Engineering Research Council of Canada. This project was developed in part at the 2019 Santa Barbara Gaia Sprint, hosted by the Kavli Institute for Theoretical Physics at the University of California, Santa Barbara. This research was supported in part at KITP by the Heising-Simons Foundation and the National Science Foundation under Grant No. NSF PHY-1748958. This work has made use of data from the Sloan Digital Sky Survey. Funding for the Sloan Digital Sky Survey IV has been provided by the Alfred P. Sloan Foundation, the U.S. Department of Energy Office of Science, and the Participating Institutions. SDSS-IV acknowledges support and resources from the Center for High-Performance Computing at the University of Utah. The SDSS web site is \url{www.sdss.org}. This work also made use of data from the European Space Agency (ESA) mission {\it Gaia} (\url{https://www.cosmos.esa.int/gaia}), processed by the {\it Gaia} Data Processing and Analysis Consortium (DPAC, \url{https://www.cosmos.esa.int/web/gaia/dpac/consortium}). Funding for the DPAC has been provided by national institutions, in particular the institutions participating in the {\it Gaia} Multilateral Agreement. This work made use of data from the LAMOST survey. The Guoshoujing Telescope (the Large Sky Area Multi-Object Fiber Spectroscopic Telescope LAMOST) is a National Major Scientific Project built by the Chinese Academy of Sciences. Funding for the project has been provided by the National Development and Reform Commission. LAMOST is operated and managed by the National Astronomical Observatories, Chinese Academy of Sciences. 

\bibliographystyle{apj}
\bibliography{references.bib}

\begin{thebibliography}{}
\expandafter\ifx\csname natexlab\endcsname\relax\def\natexlab#1{#1}\fi

\bibitem[{{Abolfathi} {et~al.}(2018){Abolfathi}, {Aguado}, {Aguilar}, {Allende
  Prieto}, {Almeida}, {Ananna}, {Anders}, {Anderson}, {Andrews}, {Anguiano},
  {Arag{\'o}n-Salamanca}, {Argudo-Fern{\'a}ndez}, {Armengaud}, {Ata},
  {Aubourg}, {Avila-Reese}, {Badenes}, {Bailey}, {Balland}, {Barger},
  {Barrera-Ballesteros}, {Bartosz}, {Bastien}, {Bates}, {Baumgarten},
  {Bautista}, {Beaton}, {Beers}, {Belfiore}, {Bender}, {Bernardi}, {Bershady},
  {Beutler}, {Bird}, {Bizyaev}, {Blanc}, {Blanton}, {Blomqvist}, {Bolton},
  {Boquien}, {Borissova}, {Bovy}, {Bradna Diaz}, {Brandt}, {Brinkmann},
  {Brownstein}, {Bundy}, {Burgasser}, {Burtin}, {Busca}, {Ca{\~n}as},
  {Cano-D{\'\i}az}, {Cappellari}, {Carrera}, {Casey}, {Cervantes Sodi}, {Chen},
  {Cherinka}, {Chiappini}, {Choi}, {Chojnowski}, {Chuang}, {Chung}, {Clerc},
  {Cohen}, {Comerford}, {Comparat}, {Correa do Nascimento}, {da Costa},
  {Cousinou}, {Covey}, {Crane}, {Cruz-Gonzalez}, {Cunha}, {da Silva Ilha},
  {Damke}, {Darling}, {Davidson}, {Dawson}, {de Icaza Lizaola}, {de la
  Macorra}, {de la Torre}, {De Lee}, {de Sainte Agathe}, {Deconto Machado},
  {Dell'Agli}, {Delubac}, {Diamond-Stanic}, {Donor}, {Downes}, {Drory}, {du Mas
  des Bourboux}, {Duckworth}, {Dwelly}, {Dyer}, {Ebelke}, {Davis Eigenbrot},
  {Eisenstein}, {Elsworth}, {Emsellem}, {Eracleous}, {Erfanianfar},
  {Escoffier}, {Fan}, {Fern{\'a}ndez Alvar}, {Fernandez-Trincado}, {Fernand o
  Cirolini}, {Feuillet}, {Finoguenov}, {Fleming}, {Font-Ribera}, {Freischlad},
  {Frinchaboy}, {Fu}, {G{\'o}mez Maqueo Chew}, {Galbany}, {Garc{\'\i}a
  P{\'e}rez}, {Garcia-Dias}, {Garc{\'\i}a-Hern{\'a}ndez}, {Garma Oehmichen},
  {Gaulme}, {Gelfand }, {Gil-Mar{\'\i}n}, {Gillespie}, {Goddard}, {Gonz{\'a}lez
  Hern{\'a}ndez}, {Gonzalez-Perez}, {Grabowski}, {Green}, {Grier}, {Gueguen},
  {Guo}, {Guy}, {Hagen}, {Hall}, {Harding}, {Hasselquist}, {Hawley}, {Hayes},
  {Hearty}, {Hekker}, {Hernand ez}, {Hernandez Toledo}, {Hogg},
  {Holley-Bockelmann}, {Holtzman}, {Hou}, {Hsieh}, {Hunt}, {Hutchinson},
  {Hwang}, {Jimenez Angel}, {Johnson}, {Jones}, {J{\"o}nsson}, {Jullo}, {Khan},
  {Kinemuchi}, {Kirkby}, {Kirkpatrick}, {Kitaura}, {Knapp}, {Kneib},
  {Kollmeier}, {Lacerna}, {Lane}, {Lang}, {Law}, {Le Goff}, {Lee}, {Li}, {Li},
  {Lian}, {Liang}, {Lima}, {Lin}, {Long}, {Lucatello}, {Lundgren}, {Mackereth},
  {MacLeod}, {Mahadevan}, {Maia}, {Majewski}, {Manchado}, {Maraston},
  {Mariappan}, {Marques-Chaves}, {Masseron}, {Masters}, {McDermid}, {McGreer},
  {Melendez}, {Meneses-Goytia}, {Merloni}, {Merrifield}, {Meszaros}, {Meza},
  {Minchev}, {Minniti}, {Mueller}, {Muller-Sanchez}, {Muna}, {Mu{\~n}oz},
  {Myers}, {Nair}, {Nand ra}, {Ness}, {Newman}, {Nichol}, {Nidever},
  {Nitschelm}, {Noterdaeme}, {O'Connell}, {Oelkers}, {Oravetz}, {Oravetz},
  {Ort{\'\i}z}, {Osorio}, {Pace}, {Padilla}, {Palanque-Delabrouille},
  {Palicio}, {Pan}, {Pan}, {Parikh}, {P{\^a}ris}, {Park}, {Peirani},
  {Pellejero-Ibanez}, {Penny}, {Percival}, {Perez-Fournon}, {Petitjean},
  {Pieri}, {Pinsonneault}, {Pisani}, {Prada}, {Prakash}, {Queiroz}, {Raddick},
  {Raichoor}, {Barboza Rembold}, {Richstein}, {Riffel}, {Riffel}, {Rix},
  {Robin}, {Rodr{\'\i}guez Torres}, {Rom{\'a}n-Z{\'u}{\~n}iga}, {Ross},
  {Rossi}, {Ruan}, {Ruggeri}, {Ruiz}, {Salvato}, {S{\'a}nchez}, {S{\'a}nchez},
  {Sanchez Almeida}, {S{\'a}nchez-Gallego}, {Santana Rojas}, {Santiago},
  {Schiavon}, {Schimoia}, {Schlafly}, {Schlegel}, {Schneider}, {Schuster},
  {Schwope}, {Seo}, {Serenelli}, {Shen}, {Shen}, {Shetrone}, {Shull}, {Silva
  Aguirre}, {Simon}, {Skrutskie}, {Slosar}, {Smethurst}, {Smith}, {Sobeck},
  {Somers}, {Souter}, {Souto}, {Spindler}, {Stark}, {Stassun}, {Steinmetz},
  {Stello}, {Storchi-Bergmann}, {Streblyanska}, {Stringfellow}, {Su{\'a}rez},
  {Sun}, {Szigeti}, {Taghizadeh-Popp}, {Talbot}, {Tang}, {Tao}, {Tayar},
  {Tembe}, {Teske}, {Thakar}, {Thomas}, {Tissera}, {Tojeiro}, {Tremonti},
  {Troup}, {Urry}, {Valenzuela}, {van den Bosch}, {Vargas-Gonz{\'a}lez},
  {Vargas-Maga{\~n}a}, {Vazquez}, {Villanova}, {Vogt}, {Wake}, {Wang},
  {Weaver}, {Weijmans}, {Weinberg}, {Westfall}, {Whelan}, {Wilcots}, {Wild},
  {Williams}, {Wilson}, {Wood-Vasey}, {Wylezalek}, {Xiao}, {Yan}, {Yang},
  {Ybarra}, {Y{\`e}che}, {Zakamska}, {Zamora}, {Zarrouk}, {Zasowski}, {Zhang},
  {Zhao}, {Zhao}, {Zheng}, {Zheng}, {Zhou}, {Zhu}, {Zinn}, \& {Zou}}]{SDSSDR14}
{Abolfathi}, B., {Aguado}, D.~S., {Aguilar}, G., {et~al.} 2018, \apjs, 235, 42

\bibitem[{{Akaike}(1974)}]{Akaike1974}
{Akaike}, H. 1974, IEEE Transactions on Automatic Control, 19, 716

\bibitem[{{Amarante} {et~al.}(2019){Amarante}, {Silva}, {Debattista}, \&
  {Smith}}]{Amarante2020}
{Amarante}, J. A.~S., {Silva}, L. B.~e., {Debattista}, V.~P., \& {Smith}, M.~C.
  2019, arXiv e-prints, arXiv:1912.12690

\bibitem[{Anderson \& Darling(1952)}]{adtest}
Anderson, T.~W., \& Darling, D.~A. 1952, Ann. Math. Statist., 23, 193

\bibitem[{{Belokurov} {et~al.}(2018){Belokurov}, {Deason}, {Koposov},
  {Catelan}, {Erkal}, {Drake}, \& {Evans}}]{Belokurov2018b}
{Belokurov}, V., {Deason}, A.~J., {Koposov}, S.~E., {et~al.} 2018, \mnras, 477,
  1472

\bibitem[{{Belokurov} {et~al.}(2017){Belokurov}, {Erkal}, {Deason}, {Koposov},
  {De Angeli}, {Evans}, {Fraternali}, \& {Mackey}}]{Belokurov2017}
{Belokurov}, V., {Erkal}, D., {Deason}, A.~J., {et~al.} 2017, \mnras, 466, 4711

\bibitem[{{Belokurov} {et~al.}(2019){Belokurov}, {Sanders}, {Fattahi}, {Smith},
  {Deason}, {Evans}, \& {Grand }}]{Belokurov2019}
{Belokurov}, V., {Sanders}, J.~L., {Fattahi}, A., {et~al.} 2019, arXiv
  e-prints, arXiv:1909.04679

\bibitem[{{Belokurov} {et~al.}(2007){Belokurov}, {Evans}, {Bell}, {Irwin},
  {Hewett}, {Koposov}, {Rockosi}, {Gilmore}, {Zucker}, {Fellhauer},
  {Wilkinson}, {Bramich}, {Vidrih}, {Rix}, {Beers}, {Schneider}, {Barentine},
  {Brewington}, {Brinkmann}, {Harvanek}, {Krzesinski}, {Long}, {Pan},
  {Snedden}, {Malanushenko}, \& {Malanushenko}}]{Belokurov2007}
{Belokurov}, V., {Evans}, N.~W., {Bell}, E.~F., {et~al.} 2007, \apjl, 657, L89

\bibitem[{{Beraldo e Silva} {et~al.}(2020){Beraldo e Silva}, {Debattista},
  {Khachaturyants}, \& {Nidever}}]{BeraldoeSilvia2020}
{Beraldo e Silva}, L., {Debattista}, V.~P., {Khachaturyants}, T., \& {Nidever},
  D. 2020, \mnras, 492, 4716

\bibitem[{{Blanton} {et~al.}(2017){Blanton}, {Bershady}, {Abolfathi},
  {Albareti}, {Allende Prieto}, {Almeida}, {Alonso-Garc{\'\i}a}, {Anders},
  {Anderson}, {Andrews}, {Aquino-Ort{\'\i}z}, {Arag{\'o}n-Salamanca},
  {Argudo-Fern{\'a}ndez}, {Armengaud}, {Aubourg}, {Avila-Reese}, {Badenes},
  {Bailey}, {Barger}, {Barrera-Ballesteros}, {Bartosz}, {Bates}, {Baumgarten},
  {Bautista}, {Beaton}, {Beers}, {Belfiore}, {Bender}, {Berlind}, {Bernardi},
  {Beutler}, {Bird}, {Bizyaev}, {Blanc}, {Blomqvist}, {Bolton}, {Boquien},
  {Borissova}, {van den Bosch}, {Bovy}, {Brandt}, {Brinkmann}, {Brownstein},
  {Bundy}, {Burgasser}, {Burtin}, {Busca}, {Cappellari}, {Delgado Carigi},
  {Carlberg}, {Carnero Rosell}, {Carrera}, {Chanover}, {Cherinka}, {Cheung},
  {G{\'o}mez Maqueo Chew}, {Chiappini}, {Choi}, {Chojnowski}, {Chuang},
  {Chung}, {Cirolini}, {Clerc}, {Cohen}, {Comparat}, {da Costa}, {Cousinou},
  {Covey}, {Crane}, {Croft}, {Cruz-Gonzalez}, {Garrido Cuadra}, {Cunha},
  {Damke}, {Darling}, {Davies}, {Dawson}, {de la Macorra}, {Dell'Agli}, {De
  Lee}, {Delubac}, {Di Mille}, {Diamond-Stanic}, {Cano-D{\'\i}az}, {Donor},
  {Downes}, {Drory}, {du Mas des Bourboux}, {Duckworth}, {Dwelly}, {Dyer},
  {Ebelke}, {Eigenbrot}, {Eisenstein}, {Emsellem}, {Eracleous}, {Escoffier},
  {Evans}, {Fan}, {Fern{\'a}ndez-Alvar}, {Fernandez-Trincado}, {Feuillet},
  {Finoguenov}, {Fleming}, {Font-Ribera}, {Fredrickson}, {Freischlad},
  {Frinchaboy}, {Fuentes}, {Galbany}, {Garcia-Dias},
  {Garc{\'\i}a-Hern{\'a}ndez}, {Gaulme}, {Geisler}, {Gelfand},
  {Gil-Mar{\'\i}n}, {Gillespie}, {Goddard}, {Gonzalez-Perez}, {Grabowski},
  {Green}, {Grier}, {Gunn}, {Guo}, {Guy}, {Hagen}, {Hahn}, {Hall}, {Harding},
  {Hasselquist}, {Hawley}, {Hearty}, {Gonzalez Hern{\'a}ndez}, {Ho}, {Hogg},
  {Holley-Bockelmann}, {Holtzman}, {Holzer}, {Huehnerhoff}, {Hutchinson},
  {Hwang}, {Ibarra-Medel}, {da Silva Ilha}, {Ivans}, {Ivory}, {Jackson},
  {Jensen}, {Johnson}, {Jones}, {J{\"o}nsson}, {Jullo}, {Kamble}, {Kinemuchi},
  {Kirkby}, {Kitaura}, {Klaene}, {Knapp}, {Kneib}, {Kollmeier}, {Lacerna},
  {Lane}, {Lang}, {Law}, {Lazarz}, {Lee}, {Le Goff}, {Liang}, {Li}, {Li},
  {Lian}, {Lima}, {Lin}, {Lin}, {Bertran de Lis}, {Liu}, {de Icaza Lizaola},
  {Long}, {Lucatello}, {Lundgren}, {MacDonald}, {Deconto Machado}, {MacLeod},
  {Mahadevan}, {Geimba Maia}, {Maiolino}, {Majewski}, {Malanushenko},
  {Malanushenko}, {Manchado}, {Mao}, {Maraston}, {Marques-Chaves}, {Masseron},
  {Masters}, {McBride}, {McDermid}, {McGrath}, {McGreer}, {Medina Pe{\~n}a},
  {Melendez}, {Merloni}, {Merrifield}, {Meszaros}, {Meza}, {Minchev},
  {Minniti}, {Miyaji}, {More}, {Mulchaey}, {M{\"u}ller-S{\'a}nchez}, {Muna},
  {Munoz}, {Myers}, {Nair}, {Nandra}, {Correa do Nascimento}, {Negrete},
  {Ness}, {Newman}, {Nichol}, {Nidever}, {Nitschelm}, {Ntelis}, {O'Connell},
  {Oelkers}, {Oravetz}, {Oravetz}, {Pace}, {Padilla}, {Palanque-Delabrouille},
  {Alonso Palicio}, {Pan}, {Parejko}, {Parikh}, {P{\^a}ris}, {Park}, {Patten},
  {Peirani}, {Pellejero-Ibanez}, {Penny}, {Percival}, {Perez-Fournon},
  {Petitjean}, {Pieri}, {Pinsonneault}, {Pisani}, {Poleski}, {Prada},
  {Prakash}, {Queiroz}, {Raddick}, {Raichoor}, {Barboza Rembold}, {Richstein},
  {Riffel}, {Riffel}, {Rix}, {Robin}, {Rockosi}, {Rodr{\'\i}guez-Torres},
  {Roman-Lopes}, {Rom{\'a}n-Z{\'u}{\~n}iga}, {Rosado}, {Ross}, {Rossi}, {Ruan},
  {Ruggeri}, {Rykoff}, {Salazar-Albornoz}, {Salvato}, {S{\'a}nchez}, {Aguado},
  {S{\'a}nchez-Gallego}, {Santana}, {Santiago}, {Sayres}, {Schiavon}, {da Silva
  Schimoia}, {Schlafly}, {Schlegel}, {Schneider}, {Schultheis}, {Schuster},
  {Schwope}, {Seo}, {Shao}, {Shen}, {Shetrone}, {Shull}, {Simon}, {Skinner},
  {Skrutskie}, {Slosar}, {Smith}, {Sobeck}, {Sobreira}, {Somers}, {Souto},
  {Stark}, {Stassun}, {Stauffer}, {Steinmetz}, {Storchi-Bergmann},
  {Streblyanska}, {Stringfellow}, {Su{\'a}rez}, {Sun}, {Suzuki}, {Szigeti},
  {Taghizadeh-Popp}, {Tang}, {Tao}, {Tayar}, {Tembe}, {Teske}, {Thakar},
  {Thomas}, {Thompson}, {Tinker}, {Tissera}, {Tojeiro}, {Hernandez Toledo}, {de
  la Torre}, {Tremonti}, {Troup}, {Valenzuela}, {Martinez Valpuesta},
  {Vargas-Gonz{\'a}lez}, {Vargas-Maga{\~n}a}, {Vazquez}, {Villanova}, {Vivek},
  {Vogt}, {Wake}, {Walterbos}, {Wang}, {Weaver}, {Weijmans}, {Weinberg},
  {Westfall}, {Whelan}, {Wild}, {Wilson}, {Wood-Vasey}, {Wylezalek}, {Xiao},
  {Yan}, {Yang}, {Ybarra}, {Y{\`e}che}, {Zakamska}, {Zamora}, {Zarrouk},
  {Zasowski}, {Zhang}, {Zhao}, {Zheng}, {Zheng}, {Zhou}, {Zhou}, {Zhu},
  {Zoccali}, \& {Zou}}]{SDSSIV}
{Blanton}, M.~R., {Bershady}, M.~A., {Abolfathi}, B., {et~al.} 2017, \aj, 154,
  28

\bibitem[{{Bovy}(2015)}]{Bovy2015}
{Bovy}, J. 2015, \apjs, 216, 29

\bibitem[{{Brown} {et~al.}(2014){Brown}, {Tumlinson}, {Geha}, {Simon},
  {Vargas}, {VandenBerg}, {Kirby}, {Kalirai}, {Avila}, {Gennaro}, {Ferguson},
  {Mu{\~n}oz}, {Guhathakurta}, \& {Renzini}}]{Brown2014}
{Brown}, T.~M., {Tumlinson}, J., {Geha}, M., {et~al.} 2014, \apj, 796, 91

\bibitem[{{Carlin} {et~al.}(2013){Carlin}, {DeLaunay}, {Newberg}, {Deng},
  {Gole}, {Grabowski}, {Jin}, {Liu}, {Liu}, {Luo}, {Yuan}, {Zhang}, {Zhao}, \&
  {Zhao}}]{Carlin2013}
{Carlin}, J.~L., {DeLaunay}, J., {Newberg}, H.~J., {et~al.} 2013, \apjl, 777,
  L5

\bibitem[{{Clarke} {et~al.}(2019){Clarke}, {Debattista}, {Nidever}, {Loebman},
  {Simons}, {Kassin}, {Du}, {Ness}, {Fisher}, {Quinn}, {Wadsley}, {Freeman}, \&
  {Popescu}}]{Clarke2019}
{Clarke}, A.~J., {Debattista}, V.~P., {Nidever}, D.~L., {et~al.} 2019, \mnras,
  484, 3476

\bibitem[{{Clementini} {et~al.}(2019){Clementini}, {Ripepi}, {Molinaro},
  {Garofalo}, {Muraveva}, {Rimoldini}, {Guy}, {Jevardat de Fombelle},
  {Nienartowicz}, {Marchal}, {Audard}, {Holl}, {Leccia}, {Marconi}, {Musella},
  {Mowlavi}, {Lecoeur-Taibi}, {Eyer}, {De Ridder}, {Regibo}, {Sarro},
  {Szabados}, {Evans}, \& {Riello}}]{Clementini2019}
{Clementini}, G., {Ripepi}, V., {Molinaro}, R., {et~al.} 2019, \aap, 622, A60

\bibitem[{{Deason} {et~al.}(2011){Deason}, {Belokurov}, \&
  {Evans}}]{Deason2011}
{Deason}, A.~J., {Belokurov}, V., \& {Evans}, N.~W. 2011, \mnras, 416, 2903

\bibitem[{{Deason} {et~al.}(2013){Deason}, {Belokurov}, {Evans}, \&
  {Johnston}}]{Deason2013}
{Deason}, A.~J., {Belokurov}, V., {Evans}, N.~W., \& {Johnston}, K.~V. 2013,
  \apj, 763, 113

\bibitem[{{Deason} {et~al.}(2018){Deason}, {Belokurov}, {Koposov}, \&
  {Lancaster}}]{Deason2018}
{Deason}, A.~J., {Belokurov}, V., {Koposov}, S.~E., \& {Lancaster}, L. 2018,
  \apjl, 862, L1

\bibitem[{{Deason} {et~al.}(2019){Deason}, {Belokurov}, \&
  {Sanders}}]{Deason2019}
{Deason}, A.~J., {Belokurov}, V., \& {Sanders}, J.~L. 2019, \mnras, 490, 3426

\bibitem[{Dempster {et~al.}(1977)Dempster, Laird, \& Rubin}]{Dempster1977}
Dempster, A.~P., Laird, N.~M., \& Rubin, D.~B. 1977, Journal of the Royal
  Statistical Society. Series B (Methodological), 39, 1

\bibitem[{{Deng} {et~al.}(2012){Deng}, {Newberg}, {Liu}, {Carlin}, {Beers},
  {Chen}, {Chen}, {Christlieb}, {Grillmair}, {Guhathakurta}, {Han}, {Hou},
  {Lee}, {L{\'e}pine}, {Li}, {Liu}, {Pan}, {Sellwood}, {Wang}, {Wang}, {Yang},
  {Yanny}, {Zhang}, {Zhang}, {Zheng}, \& {Zhu}}]{LEGUE}
{Deng}, L.-C., {Newberg}, H.~J., {Liu}, C., {et~al.} 2012, Research in
  Astronomy and Astrophysics, 12, 735

\bibitem[{{Donlon} {et~al.}(2019){Donlon}, {Newberg}, {Weiss}, {Amy}, \&
  {Thompson}}]{Donlon2019}
{Donlon}, Thomas, I., {Newberg}, H.~J., {Weiss}, J., {Amy}, P., \& {Thompson},
  J. 2019, arXiv e-prints, arXiv:1903.10136

\bibitem[{{Duffau} {et~al.}(2014){Duffau}, {Vivas}, {Zinn}, {M{\'e}ndez}, \&
  {Ruiz}}]{Duffau2014}
{Duffau}, S., {Vivas}, A.~K., {Zinn}, R., {M{\'e}ndez}, R.~A., \& {Ruiz}, M.~T.
  2014, \aap, 566, A118

\bibitem[{{Duffau} {et~al.}(2006){Duffau}, {Zinn}, {Vivas}, {Carraro},
  {M{\'e}ndez}, {Winnick}, \& {Gallart}}]{Duffau2006}
{Duffau}, S., {Zinn}, R., {Vivas}, A.~K., {et~al.} 2006, \apjl, 636, L97

\bibitem[{{Erkal} {et~al.}(2019){Erkal}, {Belokurov}, {Laporte}, {Koposov},
  {Li}, {Grillmair}, {Kallivayalil}, {Price-Whelan}, {Evans}, {Hawkins},
  {Hendel}, {Mateu}, {Navarro}, {del Pino}, {Slater}, {Sohn}, \& {Orphan Aspen
  Treasury Collaboration}}]{Erkal2019}
{Erkal}, D., {Belokurov}, V., {Laporte}, C.~F.~P., {et~al.} 2019, \mnras, 487,
  2685

\bibitem[{{Fardal} {et~al.}(2019){Fardal}, {van der Marel}, {Law}, {Sohn},
  {Sesar}, {Hernitschek}, \& {Rix}}]{Fardal2019}
{Fardal}, M.~A., {van der Marel}, R.~P., {Law}, D.~R., {et~al.} 2019, \mnras,
  483, 4724

\bibitem[{{Gaia Collaboration} {et~al.}(2016){Gaia Collaboration}, {Prusti},
  {de Bruijne}, {Brown}, {Vallenari}, {Babusiaux}, {Bailer-Jones}, {Bastian},
  {Biermann}, {Evans}, {Eyer}, {Jansen}, {Jordi}, {Klioner}, {Lammers},
  {Lindegren}, {Luri}, {Mignard}, {Milligan}, {Panem}, {Poinsignon},
  {Pourbaix}, {Randich}, {Sarri}, {Sartoretti}, {Siddiqui}, {Soubiran},
  {Valette}, {van Leeuwen}, {Walton}, {Aerts}, {Arenou}, {Cropper}, {Drimmel},
  {H{\o}g}, {Katz}, {Lattanzi}, {O'Mullane}, {Grebel}, {Holland}, {Huc},
  {Passot}, {Bramante}, {Cacciari}, {Casta{\~n}eda}, {Chaoul}, {Cheek}, {De
  Angeli}, {Fabricius}, {Guerra}, {Hern{\'a}ndez}, {Jean-Antoine-Piccolo},
  {Masana}, {Messineo}, {Mowlavi}, {Nienartowicz}, {Ord{\'o}{\~n}ez-Blanco},
  {Panuzzo}, {Portell}, {Richards}, {Riello}, {Seabroke}, {Tanga},
  {Th{\'e}venin}, {Torra}, {Els}, {Gracia-Abril}, {Comoretto},
  {Garcia-Reinaldos}, {Lock}, {Mercier}, {Altmann}, {Andrae}, {Astraatmadja},
  {Bellas-Velidis}, {Benson}, {Berthier}, {Blomme}, {Busso}, {Carry},
  {Cellino}, {Clementini}, {Cowell}, {Creevey}, {Cuypers}, {Davidson}, {De
  Ridder}, {de Torres}, {Delchambre}, {Dell'Oro}, {Ducourant}, {Fr{\'e}mat},
  {Garc{\'\i}a-Torres}, {Gosset}, {Halbwachs}, {Hambly}, {Harrison}, {Hauser},
  {Hestroffer}, {Hodgkin}, {Huckle}, {Hutton}, {Jasniewicz}, {Jordan},
  {Kontizas}, {Korn}, {Lanzafame}, {Manteiga}, {Moitinho}, {Muinonen},
  {Osinde}, {Pancino}, {Pauwels}, {Petit}, {Recio-Blanco}, {Robin}, {Sarro},
  {Siopis}, {Smith}, {Smith}, {Sozzetti}, {Thuillot}, {van Reeven}, {Viala},
  {Abbas}, {Abreu Aramburu}, {Accart}, {Aguado}, {Allan}, {Allasia},
  {Altavilla}, {{\'A}lvarez}, {Alves}, {Anderson}, {Andrei}, {Anglada Varela},
  {Antiche}, {Antoja}, {Ant{\'o}n}, {Arcay}, {Atzei}, {Ayache}, {Bach},
  {Baker}, {Balaguer-N{\'u}{\~n}ez}, {Barache}, {Barata}, {Barbier}, {Barblan},
  {Baroni}, {Barrado y Navascu{\'e}s}, {Barros}, {Barstow}, {Becciani},
  {Bellazzini}, {Bellei}, {Bello Garc{\'\i}a}, {Belokurov}, {Bendjoya},
  {Berihuete}, {Bianchi}, {Bienaym{\'e}}, {Billebaud}, {Blagorodnova},
  {Blanco-Cuaresma}, {Boch}, {Bombrun}, {Borrachero}, {Bouquillon}, {Bourda},
  {Bouy}, {Bragaglia}, {Breddels}, {Brouillet}, {Br{\"u}semeister},
  {Bucciarelli}, {Budnik}, {Burgess}, {Burgon}, {Burlacu}, {Busonero}, {Buzzi},
  {Caffau}, {Cambras}, {Campbell}, {Cancelliere}, {Cantat-Gaudin}, {Carlucci},
  {Carrasco}, {Castellani}, {Charlot}, {Charnas}, {Charvet}, {Chassat},
  {Chiavassa}, {Clotet}, {Cocozza}, {Collins}, {Collins}, {Costigan}, {Crifo},
  {Cross}, {Crosta}, {Crowley}, {Dafonte}, {Damerdji}, {Dapergolas}, {David},
  {David}, {De Cat}, {de Felice}, {de Laverny}, {De Luise}, {De March}, {de
  Martino}, {de Souza}, {Debosscher}, {del Pozo}, {Delbo}, {Delgado},
  {Delgado}, {di Marco}, {Di Matteo}, {Diakite}, {Distefano}, {Dolding}, {Dos
  Anjos}, {Drazinos}, {Dur{\'a}n}, {Dzigan}, {Ecale}, {Edvardsson}, {Enke},
  {Erdmann}, {Escolar}, {Espina}, {Evans}, {Eynard Bontemps}, {Fabre},
  {Fabrizio}, {Faigler}, {Falc{\~a}o}, {Farr{\`a}s Casas}, {Faye}, {Federici},
  {Fedorets}, {Fern{\'a}ndez-Hern{\'a}ndez}, {Fernique}, {Fienga}, {Figueras},
  {Filippi}, {Findeisen}, {Fonti}, {Fouesneau}, {Fraile}, {Fraser}, {Fuchs},
  {Furnell}, {Gai}, {Galleti}, {Galluccio}, {Garabato}, {Garc{\'\i}a-Sedano},
  {Gar{\'e}}, {Garofalo}, {Garralda}, {Gavras}, {Gerssen}, {Geyer}, {Gilmore},
  {Girona}, {Giuffrida}, {Gomes}, {Gonz{\'a}lez-Marcos},
  {Gonz{\'a}lez-N{\'u}{\~n}ez}, {Gonz{\'a}lez-Vidal}, {Granvik}, {Guerrier},
  {Guillout}, {Guiraud}, {G{\'u}rpide}, {Guti{\'e}rrez-S{\'a}nchez}, {Guy},
  {Haigron}, {Hatzidimitriou}, {Haywood}, {Heiter}, {Helmi}, {Hobbs},
  {Hofmann}, {Holl}, {Holland }, {Hunt}, {Hypki}, {Icardi}, {Irwin}, {Jevardat
  de Fombelle}, {Jofr{\'e}}, {Jonker}, {Jorissen}, {Julbe}, {Karampelas},
  {Kochoska}, {Kohley}, {Kolenberg}, {Kontizas}, {Koposov}, {Kordopatis},
  {Koubsky}, {Kowalczyk}, {Krone-Martins}, {Kudryashova}, {Kull}, {Bachchan},
  {Lacoste-Seris}, {Lanza}, {Lavigne}, {Le Poncin-Lafitte}, {Lebreton},
  {Lebzelter}, {Leccia}, {Leclerc}, {Lecoeur-Taibi}, {Lemaitre}, {Lenhardt},
  {Leroux}, {Liao}, {Licata}, {Lindstr{\o}m}, {Lister}, {Livanou}, {Lobel},
  {L{\"o}ffler}, {L{\'o}pez}, {Lopez-Lozano}, {Lorenz}, {Loureiro},
  {MacDonald}, {Magalh{\~a}es Fernandes}, {Managau}, {Mann}, {Mantelet},
  {Marchal}, {Marchant}, {Marconi}, {Marie}, {Marinoni}, {Marrese},
  {Marschalk{\'o}}, {Marshall}, {Mart{\'\i}n-Fleitas}, {Martino}, {Mary},
  {Matijevi{\v{c}}}, {Mazeh}, {McMillan}, {Messina}, {Mestre}, {Michalik},
  {Millar}, {Miranda}, {Molina}, {Molinaro}, {Molinaro}, {Moln{\'a}r},
  {Moniez}, {Montegriffo}, {Monteiro}, {Mor}, {Mora}, {Morbidelli}, {Morel},
  {Morgenthaler}, {Morley}, {Morris}, {Mulone}, {Muraveva}, {Musella},
  {Narbonne}, {Nelemans}, {Nicastro}, {Noval}, {Ord{\'e}novic},
  {Ordieres-Mer{\'e}}, {Osborne}, {Pagani}, {Pagano}, {Pailler}, {Palacin},
  {Palaversa}, {Parsons}, {Paulsen}, {Pecoraro}, {Pedrosa}, {Pentik{\"a}inen},
  {Pereira}, {Pichon}, {Piersimoni}, {Pineau}, {Plachy}, {Plum}, {Poujoulet},
  {Pr{\v{s}}a}, {Pulone}, {Ragaini}, {Rago}, {Rambaux}, {Ramos-Lerate},
  {Ranalli}, {Rauw}, {Read}, {Regibo}, {Renk}, {Reyl{\'e}}, {Ribeiro},
  {Rimoldini}, {Ripepi}, {Riva}, {Rixon}, {Roelens}, {Romero-G{\'o}mez},
  {Rowell}, {Royer}, {Rudolph}, {Ruiz-Dern}, {Sadowski}, {Sagrist{\`a}
  Sell{\'e}s}, {Sahlmann}, {Salgado}, {Salguero}, {Sarasso}, {Savietto},
  {Schnorhk}, {Schultheis}, {Sciacca}, {Segol}, {Segovia}, {Segransan},
  {Serpell}, {Shih}, {Smareglia}, {Smart}, {Smith}, {Solano}, {Solitro},
  {Sordo}, {Soria Nieto}, {Souchay}, {Spagna}, {Spoto}, {Stampa}, {Steele},
  {Steidelm{\"u}ller}, {Stephenson}, {Stoev}, {Suess}, {S{\"u}veges}, {Surdej},
  {Szabados}, {Szegedi-Elek}, {Tapiador}, {Taris}, {Tauran}, {Taylor},
  {Teixeira}, {Terrett}, {Tingley}, {Trager}, {Turon}, {Ulla}, {Utrilla},
  {Valentini}, {van Elteren}, {Van Hemelryck}, {van Leeuwen}, {Varadi},
  {Vecchiato}, {Veljanoski}, {Via}, {Vicente}, {Vogt}, {Voss}, {Votruba},
  {Voutsinas}, {Walmsley}, {Weiler}, {Weingrill}, {Werner}, {Wevers},
  {Whitehead}, {Wyrzykowski}, {Yoldas}, {{\v{Z}}erjal}, {Zucker}, {Zurbach},
  {Zwitter}, {Alecu}, {Allen}, {Allende Prieto}, {Amorim},
  {Anglada-Escud{\'e}}, {Arsenijevic}, {Azaz}, {Balm}, {Beck}, {Bernstein},
  {Bigot}, {Bijaoui}, {Blasco}, {Bonfigli}, {Bono}, {Boudreault}, {Bressan},
  {Brown}, {Brunet}, {Bunclark}, {Buonanno}, {Butkevich}, {Carret}, {Carrion},
  {Chemin}, {Ch{\'e}reau}, {Corcione}, {Darmigny}, {de Boer}, {de Teodoro}, {de
  Zeeuw}, {Delle Luche}, {Domingues}, {Dubath}, {Fodor}, {Fr{\'e}zouls},
  {Fries}, {Fustes}, {Fyfe}, {Gallardo}, {Gallegos}, {Gardiol}, {Gebran},
  {Gomboc}, {G{\'o}mez}, {Grux}, {Gueguen}, {Heyrovsky}, {Hoar}, {Iannicola},
  {Isasi Parache}, {Janotto}, {Joliet}, {Jonckheere}, {Keil}, {Kim},
  {Klagyivik}, {Klar}, {Knude}, {Kochukhov}, {Kolka}, {Kos}, {Kutka}, {Lainey},
  {LeBouquin}, {Liu}, {Loreggia}, {Makarov}, {Marseille}, {Martayan},
  {Martinez-Rubi}, {Massart}, {Meynadier}, {Mignot}, {Munari}, {Nguyen},
  {Nordlander}, {Ocvirk}, {O'Flaherty}, {Olias Sanz}, {Ortiz}, {Osorio},
  {Oszkiewicz}, {Ouzounis}, {Palmer}, {Park}, {Pasquato}, {Peltzer}, {Peralta},
  {P{\'e}turaud}, {Pieniluoma}, {Pigozzi}, {Poels}, {Prat}, {Prod'homme},
  {Raison}, {Rebordao}, {Risquez}, {Rocca-Volmerange}, {Rosen}, {Ruiz-Fuertes},
  {Russo}, {Sembay}, {Serraller Vizcaino}, {Short}, {Siebert}, {Silva},
  {Sinachopoulos}, {Slezak}, {Soffel}, {Sosnowska}, {Strai{\v{z}}ys}, {ter
  Linden}, {Terrell}, {Theil}, {Tiede}, {Troisi}, {Tsalmantza}, {Tur},
  {Vaccari}, {Vachier}, {Valles}, {Van Hamme}, {Veltz}, {Virtanen}, {Wallut},
  {Wichmann}, {Wilkinson}, {Ziaeepour}, \& {Zschocke}}]{Gaiamission}
{Gaia Collaboration}, {Prusti}, T., {de Bruijne}, J.~H.~J., {et~al.} 2016,
  \aap, 595, A1

\bibitem[{{Gaia Collaboration} {et~al.}(2018){Gaia Collaboration}, {Brown},
  {Vallenari}, {Prusti}, {de Bruijne}, {Babusiaux}, {Bailer-Jones}, {Biermann},
  {Evans}, {Eyer}, {Jansen}, {Jordi}, {Klioner}, {Lammers}, {Lindegren},
  {Luri}, {Mignard}, {Panem}, {Pourbaix}, {Randich}, {Sartoretti}, {Siddiqui},
  {Soubiran}, {van Leeuwen}, {Walton}, {Arenou}, {Bastian}, {Cropper},
  {Drimmel}, {Katz}, {Lattanzi}, {Bakker}, {Cacciari}, {Casta{\~n}eda},
  {Chaoul}, {Cheek}, {De Angeli}, {Fabricius}, {Guerra}, {Holl}, {Masana},
  {Messineo}, {Mowlavi}, {Nienartowicz}, {Panuzzo}, {Portell}, {Riello},
  {Seabroke}, {Tanga}, {Th{\'e}venin}, {Gracia-Abril}, {Comoretto},
  {Garcia-Reinaldos}, {Teyssier}, {Altmann}, {Andrae}, {Audard},
  {Bellas-Velidis}, {Benson}, {Berthier}, {Blomme}, {Burgess}, {Busso},
  {Carry}, {Cellino}, {Clementini}, {Clotet}, {Creevey}, {Davidson}, {De
  Ridder}, {Delchambre}, {Dell'Oro}, {Ducourant},
  {Fern{\'a}ndez-Hern{\'a}ndez}, {Fouesneau}, {Fr{\'e}mat}, {Galluccio},
  {Garc{\'\i}a-Torres}, {Gonz{\'a}lez-N{\'u}{\~n}ez}, {Gonz{\'a}lez-Vidal},
  {Gosset}, {Guy}, {Halbwachs}, {Hambly}, {Harrison}, {Hern{\'a}ndez},
  {Hestroffer}, {Hodgkin}, {Hutton}, {Jasniewicz}, {Jean-Antoine-Piccolo},
  {Jordan}, {Korn}, {Krone-Martins}, {Lanzafame}, {Lebzelter}, {L{\"o}ffler},
  {Manteiga}, {Marrese}, {Mart{\'\i}n-Fleitas}, {Moitinho}, {Mora}, {Muinonen},
  {Osinde}, {Pancino}, {Pauwels}, {Petit}, {Recio-Blanco}, {Richards},
  {Rimoldini}, {Robin}, {Sarro}, {Siopis}, {Smith}, {Sozzetti}, {S{\"u}veges},
  {Torra}, {van Reeven}, {Abbas}, {Abreu Aramburu}, {Accart}, {Aerts},
  {Altavilla}, {{\'A}lvarez}, {Alvarez}, {Alves}, {Anderson}, {Andrei},
  {Anglada Varela}, {Antiche}, {Antoja}, {Arcay}, {Astraatmadja}, {Bach},
  {Baker}, {Balaguer-N{\'u}{\~n}ez}, {Balm}, {Barache}, {Barata}, {Barbato},
  {Barblan}, {Barklem}, {Barrado}, {Barros}, {Barstow}, {Bartholom{\'e}
  Mu{\~n}oz}, {Bassilana}, {Becciani}, {Bellazzini}, {Berihuete}, {Bertone},
  {Bianchi}, {Bienaym{\'e}}, {Blanco-Cuaresma}, {Boch}, {Boeche}, {Bombrun},
  {Borrachero}, {Bossini}, {Bouquillon}, {Bourda}, {Bragaglia}, {Bramante},
  {Breddels}, {Bressan}, {Brouillet}, {Br{\"u}semeister}, {Brugaletta},
  {Bucciarelli}, {Burlacu}, {Busonero}, {Butkevich}, {Buzzi}, {Caffau},
  {Cancelliere}, {Cannizzaro}, {Cantat-Gaudin}, {Carballo}, {Carlucci},
  {Carrasco}, {Casamiquela}, {Castellani}, {Castro-Ginard}, {Charlot},
  {Chemin}, {Chiavassa}, {Cocozza}, {Costigan}, {Cowell}, {Crifo}, {Crosta},
  {Crowley}, {Cuypers}, {Dafonte}, {Damerdji}, {Dapergolas}, {David}, {David},
  {de Laverny}, {De Luise}, {De March}, {de Martino}, {de Souza}, {de Torres},
  {Debosscher}, {del Pozo}, {Delbo}, {Delgado}, {Delgado}, {Di Matteo},
  {Diakite}, {Diener}, {Distefano}, {Dolding}, {Drazinos}, {Dur{\'a}n},
  {Edvardsson}, {Enke}, {Eriksson}, {Esquej}, {Eynard Bontemps}, {Fabre},
  {Fabrizio}, {Faigler}, {Falc{\~a}o}, {Farr{\`a}s Casas}, {Federici},
  {Fedorets}, {Fernique}, {Figueras}, {Filippi}, {Findeisen}, {Fonti},
  {Fraile}, {Fraser}, {Fr{\'e}zouls}, {Gai}, {Galleti}, {Garabato},
  {Garc{\'\i}a-Sedano}, {Garofalo}, {Garralda}, {Gavel}, {Gavras}, {Gerssen},
  {Geyer}, {Giacobbe}, {Gilmore}, {Girona}, {Giuffrida}, {Glass}, {Gomes},
  {Granvik}, {Gueguen}, {Guerrier}, {Guiraud}, {Guti{\'e}rrez-S{\'a}nchez},
  {Haigron}, {Hatzidimitriou}, {Hauser}, {Haywood}, {Heiter}, {Helmi}, {Heu},
  {Hilger}, {Hobbs}, {Hofmann}, {Holland}, {Huckle}, {Hypki}, {Icardi},
  {Jan{\ss}en}, {Jevardat de Fombelle}, {Jonker}, {Juh{\'a}sz}, {Julbe},
  {Karampelas}, {Kewley}, {Klar}, {Kochoska}, {Kohley}, {Kolenberg},
  {Kontizas}, {Kontizas}, {Koposov}, {Kordopatis}, {Kostrzewa-Rutkowska},
  {Koubsky}, {Lambert}, {Lanza}, {Lasne}, {Lavigne}, {Le Fustec}, {Le
  Poncin-Lafitte}, {Lebreton}, {Leccia}, {Leclerc}, {Lecoeur-Taibi},
  {Lenhardt}, {Leroux}, {Liao}, {Licata}, {Lindstr{\o}m}, {Lister}, {Livanou},
  {Lobel}, {L{\'o}pez}, {Managau}, {Mann}, {Mantelet}, {Marchal}, {Marchant},
  {Marconi}, {Marinoni}, {Marschalk{\'o}}, {Marshall}, {Martino}, {Marton},
  {Mary}, {Massari}, {Matijevi{\v{c}}}, {Mazeh}, {McMillan}, {Messina},
  {Michalik}, {Millar}, {Molina}, {Molinaro}, {Moln{\'a}r}, {Montegriffo},
  {Mor}, {Morbidelli}, {Morel}, {Morris}, {Mulone}, {Muraveva}, {Musella},
  {Nelemans}, {Nicastro}, {Noval}, {O'Mullane}, {Ord{\'e}novic},
  {Ord{\'o}{\~n}ez-Blanco}, {Osborne}, {Pagani}, {Pagano}, {Pailler},
  {Palacin}, {Palaversa}, {Panahi}, {Pawlak}, {Piersimoni}, {Pineau}, {Plachy},
  {Plum}, {Poggio}, {Poujoulet}, {Pr{\v{s}}a}, {Pulone}, {Racero}, {Ragaini},
  {Rambaux}, {Ramos-Lerate}, {Regibo}, {Reyl{\'e}}, {Riclet}, {Ripepi}, {Riva},
  {Rivard}, {Rixon}, {Roegiers}, {Roelens}, {Romero-G{\'o}mez}, {Rowell},
  {Royer}, {Ruiz-Dern}, {Sadowski}, {Sagrist{\`a} Sell{\'e}s}, {Sahlmann},
  {Salgado}, {Salguero}, {Sanna}, {Santana-Ros}, {Sarasso}, {Savietto},
  {Schultheis}, {Sciacca}, {Segol}, {Segovia}, {S{\'e}gransan}, {Shih},
  {Siltala}, {Silva}, {Smart}, {Smith}, {Solano}, {Solitro}, {Sordo}, {Soria
  Nieto}, {Souchay}, {Spagna}, {Spoto}, {Stampa}, {Steele},
  {Steidelm{\"u}ller}, {Stephenson}, {Stoev}, {Suess}, {Surdej}, {Szabados},
  {Szegedi-Elek}, {Tapiador}, {Taris}, {Tauran}, {Taylor}, {Teixeira},
  {Terrett}, {Teyssand ier}, {Thuillot}, {Titarenko}, {Torra Clotet}, {Turon},
  {Ulla}, {Utrilla}, {Uzzi}, {Vaillant}, {Valentini}, {Valette}, {van Elteren},
  {Van Hemelryck}, {van Leeuwen}, {Vaschetto}, {Vecchiato}, {Veljanoski},
  {Viala}, {Vicente}, {Vogt}, {von Essen}, {Voss}, {Votruba}, {Voutsinas},
  {Walmsley}, {Weiler}, {Wertz}, {Wevers}, {Wyrzykowski}, {Yoldas},
  {{\v{Z}}erjal}, {Ziaeepour}, {Zorec}, {Zschocke}, {Zucker}, {Zurbach}, \&
  {Zwitter}}]{Gaiadr2contents}
{Gaia Collaboration}, {Brown}, A.~G.~A., {Vallenari}, A., {et~al.} 2018, \aap,
  616, A1

\bibitem[{{Gallart} {et~al.}(2019){Gallart}, {Bernard}, {Brook}, {Ruiz-Lara},
  {Cassisi}, {Hill}, \& {Monelli}}]{Gallart2019}
{Gallart}, C., {Bernard}, E.~J., {Brook}, C.~B., {et~al.} 2019, Nature
  Astronomy, 3, 932

\bibitem[{{Gilmore} {et~al.}(2002){Gilmore}, {Wyse}, \& {Norris}}]{Gilmore2002}
{Gilmore}, G., {Wyse}, R. F.~G., \& {Norris}, J.~E. 2002, \apjl, 574, L39

\bibitem[{{Grillmair}(2009)}]{Grillmair2009}
{Grillmair}, C.~J. 2009, \apj, 693, 1118

\bibitem[{Hartigan \& Hartigan(1985)}]{diptest}
Hartigan, J.~A., \& Hartigan, P.~M. 1985, Ann. Statist., 13, 70

\bibitem[{{Helmi} {et~al.}(2018){Helmi}, {Babusiaux}, {Koppelman}, {Massari},
  {Veljanoski}, \& {Brown}}]{Helmi2018}
{Helmi}, A., {Babusiaux}, C., {Koppelman}, H.~H., {et~al.} 2018, \nat, 563, 85

\bibitem[{{Hernquist} \& {Quinn}(1988)}]{Hernquist1988}
{Hernquist}, L., \& {Quinn}, P.~J. 1988, \apj, 331, 682

\bibitem[{Hurvich \& Tsai(1989)}]{Hurvich1989}
Hurvich, C.~M., \& Tsai, C.-L. 1989, Biometrika, 76, 297

\bibitem[{{Ibata} {et~al.}(2001){Ibata}, {Lewis}, {Irwin}, {Totten}, \&
  {Quinn}}]{Ibata2001}
{Ibata}, R., {Lewis}, G.~F., {Irwin}, M., {Totten}, E., \& {Quinn}, T. 2001,
  \apj, 551, 294

\bibitem[{{Iorio} \& {Belokurov}(2019)}]{Iorio2019}
{Iorio}, G., \& {Belokurov}, V. 2019, \mnras, 482, 3868

\bibitem[{{Ivezi{\'c}} {et~al.}(2012){Ivezi{\'c}}, {Beers}, \&
  {Juri{\'c}}}]{Ivezic2012}
{Ivezi{\'c}}, {\v{Z}}., {Beers}, T.~C., \& {Juri{\'c}}, M. 2012, \araa, 50, 251

\bibitem[{{Johnson} \& {Soderblom}(1987)}]{JohnsonSoderblom1987}
{Johnson}, D.~R.~H., \& {Soderblom}, D.~R. 1987, \aj, 93, 864

\bibitem[{{Kuijken} \& {Gilmore}(1989)}]{Kuijken1989}
{Kuijken}, K., \& {Gilmore}, G. 1989, \mnras, 239, 571

\bibitem[{Kullback \& Leibler(1951)}]{Kullback1951}
Kullback, S., \& Leibler, R.~A. 1951, Ann. Math. Statist., 22, 79

\bibitem[{{Li} {et~al.}(2016){Li}, {Balbinot}, {Mondrik}, {Marshall}, {Yanny},
  {Bechtol}, {Drlica-Wagner}, {Oscar}, {Santiago}, {Simon}, {Vivas}, {Walker},
  {Wang}, {Abbott}, {Abdalla}, {Benoit-L{\'e}vy}, {Bernstein}, {Bertin},
  {Brooks}, {Burke}, {Carnero Rosell}, {Carrasco Kind}, {Carretero}, {da
  Costa}, {DePoy}, {Desai}, {Diehl}, {Doel}, {Estrada}, {Finley}, {Flaugher},
  {Frieman}, {Gruen}, {Gruendl}, {Gutierrez}, {Honscheid}, {James}, {Kuehn},
  {Kuropatkin}, {Lahav}, {Maia}, {March}, {Martini}, {Ogando}, {Plazas},
  {Reil}, {Romer}, {Roodman}, {Sanchez}, {Scarpine}, {Schubnell},
  {Sevilla-Noarbe}, {Smith}, {Soares-Santos}, {Sobreira}, {Suchyta}, {Swanson},
  {Tarle}, {Tucker}, {Zhang}, \& {DES Collaboration}}]{Li2016}
{Li}, T.~S., {Balbinot}, E., {Mondrik}, N., {et~al.} 2016, \apj, 817, 135

\bibitem[{{Li} \& {Shen}(2019)}]{Li2019}
{Li}, Z.-Y., \& {Shen}, J. 2019, arXiv e-prints, arXiv:1904.03314

\bibitem[{{Lindegren} {et~al.}(2012){Lindegren}, {Lammers}, {Hobbs},
  {O'Mullane}, {Bastian}, \& {Hern{\'a}ndez}}]{Lindegren2012}
{Lindegren}, L., {Lammers}, U., {Hobbs}, D., {et~al.} 2012, \aap, 538, A78

\bibitem[{{Liu} {et~al.}(2020){Liu}, {Huang}, {Zhang}, {Xiang}, {Ren}, {Chen},
  {Yuan}, {Wang}, {Yang}, {Tian}, {Wang}, \& {Liu}}]{Liu2020}
{Liu}, G.~C., {Huang}, Y., {Zhang}, H.~W., {et~al.} 2020, \apjs, 247, 68

\bibitem[{{Ma} {et~al.}(2017){Ma}, {Hopkins}, {Wetzel}, {Kirby},
  {Angl{\'e}s-Alc{\'a}zar}, {Faucher-Gigu{\`e}re}, {Kere{\v s}}, \&
  {Quataert}}]{Ma2017}
{Ma}, X., {Hopkins}, P.~F., {Wetzel}, A.~R., {et~al.} 2017, \mnras, 467, 2430

\bibitem[{{Mackereth} {et~al.}(2019){Mackereth}, {Schiavon}, {Pfeffer},
  {Hayes}, {Bovy}, {Anguiano}, {Allende Prieto}, {Hasselquist}, {Holtzman},
  {Johnson}, {Majewski}, {O'Connell}, {Shetrone}, {Tissera}, \&
  {Fern{\'a}ndez-Trincado}}]{Mackereth2019}
{Mackereth}, J.~T., {Schiavon}, R.~P., {Pfeffer}, J., {et~al.} 2019, \mnras,
  482, 3426

\bibitem[{{Martin} {et~al.}(2018){Martin}, {Amy}, {Newberg}, {Shelton},
  {Carlin}, {Beers}, {Denissenkov}, \& {Willett}}]{Martin2018}
{Martin}, C., {Amy}, P.~M., {Newberg}, H.~J., {et~al.} 2018, \mnras, 477, 2419

\bibitem[{{Muraveva} {et~al.}(2018){Muraveva}, {Delgado}, {Clementini},
  {Sarro}, \& {Garofalo}}]{Muraveva2018}
{Muraveva}, T., {Delgado}, H.~E., {Clementini}, G., {Sarro}, L.~M., \&
  {Garofalo}, A. 2018, \mnras, 481, 1195

\bibitem[{{Neeley} {et~al.}(2019){Neeley}, {Marengo}, {Freedman}, {Madore},
  {Beaton}, {Hatt}, {Hoyt}, {Monson}, {Rich}, {Sarajedini}, {Seibert}, \&
  {Scowcroft}}]{Neeley2019}
{Neeley}, J.~R., {Marengo}, M., {Freedman}, W.~L., {et~al.} 2019, \mnras, 490,
  4254

\bibitem[{{Newberg} {et~al.}(2010){Newberg}, {Willett}, {Yanny}, \&
  {Xu}}]{Newberg2010}
{Newberg}, H.~J., {Willett}, B.~A., {Yanny}, B., \& {Xu}, Y. 2010, \apj, 711,
  32

\bibitem[{{Newberg} {et~al.}(2009){Newberg}, {Yanny}, \&
  {Willett}}]{Newberg2009}
{Newberg}, H.~J., {Yanny}, B., \& {Willett}, B.~A. 2009, \apjl, 700, L61

\bibitem[{Pedregosa {et~al.}(2011)Pedregosa, Varoquaux, Gramfort, Michel,
  Thirion, Grisel, Blondel, Prettenhofer, Weiss, Dubourg, Vanderplas, Passos,
  Cournapeau, Brucher, Perrot, \& Duchesnay}]{scikit-learn}
Pedregosa, F., Varoquaux, G., Gramfort, A., {et~al.} 2011, Journal of Machine
  Learning Research, 12, 2825

\bibitem[{Penarrubia {et~al.}(2008)Penarrubia, McConnachie, \&
  Navarro}]{Penarrubia2008}
Penarrubia, J., McConnachie, A.~W., \& Navarro, J.~F. 2008, The Astrophysical
  Journal, 672, 904

\bibitem[{{Pila-D{\'\i}ez} {et~al.}(2015){Pila-D{\'\i}ez}, {de Jong},
  {Kuijken}, {van der Burg}, \& {Hoekstra}}]{Pila-Diez2015}
{Pila-D{\'\i}ez}, B., {de Jong}, J.~T.~A., {Kuijken}, K., {van der Burg},
  R.~F.~J., \& {Hoekstra}, H. 2015, \aap, 579, A38

\bibitem[{{Plummer}(1911)}]{Plummer1911}
{Plummer}, H.~C. 1911, \mnras, 71, 460

\bibitem[{{Quinn} \& {Goodman}(1986)}]{Quinn1986}
{Quinn}, P.~J., \& {Goodman}, J. 1986, \apj, 309, 472

\bibitem[{{Riello} {et~al.}(2018){Riello}, {De Angeli}, {Evans}, {Busso},
  {Hambly}, {Davidson}, {Burgess}, {Montegriffo}, {Osborne}, {Kewley},
  {Carrasco}, {Fabricius}, {Jordi}, {Cacciari}, {van Leeuwen}, \&
  {Holland}}]{Riello2018}
{Riello}, M., {De Angeli}, F., {Evans}, D.~W., {et~al.} 2018, \aap, 616, A3

\bibitem[{{Rodriguez Wimberly} {et~al.}(2019){Rodriguez Wimberly}, {Cooper},
  {Fillingham}, {Boylan-Kolchin}, {Bullock}, \&
  {Garrison-Kimmel}}]{RodriguezWimberly2019}
{Rodriguez Wimberly}, M.~K., {Cooper}, M.~C., {Fillingham}, S.~P., {et~al.}
  2019, \mnras, 483, 4031

\bibitem[{{Ruiz-Lara} {et~al.}(2020){Ruiz-Lara}, {Gallart}, {Bernard}, \&
  {Cassisi}}]{Ruiz-Lara2020}
{Ruiz-Lara}, T., {Gallart}, C., {Bernard}, E.~J., \& {Cassisi}, S. 2020, Nature
  Astronomy, arXiv:2003.12577

\bibitem[{{Sanderson} {et~al.}(2017){Sanderson}, {Hartke}, \&
  {Helmi}}]{Sanderson2017}
{Sanderson}, R.~E., {Hartke}, J., \& {Helmi}, A. 2017, \apj, 836, 234

\bibitem[{{Sanderson} \& {Helmi}(2013)}]{Sanderson2013}
{Sanderson}, R.~E., \& {Helmi}, A. 2013, \mnras, 435, 378

\bibitem[{Schwarz(1978)}]{bic}
Schwarz, G. 1978, Ann. Statist., 6, 461

\bibitem[{{Sesar}(2012)}]{Sesar2012}
{Sesar}, B. 2012, \aj, 144, 114

\bibitem[{{Sesar} {et~al.}(2011){Sesar}, {Juri{\'c}}, \&
  {Ivezi{\'c}}}]{Sesar2011}
{Sesar}, B., {Juri{\'c}}, M., \& {Ivezi{\'c}}, {\v{Z}}. 2011, \apj, 731, 4

\bibitem[{Shelton(2018)}]{SheltonThesis}
Shelton, S. 2018, PhD thesis, Rensselaer Polytechnic Institute

\bibitem[{Simion {et~al.}(2019)Simion, Belokurov, \& Koposov}]{Simion2019}
Simion, I.~T., Belokurov, V., \& Koposov, S.~E. 2019, Monthly Notices of the
  Royal Astronomical Society, 482, 921

\bibitem[{{Snaith} {et~al.}(2014){Snaith}, {Haywood}, {Di Matteo}, {Lehnert},
  {Combes}, {Katz}, \& {G{\'o}mez}}]{Snaith2014}
{Snaith}, O.~N., {Haywood}, M., {Di Matteo}, P., {et~al.} 2014, \apjl, 781, L31

\bibitem[{{Sohn} {et~al.}(2016){Sohn}, {van der Marel}, {Kallivayalil},
  {Majewski}, {Besla}, {Carlin}, {Law}, {Siegel}, \& {Anderson}}]{Sohn2016}
{Sohn}, S.~T., {van der Marel}, R.~P., {Kallivayalil}, N., {et~al.} 2016, \apj,
  833, 235

\bibitem[{Stephens(1974)}]{adtest2}
Stephens, M.~A. 1974, Journal of the American Statistical Association, 69, 730

\bibitem[{Storn \& Price(1997)}]{Storn1997}
Storn, R., \& Price, K. 1997, J. of Global Optimization, 11, 341

\bibitem[{{Taylor}(2005)}]{TOPCAT}
{Taylor}, M.~B. 2005, in Astronomical Society of the Pacific Conference Series,
  Vol. 347, Astronomical Data Analysis Software and Systems XIV, ed.
  P.~{Shopbell}, M.~{Britton}, \& R.~{Ebert}, 29

\bibitem[{{Velazquez} \& {White}(1999)}]{Velazquez1999}
{Velazquez}, H., \& {White}, S. D.~M. 1999, \mnras, 304, 254

\bibitem[{{Villalobos} \& {Helmi}(2008)}]{Villalobos2008}
{Villalobos}, {\'A}., \& {Helmi}, A. 2008, \mnras, 391, 1806

\bibitem[{{Vivas} {et~al.}(2016){Vivas}, {Zinn}, {Farmer}, {Duffau}, \&
  {Ping}}]{Vivas2016}
{Vivas}, A.~K., {Zinn}, R., {Farmer}, J., {Duffau}, S., \& {Ping}, Y. 2016,
  \apj, 831, 165

\bibitem[{Vivas {et~al.}(2001)Vivas, Zinn, Andrews, Bailyn, Baltay, Coppi,
  Ellman, Girard, Rabinowitz, Schaefer, Shin, Snyder, Sofia, van Altena, Abad,
  Bongiovanni, Briceño, Bruzual, Prugna, Herrera, Magris, Mateu, Pacheco,
  Sánchez, Sánchez, Schenner, Stock, Vicente, Vieira, Ferrín, Hernandez,
  Gebhard, Honeycutt, Mufson, Musser, \& Rengstorf}]{Vivas2001}
Vivas, A.~K., Zinn, R., Andrews, P., {et~al.} 2001, \apjl, 554, L33

\bibitem[{{Watkins} {et~al.}(2009){Watkins}, {Evans}, {Belokurov}, {Smith},
  {Hewett}, {Bramich}, {Gilmore}, {Irwin}, {Vidrih}, {Wyrzykowski}, \&
  {Zucker}}]{Watkins2009}
{Watkins}, L.~L., {Evans}, N.~W., {Belokurov}, V., {et~al.} 2009, \mnras, 398,
  1757

\bibitem[{Weiss {et~al.}(2018b)Weiss, Newberg, \& Desell}]{Weiss2018b}
Weiss, J., Newberg, H.~J., \& Desell, T. 2018b, The Astrophysical Journal
  Letters, 867, L1

\bibitem[{{Widrow} {et~al.}(2012){Widrow}, {Gardner}, {Yanny}, {Dodelson}, \&
  {Chen}}]{Widrow2012}
{Widrow}, L.~M., {Gardner}, S., {Yanny}, B., {Dodelson}, S., \& {Chen}, H.-Y.
  2012, \apjl, 750, L41

\bibitem[{{Williams} {et~al.}(2013){Williams}, {Steinmetz}, {Binney},
  {Siebert}, {Enke}, {Famaey}, {Minchev}, {de Jong}, {Boeche}, {Freeman},
  {Bienaym{\'e}}, {Bland-Hawthorn}, {Gibson}, {Gilmore}, {Grebel}, {Helmi},
  {Kordopatis}, {Munari}, {Navarro}, {Parker}, {Reid}, {Seabroke}, {Sharma},
  {Siviero}, {Watson}, {Wyse}, \& {Zwitter}}]{Williams2013}
{Williams}, M.~E.~K., {Steinmetz}, M., {Binney}, J., {et~al.} 2013, \mnras,
  436, 101

\bibitem[{{Xu} {et~al.}(2015){Xu}, {Newberg}, {Carlin}, {Liu}, {Deng}, {Li},
  {Sch{\"o}nrich}, \& {Yanny}}]{Xu2015}
{Xu}, Y., {Newberg}, H.~J., {Carlin}, J.~L., {et~al.} 2015, \apj, 801, 105

\bibitem[{{Xue} {et~al.}(2015){Xue}, {Rix}, {Ma}, {Morrison}, {Bovy}, {Sesar},
  \& {Janesh}}]{Xue2015}
{Xue}, X.-X., {Rix}, H.-W., {Ma}, Z., {et~al.} 2015, \apj, 809, 144

\bibitem[{{Yanny} \& {Gardner}(2013)}]{Yanny2013}
{Yanny}, B., \& {Gardner}, S. 2013, \apj, 777, 91

\bibitem[{{Yanny} {et~al.}(2000){Yanny}, {Newberg}, {Kent},
  {Laurent-Muehleisen}, {Pier}, {Richards}, {Stoughton}, {Anderson}, {Annis},
  {Brinkmann}, {Chen}, {Csabai}, {Doi}, {Fukugita}, {Hennessy}, {Ivezi{\'c}},
  {Knapp}, {Lupton}, {Munn}, {Nash}, {Rockosi}, {Schneider}, {Smith}, \&
  {York}}]{Yanny2000}
{Yanny}, B., {Newberg}, H.~J., {Kent}, S., {et~al.} 2000, \apj, 540, 825

\bibitem[{{Zinn} {et~al.}(2020){Zinn}, {Chen}, {Layden}, \&
  {Casetti-Dinescu}}]{Zinn2020}
{Zinn}, R., {Chen}, X., {Layden}, A.~C., \& {Casetti-Dinescu}, D.~I. 2020,
  \mnras, 492, 2161

\end{thebibliography}

\appendix
\section{Uncertainty in Causticality}\label{app:cau_uncer}

Here, we derive the uncertainty in causticality as defined in Equation \ref{eq:causticality}. Assuming Poisson error for each particular $N_i$ = $N_j$, \begin{align}
    \sigma_C = \sqrt{\sum_{j=2}^{n-1} \Big( \frac{\partial C}{\partial N_j} \Big)^2 \sigma_{N_j}^2}, \textrm{ where } \sigma^2_{N_j} = N_j,
\end{align} and the partial derivative of the causticality with respect to each $N_j$ can be written as \begin{align}
     \frac{\partial C}{\partial N_j} = \frac{4(1-C)N_j - 2(1+C)(N_{j-1} + N_{j+1}) }{\sum\limits_{i=2}^n (N_i + N_{i-1})^2}
\end{align} Note that if all bins are multiplied by a factor of $m$, the uncertainty in causticality becomes $\sigma_C/\sqrt{m}$. Thus, while the value of causticality remains unchanged when the overall number of data points is increased, the corresponding uncertainty in the causticality decreases. Note that the first and final bins are omitted in the calculation of the uncertainty in causticality. This is because $N_{j-1}$ is ill-defined for the first bin, and $N_{j+1}$ is likewise ill-defined for the final bin. This rarely impacts the data in practice, as the first and final bins are typically zero. 

\begin{figure}
\center
\includegraphics[width=0.5\linewidth]{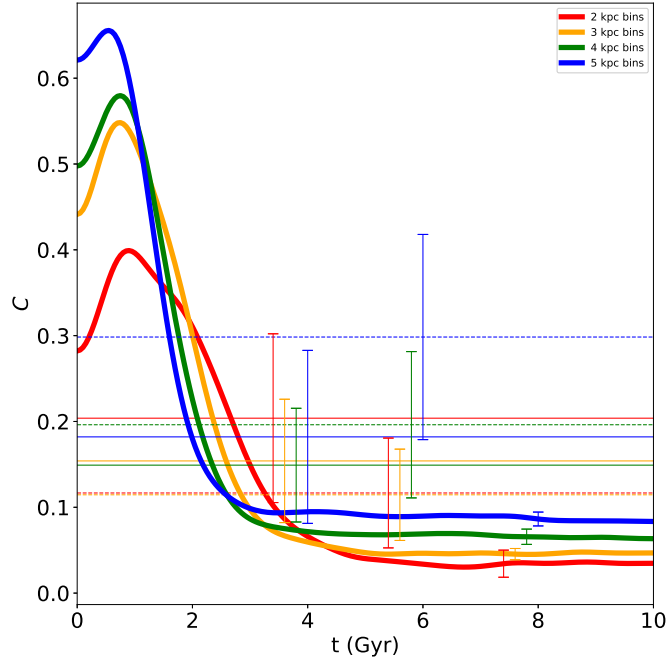}
\caption{Causticality vs. time for the radial merger simulation with $i$ = 40$^\circ$, $r_0$ = 30 kpc, and a mass of $10^9M_\odot$. The color of each line denotes the bin size used to calculate causticality. Thick solid lines are causticality for the simulated data, while the thin solid and dashed horizontal lines are the measured causticality of the VOD and HAC region data, respectively. The shape of the causticality curve does not change much with bin size, but the location of the curve moves up and to the left as bin size increases. The left group of error bars provide the Poisson errors in the measured causticality for the VOD region, the center group provide Poisson errors for the HAC region, and the right group provides the Poisson errors for the simulated data at the time where each error bar is located. The Poisson errors in the simulated data are much smaller than the observed data, as there are more objects in the simulated data.  \label{fig:cau_errors}}
\end{figure}

It is beneficial to get an idea of the effect of bin size on our measurements of causticality. Figure \ref{fig:cau_errors} shows causticality for a single radial merger simulation and the observed data, calculated with different histogram bin sizes. Also shown in this figure are the one sigma uncertainties of the causticality for the observed data. We only explore bin sizes $\geq2$ kpc, since the uncertainty in the distances of the observed data is $\sim$2 kpc at larger distances. For larger bin widths, the width of a bin is much larger than the expected width of a shell ($\sim$2-3 kpc), and information is lost regarding the shape of the shell structure. We support 2 kpc as a bin width, as it is large enough to eliminate many errors due to observational uncertainties, but is small enough to resolve shell structure.

Note that the general shape of the causticality curve is similar regardless of the bin size. A smaller bin size tends to push the equilibrium causticality value closer to zero, and the overall position of the curve is moved to the right as bin size decreases. Curves with larger bin sizes tend to ``bottom out'' and reach apparent equilibrium more quickly, as smaller shell structures are lost in wide bins.

Figure \ref{fig:cau_errors} also shows the one sigma error bars in the causticality values for each region, for each bin size. The uncertainties in the observed data (left and center error bars) are fairly large due to small number statistics. The uncertainties in the simulated data are much smaller, as the simulated data has many more objects than the observed data. The merger time estimates from both observed data sets are consistent with the 2.7 Gyr ago result in this work for this single simulation. The uncertainties in the data correspond to an uncertainty in merger time of around $\pm1$ Gyr. The measured causticality values for the observed data also vary substantially for different bin sizes, which is likely due to the small number of objects in the histograms of the observed data.

\section{Kullback-Leibler Divergence as a Metric for Phase Mixing}\label{app:kld}

In Section \ref{sec:constraint}, we use causticality as a metric for measuring the amount of phase mixing in simulations of radial mergers. Causticality is introduced in this publication, and has not been widely studied. For this reason, we repeat the analysis of the phase mixing constraints on the VRM using the Kullback-Leibler Divergence \citep[KLD,][]{Kullback1951} of the distribution. The KLD is a canonical, widely-studied statistic which measures how different one distribution is from another, and is defined as \begin{equation}\label{eq:kld}
D_{KL}(P || Q) = \sum P(r) \ln\Big(\frac{P(r)}{Q(r)} \Big)
\end{equation} for two discrete probability distributions, $P(r)$ and $Q(r)$. Each term in the KLD is defined to be equal to zero when $P(r) = 0$ or $Q(r)$ = 0, in order to avoid singularities in the natural logarithm. 

In this case, $P(r)$ is the observed histogram of shell stars as a function of Galactocentric radius, and $Q(r)$ is a uniform distribution equal to the mean of $P(r)$. The KLD is then a measure of how much information is gained by using $P(r)$ instead of a uniform distribution, or in other words, the KLD is a measurement of how sharp the peaks in $P(r)$ are. The value of the KLD will be larger when the peaks in $P(r)$ are large, and small for a relatively flat distribution of $P(r)$. A more phase mixed distribution will have smaller peaks than a distribution that has undergone less phase mixing, and therefore will have a smaller value of the KLD. KLD is analagous to causticality in this respect, as smaller values of both quantities correspond to measurements of more phase mixed distributions. We similarly expect the KLD to decrease over time for each simulation of a radial merger, until the merger is completely phase mixed.

\begin{figure}
\center
\includegraphics[width=\linewidth]{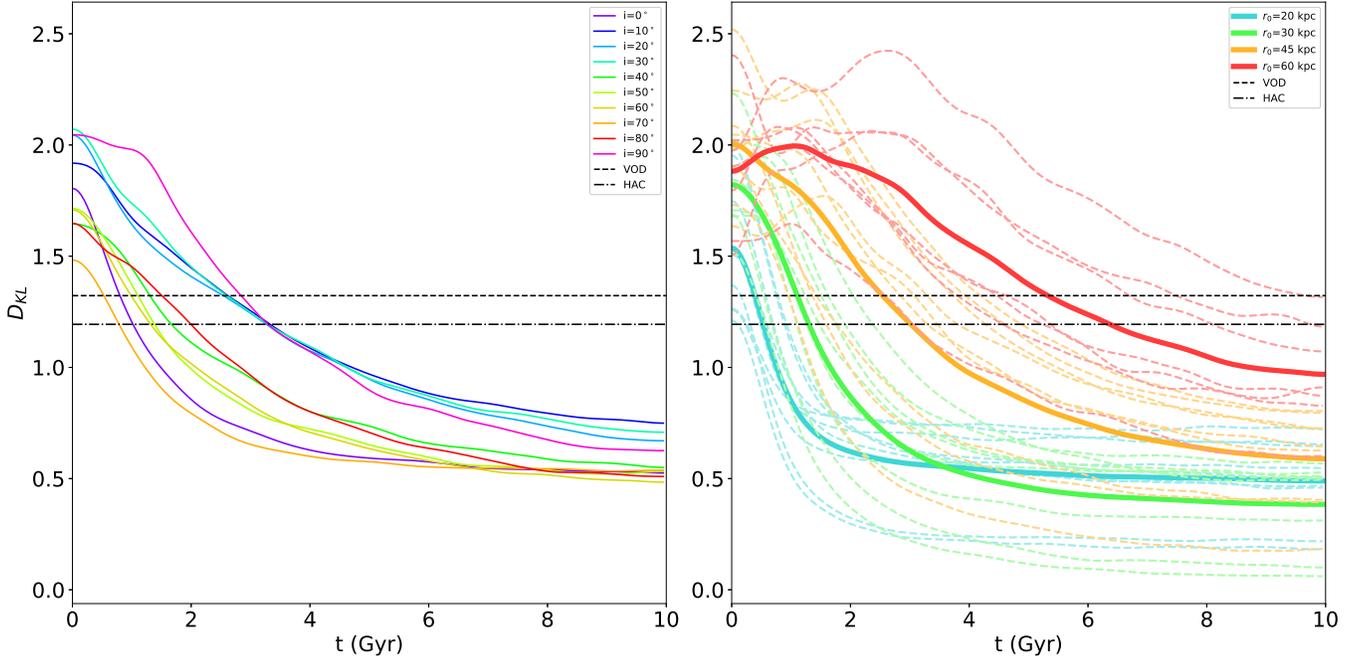}
\caption{Values of the KLD vs. time for all forty simulated radial mergers with a mass of $10^9M_\odot$. The KLD was calculated for each timestep for a radial histogram of all stars in the simulation, cut identically as in Figure \ref{fig:causticality}. Dashed horizontal lines give the values of the KLD for the observed data in the VOD region (1.32) and the HAC region (1.19). Observational errors will move the measured KLD of the observed data downwards, making the observed data appear older than it actually is. \textit{LEFT:} Mean KLD for simulations with a given initial inclination angle, averaged over the four initial distances.  Inclination angle does not appear to have a monotonic effect on how long shells take to phase mix; increasing the inclination angle could increase or decrease the mixing time. \textit{RIGHT:} Values of the KLD for simulations with a given initial distance. Mean values of the KLD are shown with thick solid lines, and are averaged over all inclination angles. Individual simulations are shown in thin dashed lines, and are colored by their initial distance. Simulations with larger initial distances take longer to phase mix on average than simulations with smaller initial distances.  \label{fig:kld}}
\end{figure}

Figure \ref{fig:kld} shows the measured values of the KLD over time for our radial merger simulations with a mass of $10^9M_\odot$, similarly to Figure \ref{fig:causticality}. As in Section \ref{sec:constraint}, we are primarily interested in the simulations $r_0$ = 20 kpc and $r_0$ = 30 kpc in Figure \ref{fig:kld}. 

One notices that the KLD of each simulation ``bottoms out'' at a lower bound and remains fairly constant for the rest of the simulation. The KLD does not bottom out at zero for our radial merger simulations because the KLD assumes that a mixed distribution is uniform, while histograms of relaxed radial merger simulations tend towards smooth Gaussian-like distributions. In theory, one could use a Gaussian distribution centered on $r_0$ as the denominator for the KLD in order to get a more accurate measurement of how phase mixed the distribution is. However, this requires knowledge of the phase-mixed distribution, and it may turn out that a Gaussian is not actually a good fit to the phase-mixed distribution. Simulations with smaller values of $r_0$ appear to bottom out more quickly than simulations with larger values of $r_0$. We anticipate that once a simulation approaches this lower bound, its overall structure does not change much, and it has effectively reached equilibrium. When a merger event reaches equilibrium, shells are no longer idenfitiable. Since we see shell structure in the observed data, the KLD of the VRM cannot have bottomed out yet. 

In Section \ref{sec:constraint}, we argued that the value of $r_0$ for the VRM is likely between 20 and 30 kpc. The radial merger simulations with these values of $r_0$ all bottom out before 5 Gyr. Since the VRM cannot have bottomed out yet, we claim that the VRM merger time occurred within the last 5 Gyr.

We seek to evaluate whether our choice of bin size has an adverse effect on our measurements of the KLD. In order to do this, we analytically calculate the value of the KLD for a continuous Gaussian distribution in our model, and then compare this value to the KLD for a binned approximation of a Gaussian in our model. We do this over a variety of Gaussian shapes and bin sizes.

Over a pair of continuous probability distributions, the KLD becomes\begin{align}
    D_{KL}(P||Q) = \int_{-\infty}^{\infty} P(r) \ln\Big(\frac{P(r)}{Q(r)}\Big) dr.
\end{align} In our case, our probability distributions are defined over $r=[0,l]$. Elsewhere, we take $P(r)$ to be zero. So, the KLD becomes:\begin{align}
    D_{KL}(P||Q) = \int_{0}^{l} P(r) \ln(l\cdot P(r)) dr,
\end{align} where we have used \begin{align}
	Q(r) = \frac{1}{l},
\end{align} in order to satisfy the normalization requirement \begin{align}
	\int_0^l Q(r) dr = 1.
\end{align}

Our model takes $P(r)$ to be a normalized Gaussian distribution:\begin{align}
    P(r) = \frac{1}{c\sqrt{2\pi}} \exp\Big[-\frac{1}{2}\Big(\frac{r-b}{c}\Big)^2\Big].
\end{align} The tails of $P(r)$ quickly approach zero as one moves away from its peak. This makes our KLD \begin{align}\label{eq:app_1}
     D_{KL}(P||Q) = \int_{0}^{l} P(r) \ln(l\cdot P(r)) dr \simeq \int_{-\infty}^{\infty} P(r) \ln(l\cdot P(r)) dr,
\end{align} as long as the Gaussian distribution is essentially within the discrete boundaries. Evaluating this integral leaves us with \begin{align}\label{eq:app_value}
     D_{KL}(P||Q) = \ln\Big[\frac{l}{c\sqrt{2\pi}}\Big] - \frac{1}{2},
\end{align} which only depends on the width of the Gaussian distribution. 

This value changes slightly for $N$ Gaussian distributions that do not overlap, which is a fair approximation of the simulated data. This change is easily computed: for $N$ Gaussians, the new $P(r)$ is given as \begin{align}
P(r) = \sum_i^N C_i P_i(r),
\end{align} where $P_i(r)$ is the probability distribution for the $i^{\textrm{th}}$ Gaussian, and the coefficients $C_i$ satisfy \begin{align}
\sum _i ^N C_i = 1.
\end{align} The integral in Equation \ref{eq:app_1} then splits into \begin{align}\label{eq:app_2}
D_{KL}(P||Q) = \sum_i^N \int_{-\infty}^{\infty} C_i P_i(r) \ln(l\cdot P(r)) dr.
\end{align} Since the Gaussians do not overlap, \begin{align}
D_{KL}(P||Q) = \sum_i^N \int_{-\infty}^{\infty} C_i P_i(r) \ln\Big(l\cdot C_i P_i(r)\Big) dr,
\end{align} because the integrand is required to disappear for all regions where $P(r)$ does not overlap with $P_i(r)$. If this was not the case, singularities would arise in the natural logarithm. This reduces to \begin{align}
D_{KL}(P||Q) = \sum_i^N \Big\lbrace C_i \int_{-\infty}^{\infty} P_i(r) \ln(l\cdot P_i(r))dr + C_i \ln (C_i)\int_{-\infty}^{\infty} P_i(r) rx\Big\rbrace.
\end{align} The first integral is equivalent to the value given in Equation \ref{eq:app_value}. The second integral is simply a normalized Gaussian integral, which is equal to unity. For $N$ Gaussian distributions with variances $c_i$, one obtains \begin{align}
D_{KL}(P||Q) = \sum_i^N C_i \Big\lbrace \ln\Big[\frac{l}{c_i\sqrt{2\pi}}\Big] - \frac{1}{2} + \ln(C_i)\Big\rbrace
\end{align}

\begin{figure}
\center
\includegraphics[width=\linewidth]{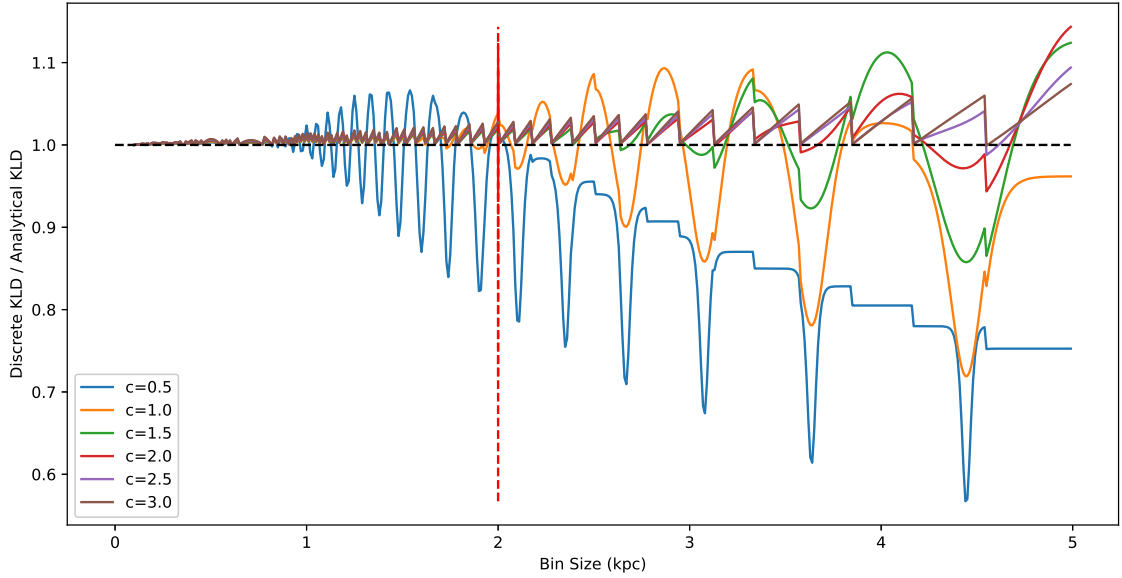}
\caption{The ratio of a binned KLD of a single Gaussian to the analytical (correct) value of the KLD with respect to bin size of the binned distribution. We mark the bin size used in this work, 2 kpc, with a dashed red line. For Gaussian widths $c \geq 1$ kpc, a 2 kpc bin size approximates the KLD within a few percent of the correct value. For smaller values of $c$, we begin to under-estimate the value of the KLD. Typical shell widths in our simulations are $\geq 0.8$ kpc, so we feel that our bin size is appropriate for the data. The variable periodicity in the values of the KLD have to do with changes in how many bins are used to approximate each Gaussian, which depends on bin size.  \label{fig:kld_bin_size}}
\end{figure}

Figure \ref{fig:kld_bin_size} shows the ratio of the KLD computed for a single binned Gaussian to the analytical KLD as a function of bin size. The bin size used in this work, 2 kpc, is marked with a dashed red line. For Gaussian models with values of $c \geq 1$ kpc, the binned KLD is equal to the analytical KLD to within a few percent when 2 kpc bins are used. As the variance of the Gaussian drops below 1 kpc, our binned KLD begins to under-approximate the actual value of the KLD by 10-20\% per Gaussian when using 2 kpc bin widths. A typical shell width in the radial merger simulations is between 0.8 and 1.2 kpc, so we conclude that our bin size is appropriate for the data.

While decreasing the bin width used in this work would provide a more accurate estimation of the KLD, we chose a bin width of 2 kpc as our average distance error was $\pm$ 0.5 kpc. Decreasing the size of the bins in the observed data would increase noise in the data, and decrease the quality of our fits to the observed data. We expect a bin width of 2 kpc to produce KLD estimates within a few percent of the actual values (Figure \ref{fig:kld_bin_size}). We also elected to keep the bin sizes of the simulated data equal to the bin sizes of the observed data in order to allow for easy comparison.

Note that if we are actually underestimating the values of the KLD calculated in Figure \ref{fig:kld}, then the distributions should be moved upwards. This will increase the time since collision at which the distribution appears similar to the observed data. However, since we show that the KLD is only off by a few percent per Gaussian component, the effect of this error on the estimated time since collision from phase mixing will be small (something like 10\%). This change in the KLD for the simulations with $r_0$ = 20 \& 30 kpc corresponds to a change of $<$1 Gyr in the upper constraint of the VRM merger time. Since we do not use phase mixing arguments to calculate the precise merger time of the VRM, this is still consistent with the conclusions of this work. Note that the addition of background halo stars in the simulations used in Figure \ref{fig:kld} would decrease the measured KLD, and partially cancel out the increase in the measured KLD due to binning. 

\newpage
\begin{deluxetable*}{lcccccccccc}
\tablecaption{Candidate shell stars in the VOD region.
\label{tab:voddata}}
\tablehead{
    Type & ID & R.A. & Dec. & $d_{hel}$ & $v_{GSR}$ & $\mu_\alpha$ & $\mu_\delta$ & $r_{gal}$ & $v_r$ & $L_z$ \\
    & & ($^\circ$) & ($^\circ$) & (kpc) & (km/s) & (mas/yr) & (mas/yr) & (kpc) & (km/s) & (kpc km/s)
}
\startdata
RRL & 2740 & 181.59 & 2.0 & 7.43 & 17 & -1.14 & 0.04 & 9.53 & 3 & 400 \\ 
RRL & 3694880517013152256 & 186.04 & -1.1 & 25.57 & -15 & -0.17 & 0.01 & 25.57 & -9 & -235 \\ 
RRL & 3681530762224477696 & 187.88 & -3.27 & 14.14 & -18 & -0.49 & -0.45 & 14.78 & -8 & 41 \\ 
RRL & 3697444371970657792 & 188.25 & 2.1 & 16.41 & 17 & 0.03 & -0.4 & 17.05 & 7 & 287 \\ 
RRL & 2744 & 188.25 & 2.1 & 15.65 & 14 & 0.1 & -0.29 & 16.21 & -3 & 95 \\ 
RRL & 3931754693600423296 & 188.42 & 12.81 & 55.15 & -1 & -0.17 & -0.04 & 55.17 & 4 & -320 \\ 
RRL & 3902831765353826176 & 188.71 & 8.83 & 16.5 & 18 & 0.15 & -0.56 & 17.52 & 2 & 384 \\ 
RRL & 3932634096743777536 & 189.76 & 14.67 & 19.19 & -17 & -0.44 & 0.06 & 20.24 & -1 & -64 \\ 
RRL & 2681 & 190.16 & -6.8 & 18.3 & -2 & -0.17 & -0.14 & 25.17 & 7 & 372 \\ 
RRL & 3696258475664594688 & 190.19 & 0.84 & 28.93 & 2 & -0.38 & -0.38 & 28.42 & 7 & -64 \\ 
RRL & 3705706686457241728 & 191.47 & 4.69 & 15.43 & 13 & -0.14 & -0.61 & 16.0 & 4 & 313 \\ 
RRL & 3689968964211832448 & 191.94 & 0.23 & 13.95 & -4 & -0.12 & -0.46 & 14.38 & -8 & 163 \\ 
RRL & 2828 & 192.34 & 7.75 & 36.36 & -27 & -0.28 & 0.38 & 36.7 & 7 & -227 \\ 
BHB & 3704495986715838208 & 193.41 & 3.16 & 8.62 & 2 & -0.48 & -0.99 & 10.33 & -2 & 268 \\ 
RRL & 2864 & 194.31 & 9.27 & 16.96 & -45 & 0.03 & -0.09 & 24.68 & -2 & -135 \\ 
BHB & 3691105756155452672 & 195.98 & 1.4 & 9.69 & -41 & -1.33 & -0.89 & 10.69 & -6 & 137 \\ 
RRL & 3617 & 196.96 & 12.56 & 19.8 & 29 & -0.15 & 0.05 & 23.51 & 1 & -6 \\ 
RRL & 3366 & 197.82 & -0.48 & 15.96 & -36 & 0.8 & 0.06 & 13.53 & 9 & 245 \\ 
RRL & 3686287623187964800 & 199.19 & -1.81 & 19.97 & 9 & 0.31 & 0.1 & 18.7 & 0 & 136 \\ 
RRL & 3687531827969154560 & 199.42 & 0.22 & 13.21 & 14 & 0.49 & 0.06 & 12.91 & -3 & 58 \\ 
RRL & 3565 & 199.58 & 9.33 & 14.78 & -44 & 0.23 & 1.37 & 13.71 & 3 & -126 \\ 
RRL & 3332 & 200.42 & -2.63 & 19.37 & 37 & -0.36 & 0.31 & 13.93 & 0 & 171 \\ 
RRL & 3488 & 202.11 & 5.86 & 19.35 & 46 & -0.62 & 0.37 & 24.38 & -5 & 147 \\ 
BHB & 3713136670641272704 & 202.69 & 4.21 & 11.06 & 46 & 1.77 & 0.63 & 11.14 & -6 & -106 \\ 
RRL & 3397 & 203.72 & 0.59 & 7.19 & 84 & -4.49 & 2.02 & 9.67 & -7 & 320 \\ 
RRL & 3637352243286265856 & 203.95 & -2.62 & 17.98 & 14 & -0.02 & -0.38 & 16.26 & 5 & 118 \\ 
BHB & 3726199074937460736 & 205.28 & 10.09 & 12.69 & 28 & 0.24 & -0.89 & 12.56 & -3 & 217 \\ 
BHB & 3662317754306076160 & 206.67 & -0.01 & 26.47 & -3 & 0.1 & 0.2 & 24.04 & -3 & 79 \\ 
RRL & 3463 & 206.78 & 4.67 & 20.6 & 35 & 0.13 & -0.08 & 25.36 & -1 & 28 \\ 
BHB & 3658276744131704832 & 207.17 & -2.31 & 15.39 & -1 & -0.49 & -0.69 & 13.64 & 1 & 35 \\ 
BHB & 3657800170264903680 & 208.75 & -2.49 & 17.28 & -10 & -0.41 & -0.35 & 15.09 & -3 & -89 \\ 
BHB & 3618850490542779776 & 209.04 & -8.23 & 9.35 & -74 & -0.22 & 1.83 & 8.44 & -6 & -464 \\ 
RRL & 3619212294292093952 & 209.55 & -8.18 & 26.89 & 0 & 0.18 & 0.51 & 23.53 & 3 & 312 \\ 
RRL & 4003 & 210.84 & -10.29 & 6.27 & -44 & 3.47 & -2.61 & 10.28 & 2 & -201 \\ 
RRL & 3959 & 211.16 & -1.79 & 15.45 & -26 & 0.19 & -0.2 & 20.3 & 5 & -3 \\ 
BHB & 3661068369794896768 & 212.08 & 1.31 & 12.24 & 2 & 0.09 & -0.1 & 10.77 & -3 & 18 \\ 
BHB & 3660278572553385216 & 212.61 & 0.96 & 12.85 & 15 & -0.41 & -1.35 & 11.16 & -5 & 183 \\ 
RRL & 3981 & 213.71 & -2.75 & 28.98 & 28 & -0.34 & 0.24 & 34.11 & 4 & -296 \\ 
RRL & 4000 & 214.21 & -8.05 & 4.79 & 44 & -2.39 & -0.05 & 11.06 & -7 & 479 \\ 
BHB & 3673720686318314240 & 215.59 & 7.79 & 41.31 & 9 & 0.22 & -0.09 & 38.1 & 0 & 235 \\ 
BHB & 3649557165951306752 & 217.41 & -1.77 & 21.04 & -23 & -0.44 & -0.08 & 17.56 & -10 & -255
\enddata

\tablecomments{All velocities and proper motions have had the solar reflex motion removed. The ID assigned by \cite{Liu2020} is provided when a star came from their catalog. Otherwise, the 19-digit \gaia \textbf{source\_id} is given as an identifier.}

\end{deluxetable*}

\newpage
\begin{deluxetable*}{lcccccccccc}
\tablecaption{Candidate shell stars in the HAC region.
\label{tab:hacdata}}
\tablehead{
    Type & ID & R.A. & Dec. & $d_{hel}$ & $v_{GSR}$ & $\mu_\alpha$ & $\mu_\delta$ & $r_{gal}$ & $v_r$ & $L_z$ \\
    & & ($^\circ$) & ($^\circ$) & (kpc) & (km/s) & (mas/yr) & (mas/yr) & (kpc) & (km/s) & (kpc km/s)
}
\startdata
RRL & 1196273946718241024 & 237.58 & 15.86 & 9.85 & -23 & -1.85 & -0.53 & 8.1 & 5 & 266 \\ 
RRL & 4711 & 238.21 & 26.91 & 25.72 & 2 & -0.05 & -0.06 & 20.43 & 5 & -26 \\ 
BHB & 1191634488685212800 & 239.18 & 13.47 & 27.78 & -7 & 0.19 & 0.1 & 23.44 & -7 & 253 \\ 
RRL & 4454989329251784064 & 239.47 & 9.59 & 11.29 & 83 & -2.03 & -3.36 & 8.19 & -7 & -315 \\ 
BHB & 1383512732452176896 & 240.16 & 41.92 & 45.42 & -8 & -1.05 & -0.26 & 44.02 & -1 & 385 \\ 
BHB & 1379462651307214208 & 240.55 & 38.77 & 26.7 & -11 & -0.11 & 0.03 & 25.38 & -8 & 174 \\ 
BHB & 1386004779852419328 & 240.79 & 43.72 & 10.53 & -7 & -0.0 & 0.38 & 11.62 & 6 & 103 \\ 
RRL & 1379489073945967744 & 241.35 & 39.08 & 8.82 & -29 & 0.29 & 0.71 & 9.9 & -1 & 128 \\ 
BHB & 4453428950452824192 & 243.02 & 9.99 & 12.97 & 101 & 4.64 & 0.21 & 9.24 & 1 & -373 \\ 
BHB & 4453487774325208448 & 243.12 & 10.59 & 23.67 & -7 & -0.26 & -0.14 & 18.94 & -6 & -233 \\ 
BHB & 1302909807058760064 & 243.53 & 24.8 & 12.8 & 50 & -6.46 & -3.73 & 10.93 & 8 & 27 \\ 
BHB & 1200510674256549120 & 243.9 & 17.85 & 8.02 & -45 & 0.97 & 0.83 & 6.97 & -8 & 43 \\ 
RRL & 1305272142150939392 & 245.12 & 27.05 & 26.08 & -5 & -0.41 & -0.28 & 23.08 & -9 & -411 \\ 
RRL & 1305220465104424960 & 245.26 & 26.94 & 8.4 & -12 & 2.58 & 1.4 & 8.14 & 5 & -260 \\ 
BHB & 4465226504059767424 & 245.28 & 16.38 & 9.33 & 26 & -2.16 & -1.58 & 7.38 & -5 & 56 \\ 
BHB & 1201084756765042944 & 245.32 & 19.29 & 24.57 & -1 & -0.45 & -0.17 & 20.68 & -1 & -289 \\ 
BHB & 4459959259247977600 & 246.45 & 12.33 & 8.33 & 5 & 0.39 & -0.12 & 6.33 & -8 & -52 \\ 
RRL & 4996 & 247.02 & 7.49 & 5.28 & -33 & 3.28 & -5.28 & 5.51 & -4 & 72 \\ 
RRL & 1297123558398503680 & 247.27 & 20.15 & 23.58 & 10 & -0.22 & -0.32 & 19.75 & 2 & -497 \\ 
RRL & 4993 & 247.4 & 7.2 & 7.63 & -163 & 4.19 & -6.07 & 3.06 & -10 & -21 \\ 
RRL & 1330968484805660800 & 248.32 & 36.8 & 28.8 & 2 & -0.24 & -0.1 & 26.83 & 1 & -150 \\ 
RRL & 1304493443103794304 & 248.82 & 27.17 & 26.29 & 1 & -0.05 & 0.04 & 23.15 & 3 & 62 \\ 
RRL & 1325177902522801664 & 248.99 & 34.01 & 30.76 & -2 & -0.13 & 0.11 & 28.34 & 4 & 351 \\ 
BHB & 4459171734045577600 & 249.38 & 11.32 & 25.49 & 6 & 0.21 & -0.06 & 20.44 & 1 & -23 \\ 
RRL & 1312598012032706688 & 249.47 & 31.33 & 14.41 & -33 & -0.34 & 0.39 & 12.84 & -9 & 399 \\ 
RRL & 1355750824060585088 & 251.03 & 39.88 & 21.61 & 0 & -0.45 & -0.06 & 20.36 & 2 & 87 \\ 
BHB & 1313324106319470080 & 252.61 & 32.27 & 14.05 & 8 & -0.44 & -0.55 & 12.56 & -10 & -218 \\ 
BHB & 4449230843258872064 & 253.31 & 13.37 & 14.27 & 4 & 0.86 & 0.41 & 10.19 & 8 & 149 \\ 
RRL & 4564859299966696704 & 253.51 & 21.24 & 20.02 & 9 & -0.32 & -0.32 & 16.32 & -0 & -390 \\ 
BHB & 4447791578241500032 & 253.61 & 11.58 & 26.67 & -9 & -0.58 & -0.06 & 21.45 & -5 & -446 \\ 
BHB & 4447770897975217792 & 253.64 & 11.28 & 8.32 & 50 & -5.08 & -2.27 & 5.86 & -5 & 89 \\ 
BHB & 4561332650781267200 & 253.92 & 19.35 & 9.71 & 25 & -3.84 & -1.67 & 7.61 & -10 & 40 \\ 
BHB & 4448291173134284288 & 254.47 & 12.09 & 4.73 & 88 & -5.24 & -0.15 & 5.54 & 9 & 52 \\ 
BHB & 1306870660259421568 & 255.4 & 27.76 & 11.41 & 50 & -0.79 & -0.88 & 9.83 & 10 & -381 \\ 
RRL & 1313544901997725824 & 255.62 & 32.66 & 25.06 & 3 & -0.1 & -0.25 & 22.5 & -5 & -496 \\ 
BHB & 1308320327684015360 & 255.77 & 27.68 & 18.58 & -5 & -0.42 & 0.24 & 15.8 & 7 & 277 \\ 
BHB & 4571771203718705536 & 257.53 & 23.93 & 17.41 & -19 & 0.14 & 0.23 & 14.2 & -9 & 272 \\ 
RRL & 5563 & 259.49 & 42.19 & 29.32 & -33 & 0.31 & -0.14 & 27.82 & -4 & -351 \\ 
BHB & 1348471331235718656 & 264.11 & 43.64 & 45.97 & -2 & -0.49 & 0.03 & 44.24 & 0 & 92 
\enddata

\tablecomments{All velocities and proper motions have had the solar reflex motion removed. The ID assigned by \cite{Liu2020} is provided when a star came from their catalog. Otherwise, the 19-digit \gaia \textbf{source\_id} is given as an identifier.}

\end{deluxetable*}

{\setlength\tabcolsep{1pt}
\FloatBarrier
\LTcapwidth=\textwidth
\begin{longtable}{rrrrrrrrrrrrrrrrrrr}
\caption{Recovered merger times and actual merger times for our series of radial merger simulations with a mass of $10^9M_\odot$. }
\label{tab:recovered} 
\endfirsthead
\hline
\hline
\multicolumn{4}{c}{Simulation} & \multicolumn{2}{c}{Newberg2010} & & \multicolumn{2}{c}{MWPot.2014} & & \multicolumn{4}{c}{Simulation} & \multicolumn{2}{c}{Newberg2010} & & \multicolumn{2}{c}{MWPot.2014} \\ 
$i$ & $r_0$ & $t_e$ & $t_m$ & $t_r$ & \# Sh.&  & $t_r$ & \# Sh. & & $i$ & $r_0$ & $t_e$ & $t_m$ & $t_r$ & \# Sh. & & $t_r$ & \# Sh.\\
($^\circ$) & (kpc) & (Gyr) & (Gyr) & (Gyr) & & & (Gyr) & & & ($^\circ$) & (kpc) & (Gyr) & (Gyr) & (Gyr) & & & (Gyr) & \\
\hline 
0 & 20 & 1.0 & 0.8 & 1.2 & 3 & & 1.3 & 3 & & 50 & 20 & 1.0 & 0.7 & 1.3 & 2 & & 0.9 & 2 \\ 
0 & 20 & 2.0 & 1.8 & 2.3 & 2 & & 2.9 & 2 & & 50 & 20 & 2.0 & 1.7 & 2.0 & 2 & & 1.9 & 2 \\ 
0 & 20 & 3.0 & 2.8 & ...   & ... & & 2.1 & 2 & & 50 & 20 & 3.0 & 2.7 & 2.9 & 2 & & 2.8 & 2 \\ 
0 & 20 & 4.0 & 3.8 & 1.9 & 2 & & ...   & ... & & 50 & 20 & 4.0 & 3.7 & 4.5 & 2 & & 4.1 & 3 \\ 
0 & 20 & 5.0 & 4.8 & 2.7 & 2 & & 1.9 & 2 & & 50 & 20 & 5.0 & 4.7 & 2.4 & 2 & & 4.8 & 2 \\ 
0 & 30 & 1.0 & ...   & ...   & ... & & ...   & ... & & 50 & 30 & 1.0 & 0.9 & 1.0 & 4 & & 0.8 & 3 \\ 
0 & 30 & 2.0 & ...   & ...   & ... & & ...   & ... & & 50 & 30 & 2.0 & 1.9 & 1.9 & 3 & & 1.8 & 4 \\ 
0 & 30 & 3.0 & ...   & ...   & ... & & ...   & ... & & 50 & 30 & 3.0 & 2.9 & 2.9 & 4 & & 2.5 & 2 \\ 
0 & 30 & 4.0 & ...   & ...   & ... & & ...   & ... & & 50 & 30 & 4.0 & 3.9 & 2.5 & 2 & & 3.8 & 2 \\ 
0 & 30 & 5.0 & ...   & ...   & ... & & ...   & ... & & 50 & 30 & 5.0 & 4.9 & 4.7 & 2 & & ...   & ... \\ 
0 & 45 & 1.0 & 0.9 & 1.4 & 2 & & 0.8 & 2 & & 50 & 45 & 1.0 & 0.9 & 1.0 & 2 & & 0.8 & 2 \\ 
0 & 45 & 2.0 & 1.9 & 1.9 & 3 & & 1.7 & 4 & & 50 & 45 & 2.0 & 1.9 & 1.9 & 4 & & 1.6 & 2 \\ 
0 & 45 & 3.0 & 2.9 & 1.3 & 3 & & 2.5 & 3 & & 50 & 45 & 3.0 & 2.9 & 2.9 & 2 & & 3.0 & 3 \\ 
0 & 45 & 4.0 & 3.9 & 3.9 & 3 & & 3.9 & 4 & & 50 & 45 & 4.0 & 3.9 & 4.2 & 2 & & 4.0 & 4 \\ 
0 & 45 & 5.0 & 4.9 & 4.9 & 3 & & 2.8 & 3 & & 50 & 45 & 5.0 & 4.9 & 3.0 & 3 & & ...   & ... \\ 
0 & 60 & 1.0 & ...   & ...   & ... & & ...   & ... & & 50 & 60 & 1.0 & ...   & ...   & ... & & ...   & ... \\ 
0 & 60 & 2.0 & ...   & ...   & ... & & ...   & ... & & 50 & 60 & 2.0 & 1.9 & 1.8 & 2 & & 2.1 & 2 \\ 
0 & 60 & 3.0 & ...   & ...   & ... & & ...   & ... & & 50 & 60 & 3.0 & 2.9 & 2.8 & 3 & & 2.5 & 3 \\ 
0 & 60 & 4.0 & ...   & ...   & ... & & ...   & ... & & 50 & 60 & 4.0 & 3.9 & 2.5 & 2 & & ...   & ... \\ 
0 & 60 & 5.0 & ...   & ...   & ... & & ...   & ... & & 50 & 60 & 5.0 & 4.9 & 4.7 & 2 & & 4.0 & 2 \\ 
10 & 20 & 1.0 & 0.7 & 0.9 & 4 & & 1.1 & 4 & & 60 & 20 & 1.0 & 0.7 & 2.0 & 2 & & ...   & ... \\ 
10 & 20 & 2.0 & ...   & ...   & ... & & ...   & ... & & 60 & 20 & 2.0 & 1.7 & 2.0 & 2 & & ...   & ... \\ 
10 & 20 & 3.0 & 2.7 & 3.1 & 2 & & 2.9 & 2 & & 60 & 20 & 3.0 & 2.7 & 3.4 & 2 & & 2.1 & 2 \\ 
10 & 20 & 4.0 & 3.7 & 3.1 & 2 & & 4.1 & 1 & & 60 & 20 & 4.0 & 3.7 & 3.8 & 2 & & 2.8 & 2 \\ 
10 & 20 & 5.0 & 4.7 & 4.9 & 2 & & 4.7 & 2 & & 60 & 20 & 5.0 & 4.7 & 4.7 & 2 & & ...   & ... \\ 
10 & 30 & 1.0 & 0.9 & 0.9 & 2 & & 0.9 & 2 & & 60 & 30 & 1.0 & 0.9 & 1.0 & 2 & & 0.9 & 2 \\ 
10 & 30 & 2.0 & 1.9 & 1.8 & 4 & & 1.6 & 4 & & 60 & 30 & 2.0 & 1.9 & ...   & ... & & 1.6 & 2 \\ 
10 & 30 & 3.0 & 2.9 & 1.7 & 3 & & ...   & ... & & 60 & 30 & 3.0 & 2.9 & ...   & ... & & 3.1 & 4 \\ 
10 & 30 & 4.0 & ...   & ...   & ... & & ...   & ... & & 60 & 30 & 4.0 & 3.9 & 3.9 & 2 & & 4.6 & 2 \\ 
10 & 30 & 5.0 & 4.9 & ...   & ... & & 3.4 & 2 & & 60 & 30 & 5.0 & 4.9 & 4.5 & 3 & & ...   & ... \\ 
10 & 45 & 1.0 & 0.7 & ...   & ... & & 1.1 & 3 & & 60 & 45 & 1.0 & 0.9 & 1.1 & 2 & & ...   & ... \\ 
10 & 45 & 2.0 & 1.7 & 1.8 & 2 & & 1.9 & 2 & & 60 & 45 & 2.0 & 1.9 & 2.1 & 3 & & 1.7 & 3 \\ 
10 & 45 & 3.0 & 2.7 & 2.7 & 4 & & 2.9 & 4 & & 60 & 45 & 3.0 & 2.9 & 3.1 & 3 & & 2.6 & 3 \\ 
10 & 45 & 4.0 & 3.7 & ...   & ... & & 3.7 & 4 & & 60 & 45 & 4.0 & 3.9 & 3.3 & 2 & & 2.7 & 2 \\ 
10 & 45 & 5.0 & 4.7 & 4.9 & 3 & & 4.7 & 3 & & 60 & 45 & 5.0 & 4.9 & 2.7 & 2 & & 2.3 & 2 \\ 
10 & 60 & 1.0 & 0.7 & 1.2 & 3 & & 1.6 & 3 & & 60 & 60 & 1.0 & 0.9 & 1.2 & 2 & & ...   & ... \\  
10 & 60 & 2.0 & ...   & ...   & ... & & ...   & ... & & 60 & 60 & 2.0 & ...   & ...   & ... & & ...   & ... \\  
10 & 60 & 3.0 & ...   & ...   & ... & & ...   & ... & & 60 & 60 & 3.0 & 2.9 & 2.8 & 3 & & 3.0 & 2 \\  
10 & 60 & 4.0 & 3.7 & 3.7 & 4 & & 3.6 & 4 & & 60 & 60 & 4.0 & 3.9 & 3.9 & 3 & & ...   & ... \\  
10 & 60 & 5.0 & 4.7 & 3.5 & 3 & & 3.7 & 3 & & 60 & 60 & 5.0 & 4.9 & 4.9 & 3 & & 4.7 & 3 \\  
20 & 20 & 1.0 & 0.7 & 0.9 & 4 & & 0.8 & 4 & & 70 & 20 & 1.0 & 0.5 & 0.8 & 2 & & 0.5 & 2 \\ 
20 & 20 & 2.0 & 1.7 & ...   & ... & & 2.0 & 2 & & 70 & 20 & 2.0 & 1.5 & ...   & ... & & 1.8 & 2 \\ 
20 & 20 & 3.0 & 2.7 & 2.9 & 2 & & 2.7 & 2 & & 70 & 20 & 3.0 & 2.5 & 3.8 & 2 & & ...   & ... \\ 
20 & 20 & 4.0 & 3.7 & 2.0 & 2 & & 2.3 & 2 & & 70 & 20 & 4.0 & 3.5 & 4.1 & 2 & & 3.9 & 2 \\ 
20 & 20 & 5.0 & 4.7 & 4.8 & 2 & & ...   & ... & & 70 & 20 & 5.0 & 4.5 & 4.7 & 2 & & 3.8 & 2 \\ 
20 & 30 & 1.0 & 0.8 & 1.0 & 2 & & 0.8 & 2 & & 70 & 30 & 1.0 & 0.7 & 1.1 & 2 & & 0.9 & 2 \\ 
20 & 30 & 2.0 & 1.8 & 2.8 & 2 & & 1.7 & 3 & & 70 & 30 & 2.0 & 1.7 & 2.1 & 3 & & 1.8 & 4 \\  
20 & 30 & 3.0 & 2.8 & 2.7 & 2 & & 2.7 & 2 & & 70 & 30 & 3.0 & 2.7 & 3.7 & 3 & & 3.0 & 4 \\  
20 & 30 & 4.0 & 3.8 & 3.4 & 2 & & 2.3 & 2 & & 70 & 30 & 4.0 & 3.7 & 3.9 & 3 & & 3.9 & 3 \\  
20 & 30 & 5.0 & 4.8 & 4.9 & 3 & & 2.4 & 2 & & 70 & 30 & 5.0 & 4.7 & 4.9 & 2 & & ...   & ... \\  
20 & 45 & 2.0 & ...   & ...   & ... & & ...   & ... & & 70 & 45 & 1.0 & ...   & ...   & ... & & ...   & ... \\  
20 & 45 & 2.0 & 1.8 & 1.8 & 2 & & 2.0 & 2 & & 70 & 45 & 2.0 & ...   & ...   & ... & & ...   & ... \\  
20 & 45 & 3.0 & 2.8 & 2.7 & 4 & & 2.6 & 4 & & 70 & 45 & 3.0 & ...   & ...   & ... & & ...   & ... \\  
20 & 45 & 4.0 & 3.8 & 3.7 & 4 & & 3.7 & 4 & & 70 & 45 & 4.0 & 3.9 & 2.5 & 2 & & ...   & ... \\  
20 & 45 & 5.0 & 4.8 & 4.8 & 3 & & 4.6 & 4 & & 70 & 45 & 5.0 & ...   & ...   & ... & & ...   & ... \\  
20 & 60 & 1.0 & 0.8 & 0.8 & 2 & & 1.2 & 2 & & 70 & 60 & 1.0 & ...   & ...   & ... & & ...   & ... \\  
20 & 60 & 2.0 & 0.8 & 0.8 & 2 & & ...   & ... & & 70 & 60 & 2.0 & ...   & ...   & ... & & ...   & ... \\  
20 & 60 & 3.0 & 2.8 & 2.6 & 3 & & 2.9 & 3 & & 70 & 60 & 3.0 & ...   & ...   & ... & & ...   & ... \\  
20 & 60 & 4.0 & 3.8 & 3.6 & 3 & & 3.7 & 3 & & 70 & 60 & 4.0 & ...   & ...   & ... & & ...   & ... \\  
20 & 60 & 5.0 & 4.8 & 4.6 & 3 & & 4.5 & 3 & & 70 & 60 & 5.0 & ...   & ...   & ... & & ...   & ... \\  
30 & 20 & 1.0 & 0.7 & 0.7 & 4 & & ...   & ... & & 80 & 20 & 1.0 & ...   & ...   & ... & & ...   & ... \\ 
30 & 20 & 2.0 & 1.7 & 3.1 & 2 & & 1.8 & 2 & & 80 & 20 & 2.0 & ...   & ...   & ... & & ...   & ... \\ 
30 & 20 & 3.0 & 2.7 & ...   & ... & & 4.5 & 2 & & 80 & 20 & 3.0 & ...   & ...   & ... & & ...   & ... \\ 
30 & 20 & 4.0 & 3.7 & 2.2 & 2 & & 2.5 & 2 & & 80 & 20 & 4.0 & ...   & ...   & ... & & ...   & ... \\ 
30 & 20 & 5.0 & 4.7 & 4.8 & 2 & & ...   & ... & & 80 & 20 & 5.0 & ...   & ...   & ... & & ...   & ... \\ 
30 & 30 & 1.0 & 0.9 & 1.0 & 2 & & 1.0 & 2 & & 80 & 30 & 1.0 & 0.9 & 0.8 & 2 & & 0.7 & 2 \\  
30 & 30 & 2.0 & 1.9 & 1.7 & 3 & & 1.5 & 3 & & 80 & 30 & 2.0 & 1.9 & 1.8 & 3 & & 1.5 & 3 \\  
30 & 30 & 3.0 & 2.9 & 3.1 & 3 & & 3.3 & 3 & & 80 & 30 & 3.0 & 2.9 & 2.7 & 3 & & 3.2 & 4 \\  
30 & 30 & 4.0 & 3.9 & 3.7 & 4 & & ...   & ... & & 80 & 30 & 4.0 & 3.9 & 3.3 & 4 & & 3.8 & 4 \\  
30 & 30 & 5.0 & 4.9 & 4.7 & 4 & & 4.5 & 4 & & 80 & 30 & 5.0 & 4.9 & 4.2 & 4 & & 4.5 & 4 \\  
30 & 45 & 1.0 & 0.9 & 0.9 & 2 & & 1.2 & 2 & & 80 & 45 & 1.0 & 0.4 & 1.1 & 3 & & 1.5 & 2 \\  
30 & 45 & 2.0 & 1.9 & 1.9 & 3 & & 2.1 & 2 & & 80 & 45 & 2.0 & 1.4 & 1.4 & 2 & & 1.7 & 2 \\  
30 & 45 & 3.0 & 2.9 & 2.9 & 4 & & 2.6 & 4 & & 80 & 45 & 3.0 & 2.4 & 3.0 & 2 & & 3.0 & 2 \\  
30 & 45 & 4.0 & 3.9 & 3.9 & 4 & & 3.4 & 3 & & 80 & 45 & 4.0 & 3.4 & 3.4 & 2 & & 4.0 & 2 \\  
30 & 45 & 5.0 & 4.9 & 4.8 & 4 & & 4.7 & 4 & & 80 & 45 & 5.0 & 4.4 & 5.0 & 2 & & 4.3 & 2 \\  
30 & 60 & 1.0 & 0.8 & 1.4 & 2 & & 1.2 & 3 & & 80 & 60 & 1.0 & 0.9 & 1.0 & 2 & & ...   & ... \\  
30 & 60 & 2.0 & 1.8 & 1.7 & 4 & & ...   & ... & & 80 & 60 & 2.0 & 1.9 & 1.9 & 3 & & 2.2 & 2 \\  
30 & 60 & 3.0 & 2.8 & 2.8 & 4 & & 2.4 & 3 & & 80 & 60 & 3.0 & 2.9 & 2.9 & 2 & & 3.0 & 2 \\  
30 & 60 & 4.0 & 3.8 & 3.6 & 4 & & 3.6 & 3 & & 80 & 60 & 4.0 & 3.9 & 3.9 & 3 & & ...   & ... \\  
30 & 60 & 5.0 & 4.8 & 4.4 & 4 & & 4.6 & 2 & & 80 & 60 & 5.0 & 4.9 & 4.5 & 3 & & 4.6 & 2 \\  
40 & 20 & 1.0 & ...   & ...   & ... & & ...   & ... & & 90 & 20 & 1.0 & ...   & ...   & ... & & ...   & ... \\ 
40 & 20 & 2.0 & ...   & ...   & ... & & ...   & ... & & 90 & 20 & 2.0 & ...   & ...   & ... & & ...   & ... \\ 
40 & 20 & 3.0 & ...   & ...   & ... & & ...   & ... & & 90 & 20 & 3.0 & ...   & ...   & ... & & ...   & ... \\ 
40 & 20 & 4.0 & ...   & ...   & ... & & ...   & ... & & 90 & 20 & 4.0 & ...   & ...   & ... & & ...   & ... \\ 
40 & 20 & 5.0 & ...   & ...   & ... & & ...   & ... & & 90 & 20 & 5.0 & ...   & ...   & ... & & ...   & ... \\ 
40 & 30 & 1.0 & 0.7 & 1.2 & 2 & & 1.0 & 2 & & 90 & 30 & 1.0 & ...   & ...   & ... & & ...   & ... \\  
40 & 30 & 2.0 & 1.7 & ...   & ... & & 2.8 & 3 & & 90 & 30 & 2.0 & ...   & ...   & ... & & ...   & ... \\  
40 & 30 & 3.0 & 2.7 & 2.9 & 2 & & 3.2 & 3 & & 90 & 30 & 3.0 & ...   & ...   & ... & & ...   & ... \\  
40 & 30 & 4.0 & 3.7 & 3.6 & 3 & & 4.3 & 3 & & 90 & 30 & 4.0 & ...   & ...   & ... & & ...   & ... \\  
40 & 30 & 5.0 & 3.7 & 3.6 & 3 & & ...   & ... & & 90 & 30 & 5.0 & ...   & ...   & ... & & ...   & ... \\  
40 & 45 & 1.0 & 0.8 & ...   & ... & & 0.9 & 2 & & 90 & 45 & 1.0 & 0.7 & ...   & ... & & 0.7 & 2 \\  
40 & 45 & 2.0 & 1.8 & 1.9 & 2 & & 2.2 & 3 & & 90 & 45 & 2.0 & 1.7 & 1.7 & 2 & & 1.9 & 2 \\  
40 & 45 & 3.0 & 2.8 & 3.0 & 3 & & 3.1 & 3 & & 90 & 45 & 3.0 & 2.7 & 2.6 & 3 & & 2.7 & 3 \\  
40 & 45 & 4.0 & 3.8 & 4.0 & 3 & & 4.0 & 3 & & 90 & 45 & 4.0 & 3.7 & 4.3 & 4 & & 3.6 & 4 \\  
40 & 45 & 5.0 & 4.8 & 4.3 & 3 & & 4.6 & 3 & & 90 & 45 & 5.0 & 4.7 & 3.1 & 2 & & 4.5 & 4 \\  
40 & 60 & 1.0 & 0.8 & 1.0 & 2 & & 2.9 & 3 & & 90 & 60 & 1.0 & ...   & ...   & ... & & ...   & ... \\  
40 & 60 & 2.0 & 1.8 & ...   & ... & & 2.1 & 2 & & 90 & 60 & 2.0 & ...   & ...   & ... & & ...   & ... \\  
40 & 60 & 3.0 & 2.8 & ...   & ... & & 3.0 & 2 & & 90 & 60 & 3.0 & ...   & ...   & ... & & ...   & ... \\  
40 & 60 & 4.0 & 3.8 & 4.3 & 2 & & 3.3 & 2 & & 90 & 60 & 4.0 & ...   & ...   & ... & & ...   & ... \\  
40 & 60 & 5.0 & 4.8 & 4.8 & 3 & & 4.7 & 3 & & 90 & 60 & 5.0 & ...   & ...   & ... & & ...   & ... \\ \hline
\end{longtable}

A description of the simulations is provided in Section \ref{sec:simulations}. A range of inclination angles ($i$), evolve times ($t_e$), and initial distances ($r_0$) were tested, and the results of the method described in Section \ref{sec:dating} are listed above. The actual merger time calculated by when the dwarf galaxy's center of mass passes through the Galactic center is $t_m$, and the recovered merger time calculated with our method is $t_r$. These values are also shown in Figure \ref{fig:recovery}. We show values calculated by both the \cite{Newberg2010} model potential, and the \textit{MWPotential2014} model. Simulations that did not generate shell structure and timesteps where the method was unable to recover a single merger time are given as blank rows. 74.5\% of all trials recovered a merger time in at least one of the model potentials. The majority of instances where merger times were not recovered was due to a lack of shell structure in the simulated data.
}

\newpage
\begin{deluxetable*}{rrrrrr}
\tablecaption{Times and values of local minima of $\delta(t)$, and number of model particles with $dr/dt>0$ at each time in Figure \ref{fig:osc}. \label{tab:times}}
\tablehead{
    \multicolumn{3}{c}{\textbf{Orphan Stream Model 5}} & \multicolumn{3}{c}{\textbf{MWPotential2014}} \\
    Time & $\delta(t)$ & \# $dr/dt>0$ & Time & $\delta(t)$ & \# $dr/dt>0$ \\
    (Gyr) & (kpc) & & (Gyr) & (kpc) &
}
\startdata
  0.2 & 51 & 1 & 0.2 & 54 & 1 \\
  0.5 & 65 & 2 & 0.5 & 58 & 2 \\
  2.4 & 62 & 1 & 2.4 & 47 & 3 \\
  \textbf{2.7} & \textbf{30} & \textbf{4} & \textbf{2.7} & \textbf{58} & \textbf{0} \\
  3.1 & 58 & 1 & 2.9 & 30 & 3 \\
  5.3 & 52 & 1 & 5.5 & 39 & 1 \\
  5.6 & 60 & 2 & 8.0 & 60 & 2 \\
  \textbf{8.2} & \textbf{49} & \textbf{0} & 8.3 & 45 & 1 \\
  8.5 & 33 & 3 & 8.9 & 60 & 1 \\
   & & & 8.9 & 57 & 2
\enddata

\tablecomments{A minimum is significant if it is at least two standard deviations from the mean of $\delta(t)$. Minima are given for both Milky Way potentials used in this work. Local minima with $dr/dt>0$ = 0 or 4 correspond to likely merger times, and are given in bold font. Both model potentials suggest 2.7 Gyr as a likely merger time. The 8.2 Gyr merger time in the Orphan Stream Model 5 potential is not a likely merger time, as it was not recovered in the second potential and it is outside of the phase mixing constraints for the merger time of the VRM.}

\end{deluxetable*}

\end{document}